\documentclass{aa}

\usepackage[hidelinks]{hyperref}
\usepackage{subcaption}
\usepackage{textgreek}
\usepackage{amsmath}
\usepackage{pdflscape}
\usepackage{caption}
\usepackage{bm}

\usepackage{subcaption}
\hypersetup{
  colorlinks   = true, 
  urlcolor     = blue, 
  linkcolor    = blue, 
  citecolor   = blue 
}
\usepackage{graphicx}
\usepackage{listings}
\usepackage{float}
\usepackage{txfonts}
\usepackage[]{hyperref}
\AtBeginDocument{\renewcommand{\textbf}[1]{#1}}

\begin{document}

   \title{Astrometric view of companions in the inner dust cavities of protoplanetary disks}


   \author{Miguel Vioque
          \inst{1}
          \and
          Richard A. Booth
          \inst{2}
          \and
          Enrico Ragusa
          \inst{3}  
          \and
          Álvaro Ribas
          \inst{4}          
          \and
          Nicolás T. Kurtovic
          \inst{5,6}          
          \and
          Giovanni P. Rosotti
          \inst{3}      
          \and
          Zephyr Penoyre
          \inst{7}                
          \and
          Stefano Facchini
          \inst{3}               
          \and
          Antonio Garufi
          \inst{8} 
          \and
          Carlo F. Manara
          \inst{1}                        
          \and
          Nuria Huélamo
          \inst{13} 
          \and
          Andrew Winter
          \inst{9}          
          \and
          Sebastián Pérez
          \inst{10,11,12}       
          \and
          Myriam Benisty
          \inst{6}              
          \and
          Ignacio Mendigutía
          \inst{13}                     
          \and
          Nicolás Cuello
          \inst{14}
          \and
          Anna B. T. Penzlin
          \inst{15}
          \and
          Alfred Castro-Ginard
          \inst{16}            
          \and
          Richard Teague
          \inst{17} }

   \institute{European Southern Observatory, Karl-Schwarzschild-Str. 2, 85748 Garching bei München, Germany\\
              \email{miguel.vioque@eso.org}
         \and
             School of Physics and Astronomy, University of Leeds, Leeds LS2 9JT, UK
         \and
             Dipartimento di Fisica, Università degli Studi di Milano, Via Celoria 16, 20133 Milano, Italy 
         \and
             Institute of Astronomy, University of Cambridge, Cambridge, UK      
         \and
             Max Planck Institute for Extraterrestrial Physics, Giessenbachstrasse 1, D-85748 Garching, Germany
         \and
             Max-Planck-Institut für Astronomie (MPIA), Königstuhl 17, 69117 Heidelberg, Germany
         \and
             Leiden Observatory, Leiden University, P.O. Box 9513, 2300 RA Leiden, the Netherlands   
         \and
             INAF - Istituto di Radioastronomia, Via Gobetti 101, I-40129, Bologna, Italy
         \and
             Astronomy Unit, School of Physics and Astronomy, Queen Mary University of London, London E1 4NS, UK           
         \and
             Millennium Nucleus on Young Exoplanets and their Moons (YEMS), Chile
         \and    
             Departamento de Física, Universidad de Santiago de Chile, Av. Víctor Jara 3659, Santiago, Chile
         \and    
             Center for Interdisciplinary Research in Astrophysics and Space Exploration (CIRAS), Universidad de Santiago de Chile, Chile
         \and
             Centro de Astrobiología (CAB), CSIC-INTA, ESAC Campus, Camino bajo del Castillo s/n, E-28692, Madrid, Spain
         \and
             Univ. Grenoble Alpes, CNRS, IPAG, 38000 Grenoble, France
         \and
              University Observatory, Faculty of Physics, Ludwig-Maximilians-Universität München, Munich, Germany
         \and
             Departament de Física Quàntica i Astrofísica (FQA), Universitat de Barcelona (UB), Martí i Franquès, 1, 08028 Barcelona, Spain
         \and
             Department of Earth, Atmospheric, and Planetary Sciences, Massachusetts Institute of Technology, Cambridge, MA 02139, USA 
                     }

   \date{Submitted on September 2, 2025; accepted for publication in A\&A on November 26, 2025.}

 
  \abstract
{Protoplanetary disks with inner dust cavities (often referred to as ``transition disks'') are potential signposts of planet formation. However, few companions have been identified within these cavities, and the role of companions in shaping them remains unclear.}
{We used \textit{Gaia} astrometry to search for planetary and stellar companions in a sample of \textbf{98} transition disks, assessing the occurrence rate of such companions and their potential influence on cavity formation.}
{For the \textbf{98} young stellar objects (YSOs) with inner dust cavities, we computed \textit{Gaia} proper motion anomalies, \textbf{which together with the RUWE,} identify companions with mass ratios $q \gtrsim 0.01$ at $\sim$0.1--30 au. We assessed the impact of disk gravity, accretion, disk-scattered light, dippers, \textbf{starspots}, jets, and outflows on the measured proper motion anomalies, concluding that these effects are unlikely to affect our analyses and that astrometric techniques such as the one of this work can be robustly applied to YSOs.}
{Significant \textbf{proper motion} anomalies are found in 31 transition disks (32\% of the sample), indicative of companions. We recovered 85\% of the known companions within our sensitivity range. Assuming \textbf{that the astrometry of each system} is dominated by a single companion, we modelled the semi-major axis and mass required to reproduce the observed astrometric signals. Most inferred companions have $M > 30$ M$\rm{_{J}}$, placing many within or near the stellar mass regime. Seven sources host companions compatible with a planetary mass ($M < 13$ M$\rm{_{J}}$, HD 100453, J04343128+1722201, J16102955-3922144, MHO6, MP Mus, PDS 70, and Sz 76). For the non-detections, we provide the companion masses and semi-major axes that can be excluded in future searches. About half (53\%) of detected companions cannot be reconciled with having carved the observed dust cavities.}
{We have gathered evidence of the presence of companions in a large sample of transition disks. However, we find that the population of transition disks cannot be fully described as a circumbinary population. Transition disks host as many companions within our sensitivity range as do randomly sampled groups of YSOs and main-sequence stars. If dust cavities are shaped by companions, such companions must reside at larger orbital separations than those of the companions detected here, and we predict them to be of planetary mass.}


   \maketitle
%

\section{Introduction}

\begin{figure*}[ht!]
    \centering
    \includegraphics[width=1.0\textwidth]{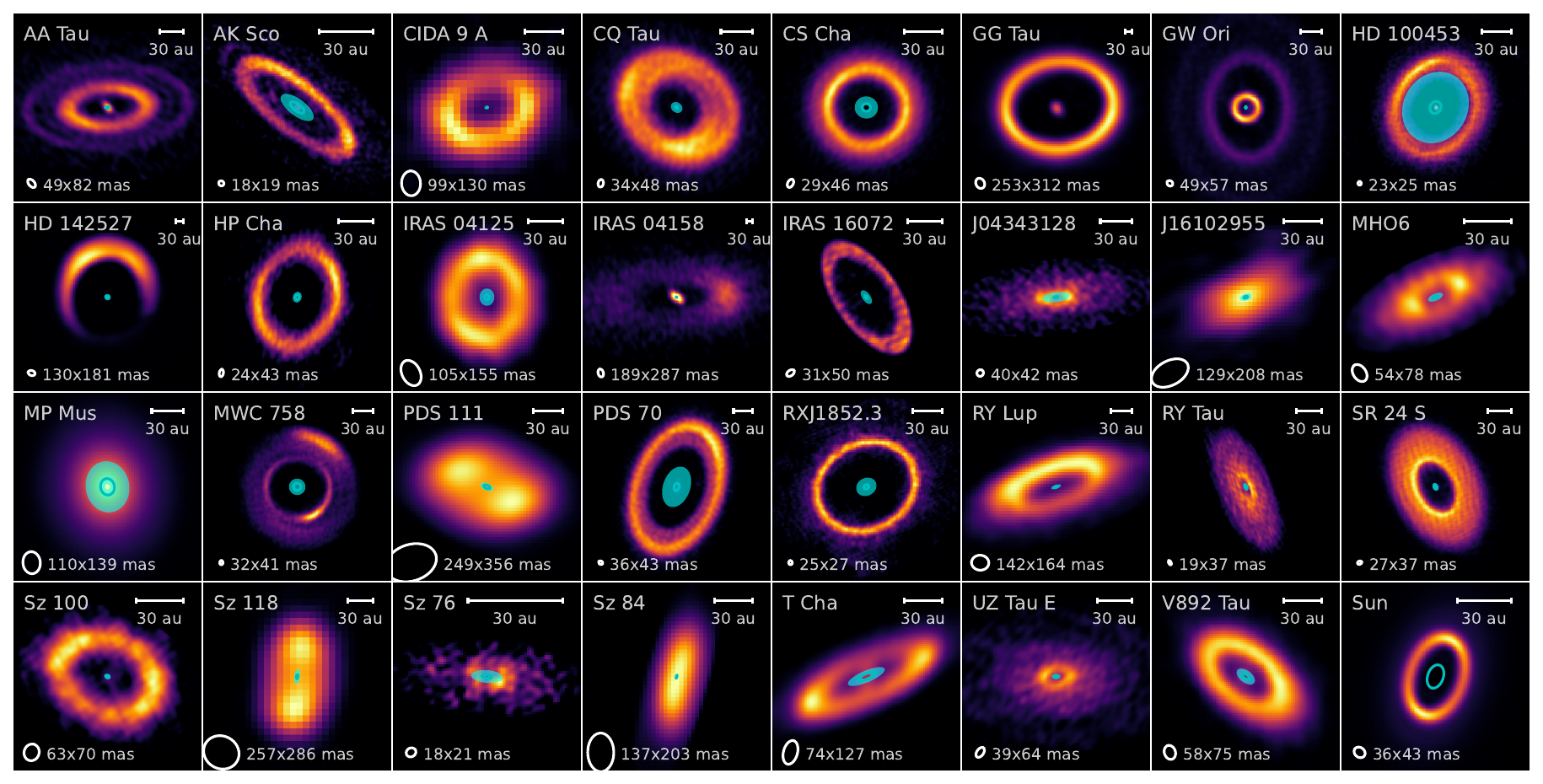}
    \caption{ALMA continuum images of the sample of 31 disks with inner dust cavities (``transition disks'') for which we find a significant proper motion anomaly ($|\Delta \mu|/\sigma_{|\Delta \mu|}\geq3$) indicative of the presence of companions. Assuming one companion dominates the proper motion anomaly (Sect. \ref{sec_methodology}), the solid cyan line indicates the 50th percentile of the companion location (the cyan coloured areas are the 10th and 90th percentiles) as derived in Sect. \ref{s_significant} (see Figs. \ref{fig_HD142527} and \ref{fig:all_detect}, exception is GG Tau, whose companion could not be modelled). Bottom-right corner: Image of the Sun,  for reference, as it is predicted to look at 1 Myr (\citealp{2022A&A...659A...6B}) with Jupiter's orbit in \textbf{cyan}. The ALMA synthesised beams are included at the bottom left of each panel. \textbf{Non-detections can be seen in Fig. \ref{big_mosaic_no_detections}}. }
    \label{big_mosaic_detections}
\end{figure*}

Detecting planetary or stellar companions around forming stars, or young stellar objects (YSOs), is crucial for understanding how multiple systems interact with their natal environments and how companions and protoplanetary disks co-evolve as well as for placing observational constraints on the formation of planets and stellar multiples. In addition, the origin and evolution of companions in YSOs are key to understanding the exoplanet population. 

However, the fraction of YSOs with companions, and their typical separation and mass, are largely uncharacterised (\citealp{2010ApJS..190....1R,2014prpl.conf..267R,2025Symm...17..344C}). Despite the abundance of dust substructures that has been found in protoplanetary disks (\citealp{2020ARA&A..58..483A,2023ASPC..534..423B}), the role of multiplicity in planet formation and disk architectures is poorly understood. One reason for these unknowns is the difficulty involved in identifying companions in the $\sim$ 0.1-30 au separation range from the central star (\citealp{2023ASPC..534..605B}). These companions are often too distant for spectroscopic identification, but they are also too close for direct imaging or millimetre-interferometry due to the protoplanetary disk presence and the contrast luminosity of the central star (e.g. \citealp{2023A&A...680A.114R,2025ApJ...989..134R}). In this work, we use \textit{Gaia} astrometry to look for companions in this $\sim$ 0.1-30 au separation range, bridging the gap where neither direct imaging nor spectroscopy can efficiently detect them.


We focus on the population of YSOs hosting protoplanetary disks with inner dust cavities (often called ``transition'' or ``pre-transition'' disks). These inner dust cavities have often been explained by companions carving the dust-disk, and thus these disks have the highest chances of hosting massive companions (e.g. \citealp{2017MNRAS.464.1449R, 2018ApJ...859...32P, 2018ApJ...854..177V, 2020ApJ...892..111F, 2023A&A...671A.140G}). However, to date, only a few companions have been identified within the cavities of transition disks (\citealp{2023ASPC..534..799C,2023ASPC..534..605B,2023EPJP..138..225V}), and these are often too close ($<$1 au) to explain the cavity size. The small number of known circumbinary disks at the YSO stage contrasts with the high multiplicity fraction observed among main-sequence stars (approximately 20–40\% for stars with masses between 0.3 and 1 M$_\odot$, and up to 100\% for O-type stars, \citealp{2023ASPC..534..275O}). It is unknown if this is caused by an observational limitation. \citet{2025A&A...698A.102R} conclude that in 40\% of the systems they analysed, the hypothesis that a still undetected stellar binary companion is responsible for carving the cavity cannot be ruled out. In contrast, the possibility that still undetected planetary companions are responsible for the cavities cannot be excluded in any system. Similarly, the analyses of \citet{2021AJ....161...33V} and \citet{2023A&A...670A.154W} show that, with current observational constraints, still undetected massive substellar companions are possible at close radii (typically of a few tens of astronomical units). We note other processes such as grain growth, dead zones, and photoevaporation have also been proposed for producing inner dust cavities (e.g. \citealp{2016A&A...596A..81P,2017RSOS....470114E,2024A&A...691A.155H}).

Different techniques leveraging the \textit{Gaia} data have proven to be successful at tracing companions (e.g. \citealp{2021ApJ...907L..33S,2023A&A...674A..34G, 2023A&A...674A..10H,2024A&A...688A...1C}). Among these, the one called ``proper motion anomaly'' is particularly suited to infer the presence of companions in YSOs. This technique traces astrometric accelerations of the system photocentre due to the presence of unresolved companions by comparing proper motions measured across different epochs (\citealp{2014ApJ...797...14P,2018ApJS..239...31B,2019A&A...623A..72K}). It is mostly sensitive to asymmetric perturbations, and hence it is less affected by the different types of azimuthally symmetric variability typical of YSOs. Proper motion anomalies have already been used to identify exoplanets and brown dwarfs around main-sequence stars \citep[later confirmed with direct imaging, e.g.][]{2022MNRAS.513.5588B,2023Sci...380..198C,2023A&A...672A..94D, 2023ApJ...950L..19F, 2023A&A...672A..93M, 2025A&A...702A..77K}. \textbf{In particular,} the combination of the \textit{Hipparcos} astrometric survey with \textit{Gaia} has provided the community with large catalogues of proper motion anomalies (e.g. \citealp{2021ApJS..254...42B,2022A&A...657A...7K,2025A&A...702A..77K}). However, because YSOs are typically faint, most of them do not appear in the \textit{Hipparcos} catalogue, and thus in this work we focus on \textit{Gaia}-only proper motion anomalies (e.g. \citealp{2022MNRAS.513.2437P,2022MNRAS.513.5270P,2024MNRAS.527.3076D}). This \textit{Gaia}-only approach was already successfully used to infer the presence of a gas giant in the protoplanetary disk around MP Mus (\citealp{2025NatAs...9.1176R}).


In this work we use \textit{Gaia} proper motion anomalies to survey the population of companions in the $\sim$ 0.1-30 au separation range of protoplanetary disks with inner dust cavities (``transition disks''). We describe the methodology in Sect. \ref{sec_methodology}. Our results for individual transition disks are presented in Sect. \ref{s_individual_sources}. We present a population analysis in Sect. \ref{S_analysis}, describe possible sources of astrometric noise in Sect. \ref{S_other_sources}, and conclude in Sect. \ref{S_conclusions}.

\section{Methodology}\label{sec_methodology}

We selected a sample of ``transition disks'' from the literature. We define ``transition disk'' as any source with an identification of an inner-dust cavity at millimetre wavelengths. From this compilation, there is enough \textit{Gaia} DR3 and DR2 astrometry to derive proper motion anomalies for \textbf{98} sources (Figs. \ref{big_mosaic_detections}, \ref{big_mosaic_no_detections}, Table \ref{Table_1}). Only 3 of the compiled transition disks do not have enough \textit{Gaia} DR3 and DR2 data for deriving proper motion anomalies: XZ Tau B, J16070384-3911113, and J17110392-2722551 ([BHB2007] 1).



We used the proper motion anomaly defined in \citet{2022MNRAS.513.2437P} as the change in velocity between \textit{Gaia} DR2 and \textit{Gaia} DR3 (\citealp{2016A&A...595A...1G,2018A&A...616A...1G,2021A&A...649A...1G}):

\begin{equation}
    |\Delta \mu| = \sqrt{(\mu_{\alpha,DR3}-\mu_{\alpha,DR2})^2 + (\mu_{\delta,DR3}-\mu_{\delta,DR2})^2},\noindent
\end{equation}

where $\mu_{\alpha}$ and $\mu_{\delta}$ are the proper motion in right ascension and declination, respectively.\footnote{\textbf{We explore applying the \textit{Gaia} DR3 corrections of \citet{2021A&A...649A.124C} and find our results are mostly unchanged. We decide not to apply these corrections for consistency with \textit{Gaia} DR2.
}} We note $|\Delta \mu|$ does not depend on the proper motion of the centre of mass of the system. The uncertainty of this proper motion anomaly is defined in \citet{2022MNRAS.513.5270P} as

\begin{equation}
\sigma_{|\Delta \mu|}=\frac{\sqrt{\Delta \mu_{\alpha}^2 (\sigma_{\mu_{\alpha,DR3}}^2 + \sigma_{\mu_{\alpha,DR2}}^2) + \Delta \mu_{\delta}^2 (\sigma_{\mu_{\delta,DR3}}^2 + \sigma_{\mu_{\delta,DR2}}^2)}}{|\Delta \mu|},\noindent
\end{equation}

where $\sigma$ indicates the uncertainty associated with each quantity. The significance of the proper motion anomaly is defined as $|\Delta \mu|/\sigma_{|\Delta \mu|}$.  This technique is most sensitive to tracing companions at intermediate orbital periods that are not very different from the \textit{Gaia} DR2 and DR3 time baseline. \textbf{When combined with the RUWE (renormalised unit weight error, a goodness-of-fit measure contained in \textit{Gaia} DR3, \citealp{2021A&A...649A...2L})} simulations suggest most detections occur for orbital separations between $\sim0.1$ to $30$ au, approximately (or periods between 0.03 and 160 years for a 1 M$_{\odot}$ star, \textbf{Sect. \ref{s_significant}} and \citealp{2022MNRAS.513.2437P,2022MNRAS.513.5270P}). The measured $|\Delta \mu|$ and $\sigma_{|\Delta \mu|}$ for all \textbf{98} transition disk sources considered in this work are presented in Table \ref{Table_1}.


\begin{figure}
    \centering
    \includegraphics[width=0.8\columnwidth]{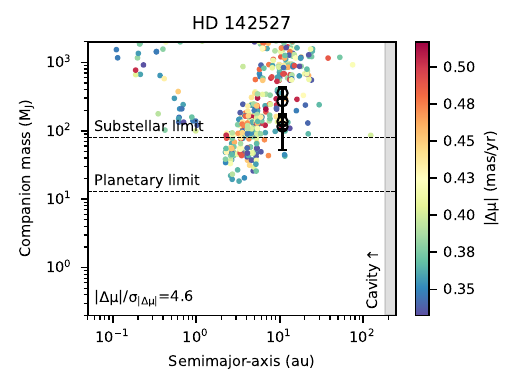}\vspace{-8pt}
    \includegraphics[width=0.8\columnwidth]{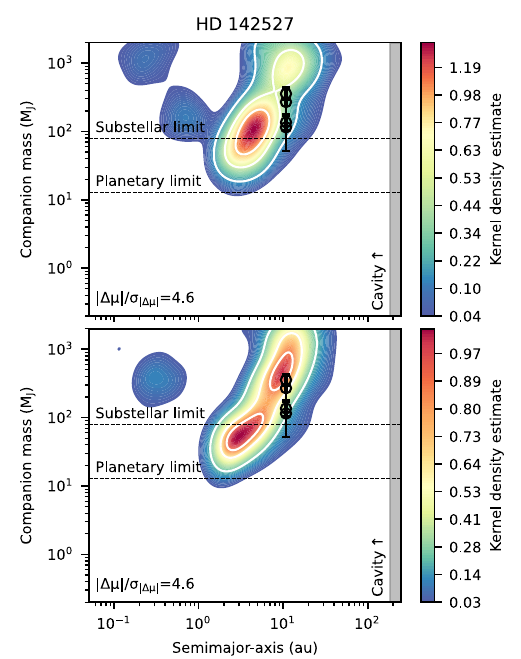}\vspace{-8pt}
    \caption{Orbital separation and mass of the companion that would produce the observed \textit{Gaia} astrometry of HD 142527 ($0.33<|\Delta \mu|<0.52$ mas~yr$^{-1}$, $|\Delta \mu|/\sigma_{|\Delta \mu|}>4$, and $\rm{RUWE_{DR3}}<1.25$, from Sect. \ref{s_individual_sources} Eqs. \ref{eq_3}, \ref{eq_4}, \ref{eq_5}). The top panel shows the individual \textit{Gaia} simulations for this system. The middle panel presents the same simulations but smoothed with a kernel density estimate (contours indicate the 20, 50, and 80\% levels of the normalised area). Black circles mark the location and the proposed masses for the known companion in this system (see Sect. \ref{S_individual_sources}). The bottom panel is the same as the middle panel but assuming the companion responsible for the astrometric signal is in the plane of the disk. Similar plots for all other 30 systems with significant astrometric accelerations are shown in Fig. \ref{fig:all_detect}.}
    \label{fig_HD142527}
\end{figure}

The sensitivity to detect astrometric accelerations via proper motion anomalies decays linearly with distance from Earth, as the observed proper motion anomaly equals the real tangential velocity of the moving photocentre ($v_t$) times the parallax ($\varpi$, $|\Delta \mu| = \varpi\cdot v_t$). In addition, the uncertainties of \textit{Gaia} astrometry are heavily dependent on source brightness (\citealp{2021A&A...649A...2L}). Hence $|\Delta \mu|/\sigma_{|\Delta \mu|}$ varies for otherwise identical systems if located at different distances. Orientation also plays a role, as the signal is smaller for edge-on orbits. In Sect. \ref{S_other_sources}, we evaluate the potential impact of other sources of astrometric signals on the proper motion anomaly in YSOs and conclude that this methodology can be robustly applied to companion searches in YSOs.

We compiled effective temperatures ($\rm{T_{eff}}$), stellar luminosities ($\rm{L_{\star}}$, \textbf{updated to \textit{Gaia} DR3 distances when needed}), disk geometries (inclinations and position angles), and disk cavity sizes from the literature for all \textbf{98} sources. Using $\rm{T_{eff}}$ and $\rm{L_{\star}}$, we derive stellar masses homogeneously for all sources, using \citet{2015A&A...577A..42B} evolutionary tracks for stars less than 1.35 $\rm{M_{\odot}}$ and \citet{2022A&A...665A.126N} PARSEC V2.0 for masses above 1.35 $\rm{M_{\odot}}$. This approach has proven to give accurate stellar masses in YSOs (\citealp{2026arXiv260303422Z}). All stellar and disk properties for the considered transition disks can be found in Table \ref{Table_1}, with references.

\begin{figure*}[ht!]
    \centering
    \includegraphics[width=0.335\textwidth]{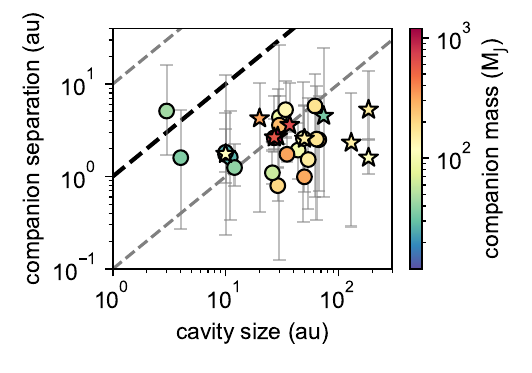}   
    \hspace{-5pt}\includegraphics[width=0.335\textwidth]{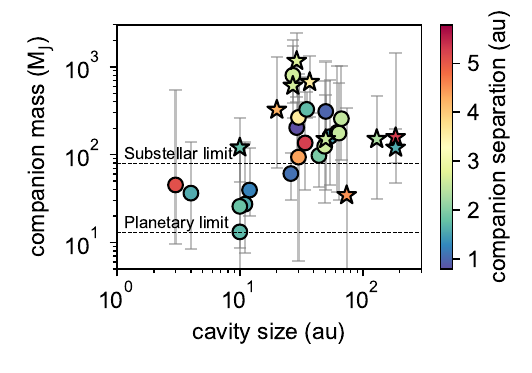}  
    \hspace{-5pt}\includegraphics[width=0.335\textwidth]{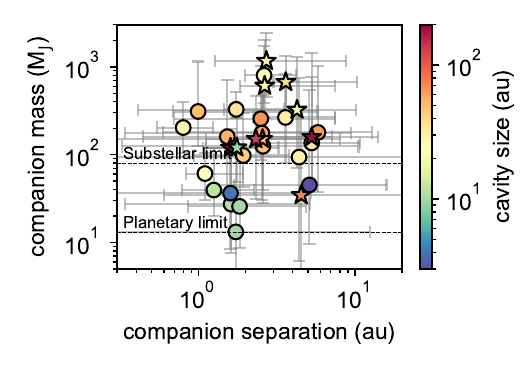}        
    \caption{Predicted space of possible companions. Central values and uncertainties indicate the 10th, 50th, and 90th percentiles of the distribution of companion masses and separations (semi-major axes) for the transition disks with significant astrometric acceleration (30 sources, as no two-body solution was found for GG Tau). Sources with known companions are shown with a star symbol. Left panel: Cavity size at millimetre wavelengths vs companion semi-major axis. One-to-one line is shown in black and $\pm1$ dex lines are shown in grey. All companions are consistent with being inside the dust cavity (MP Mus is the source appearing above the one-to-one line).  Centre panel: Cavity size vs companion mass. Right panel: Companion semi-major axis vs companion mass.}
    \label{fig_detection_sample}
\end{figure*}

For each transition disk in the sample, we performed $40\,000$ simulations modelling a two-body interaction to study the parameter space of semi-major axis and companion mass that produces the observed $|\Delta \mu|$, $\sigma_{|\Delta \mu|}$, and RUWE. The procedure is as follows: we first adopt the scanning law of \textit{Gaia}~DR2 (\citealp{2021MNRAS.501.2954B}) and \textit{Gaia}~DR3\footnote{\href{https://gea.esac.esa.int/archive/documentation/GDR3/Gaia_archive/chap_datamodel/sec_dm_auxiliary_tables/ssec_dm_commanded_scan_law.html}{Gaia Archive scan law}} to retrieve the epochs and scan angles at which \textit{Gaia} observed each source. Using these epochs and scan angles, along with realistic \textit{Gaia} uncertainties (\citealp{2018A&A...616A...2L,2021A&A...649A...2L}), we simulate the \textit{Gaia}~DR2 and DR3 astrometric observations of each source under the assumption of a two-body system with varying mass ratios and orbital periods\footnote{These simulations are performed using the \texttt{astromet} package developed by \citet{2022MNRAS.513.2437P}: \href{https://github.com/zpenoyre/astromet.py}{https://github.com/zpenoyre/astromet.py}}. \textbf{\textit{Gaia} synthetic observables, including the RUWE, are calculated via a close emulation of the original astrometric pipeline of the \textit{Gaia} mission (\citealp{2012A&A...538A..78L,2022MNRAS.513.2437P})}. \textbf{Underlying correlations between \textit{Gaia} DR2 and DR3 which could have a small effect on $\sigma_{|\Delta \mu|}$ are not modelled.}

For each of the $40\,000$ simulations per system, we randomly sampled the log-uniform space of mass ratios (from 0.0001 to 1) and periods (from 0.01 to 2512 yr). The mass of the primary source is sampled at random every time following the measured values and uncertainties of stellar mass (see Table \ref{Table_1}). Eccentricity is  sampled at random in each simulation uniformly between 0 and 1 (\citealp{2013ARA&A..51..269D, 2017ApJS..230...15M}). Viewing angles and orbital phases are also taken at random. Alternatively, viewing angles and orbital phases can be sampled within the measured values and uncertainties of the disk inclinations and position angles. This adds the assumption that possible companions are contained in the plane of the disk. However, we find that this assumption does not reduce the space of companion solutions significantly (e.g. Fig. \ref{fig_HD142527}), and hence we do not consider it in this work to account for companions outside the disk plane (see \citealp{2024Natur.635..574B,2025Natur.644..356B}). 

We model each system as a two-body interaction because the interpretation of the astrometric accelerations becomes more complex for higher-order multiple systems. Therefore, for these simulations we assume that, even if multiple companions are present, the astrometric signal is dominated by a single companion. Additionally, we adopt a relation between the mass ratio ($q = M_{\rm{comp}} / M_{\rm{primary}}$) and light ratio ($l = L_{\rm{comp}} / L_{\rm{primary}}$) of the components, given by $l = q^{3.5}$. We extend this commonly used heuristic for main-sequence stars to Class-II YSOs, which are optically bright and have almost reached their final mass (\citealp{2016ARA&A..54..135H}).

From each of the $40\,000$ simulations we retrieve the $|\Delta \mu|$, $\sigma_{|\Delta \mu|}$, and RUWE each simulated companion would produce in each considered transition disk system.






\section{Analysis of individual transition disks}\label{s_individual_sources}

\subsection{Sources with significant astrometric accelerations}\label{s_significant}

We consider $|\Delta \mu|/\sigma_{|\Delta \mu|}\geq3$ a significant detection of astrometric acceleration, \textbf{or proper motion anomaly} ($|\Delta \mu|$),  in a system. In our sample of transition disks, 31 sources show significant astrometric acceleration (Fig. \ref{big_mosaic_detections}), 15 of them having $|\Delta \mu|/\sigma_{|\Delta \mu|}\geq5$. These sources are indicated in Table \ref{Table_1} and Fig. \ref{fig:all_detect}. Only 3 sources have significant (>3$\sigma$) $|\Delta \mu|$ in the independent \textit{Hipparcos}-\textit{Gaia} proper motion analysis of \citet[RY Tau, GW Ori, and HD 142527]{2022A&A...657A...7K}, which we also retrieve with our methodology.

For the 31 detections, we considered the simulated companions that produce both a proper motion anomaly and a RUWE consistent with the observations. This defines the range of companion separations and masses capable of reproducing the observed astrometry (under the assumptions of Sect. \ref{sec_methodology}). In particular, we considered the simulations with $|\Delta \mu|_{\text{sim}}$, $\sigma_{|\Delta \mu|_{\text{sim}}}$, and $\text{RUWE}_{\text{sim}}$ satisfying all the following conditions:

\begin{equation}\label{eq_3}
|\Delta \mu|_{\text{sim}}-\sigma_{|\Delta \mu|_{\text{sim}}}<|\Delta \mu|<|\Delta \mu|_{\text{sim}}+\sigma_{|\Delta \mu|_{\text{sim}}},\noindent
\end{equation}

\begin{equation}\label{eq_4}
\text{int}[|\Delta \mu|/\sigma_{|\Delta \mu|}]<|\Delta \mu|_{\text{sim}}/\sigma_{|\Delta \mu|_{\text{sim}}},\noindent
\end{equation}

\begin{equation}\label{eq_5}
\begin{cases}
    0.5\,\text{RUWE}<\text{RUWE}_{\text{sim}}<1.5\,\text{RUWE} & \text{if } \text{RUWE} \geq1.25,\\
    
    \text{RUWE}_{\text{sim}}<1.25              & \text{if } \text{RUWE} <1.25.\noindent
\end{cases}
\end{equation}

We use the index ``sim'' to differentiate simulated from observed quantities. The 1.25 threshold for the RUWE was proposed as the upper limit for a single star solution by \citet{2022MNRAS.513.5270P} and \citet{2024A&A...688A...1C}. We consider a broad range around the observed RUWE as it has been suggested protoplanetary disks can impact the RUWE (e.g. \citealp{2022RNAAS...6...18F}).

Fig.~\ref{fig_HD142527} exemplifies this methodology for the system HD 142527. It shows the orbital semi-major axes and masses a companion should have to induce the $|\Delta \mu|$, $\sigma_{|\Delta \mu|}$, and RUWE observed in this system. Equivalent plots for the other 30 transition disks with significant astrometric accelerations are shown in Fig. \ref{fig:all_detect}. In general, our simulations show that beyond $\sim$1 au the companion mass needed to produce the observed astrometry increases with increasing semi-major axis. Under $\sim1$ au, the required companion mass increases with decreasing semi-major axis (similar behaviour was reported in \citealp{2025A&A...702A..76K}).


From this space of possible companions (e.g. Fig.~\ref{fig_HD142527}), we derive the 10th, 50th, and 90th percentiles of the distribution of possible companion masses and semi-major axes (which we call ``separations'') for every source with a significant astrometric acceleration. These are presented in Fig. \ref{fig_detection_sample} (and tabulated in Table \ref{Table_1}). Despite its significant proper motion anomaly, no two-body simulation could reproduce the astrometry of GG Tau (in this work we consider GG Tau A, which is a known hierarchical triple stellar system, \citealp{2014A&A...565L...2D,2020A&A...639A..62K,2024A&A...688A.102T}), and thus we report no companion mass and separation for GG Tau. 



All the companions identified in this work are consistent with being within the dust cavity as seen at millimetre wavelengths (Fig. \ref{fig_detection_sample}). Fig. \ref{fig_detection_sample} also shows we mostly recover $M>30$ M$\rm{_{J}}$ companions, many of which are largely compatible with the stellar mass regime. Only eight companions have median masses in the brown-dwarf regime. Seven sources host companions compatible with a planetary mass within uncertainties ($<$13 M$\rm{_{J}}$, HD 100453, J04343128+1722201, J16102955-3922144, MHO6, MP Mus, PDS 70, and Sz 76), although J04343128+1722201 is the only one with a companion median mass close to the planetary limit.



\subsection{Constraints from non-detections}\label{s_nondetections}

\begin{figure}
    \centering
    \includegraphics[width=1.0\columnwidth]{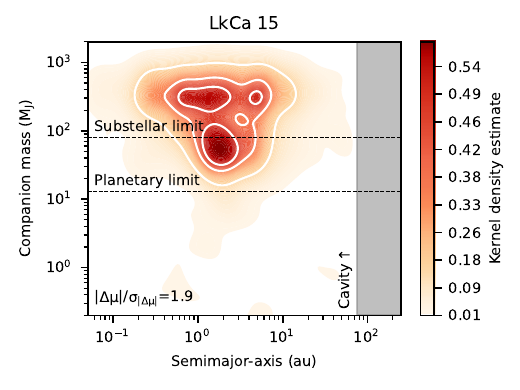} \includegraphics[width=1.0\columnwidth]{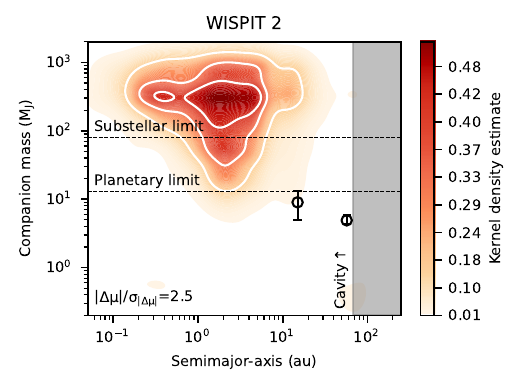}
    \caption{Parameter space of orbital separations and companion masses that would produce either a significant \textit{Gaia} proper motion anomaly or a \textbf{higher-than-observed} RUWE value in LkCa 15 \textbf{and WISPIT 2}, smoothed with a kernel density estimate (contours indicate the 20, 50, and 80\% levels of the normalised area). These regions can be discarded as hosting a companion \textbf{dominating the astrometry} (Sect. \ref{s_nondetections}, Eqs. \ref{eq_6}, \ref{eq_7}).\textbf{ For WISPIT 2, the confirmed (at $\sim$57 au, \citealp{2025ApJ...990L...8V}) and proposed (at $\sim$15 au, \citealp{2025ApJ...990L...9C}) planetary mass companions are indicated}. Similar plots for all other \textbf{65} \textbf{systems with} non-detections are shown in Fig. \ref{fig:all_non_detect}.}
    \label{fig_LkCa15}
\end{figure}

The remaining \textbf{67} transition disk sources considered in this work do not have a significant astrometric acceleration (i.e. $|\Delta \mu|/\sigma_{|\Delta \mu|}<3$). For these, we consider the simulated companions that should have produced a significant proper motion anomaly or a RUWE higher than the one observed. This allows us to define the ranges of companion masses and separations that can be discarded in these systems (\textbf{assuming each system’s astrometry is dominated by a single companion}, Sect. \ref{sec_methodology}). In particular, we considered the simulations with $|\Delta \mu|_{\text{sim}}$, $\sigma_{|\Delta \mu|_{\text{sim}}}$, and $\text{RUWE}_{\text{sim}}$ that satisfied either of the following conditions:

\begin{equation}\label{eq_6}
|\Delta \mu|_{\text{sim}}/\sigma_{|\Delta \mu|_{\text{sim}}}\geq3,\noindent
\end{equation}

\begin{equation}\label{eq_7}
\begin{cases}
    \text{RUWE}<\text{RUWE}_{\text{sim}},& \text{if } \text{RUWE} \geq1.25,\\
    \text{RUWE}_{\text{sim}}\geq1.25,              & \text{if } \text{RUWE} <1.25.\noindent
\end{cases}
\end{equation}

Fig.~\ref{fig_LkCa15} shows this methodology for the systems LkCa 15 \textbf{and WISPIT 2}. Equivalent plots for the other \textbf{65} transition disks with non-detections are shown in Fig. \ref{fig:all_non_detect}.

A non-detection does not imply the absence of companions in a system. In fact, as exemplified in Fig.~\ref{fig_LkCa15}, even when considering the ideal two-body scenario only a narrow range of companion masses and separations can be discarded.  These are typically masses >40 M$\rm{_{J}}$ at separations of 0.3 to 10 au, but it varies from source to source and we encourage the reader to check the plots in Fig. \ref{fig:all_non_detect} for any particular object. In addition, different sources of astrometric noise (e.g. variable extinction, jets, starspots, crowding, scattered light, see Sect. \ref{S_other_sources}) might have diluted potential companion astrometric signals, expanding the parameter space of possible companions in the non-detection scenario beyond what is described by our simulations. 


\subsection{Comment on individual sources}\label{S_individual_sources}

In this section we touch on how our companion detections and non-detections compare with the results of other works (see Figs. \ref{fig:all_detect} and \ref{fig:all_non_detect} and Table \ref{Table_1}). 

There are 14 known circumbinaries in the complete considered sample of \textbf{98} transition disks (see \citealp{2023EPJP..138..225V}, \citealp{2025Symm...17..344C}, and references therein): AK Sco, AS 205 S, CS Cha, GG Tau (circumtriple), GW Ori, HD 142527, HD 34700, HP Cha, IRAS 04158+2805, MHO2, RXJ1633.9-2442, UZ Tau E, V4046 Sgr, and V892 Tau. To this we add PDS 70, IRAS 04125+2902, \textbf{and WISPIT 2} with confirmed substellar companions (\citealp{2018A&A...617A..44K,2019NatAs...3..749H,2024Natur.635..574B,2025ApJ...990L...8V,2025ApJ...990L...9C}). We detect significant astrometric accelerations in 11 of these sources, with the exceptions being AS 205 S, HD 34700, MHO2, RXJ1633.9-2442, V4046 Sgr, \textbf{and WISPIT 2}. However, AS 205 S, HD 34700, and V4046 Sgr are spectroscopic binaries with short (a few days) orbital periods (\citealp{2018ApJ...869L..44K, 2004AJ....127.1187T, 2004A&A...421.1159S}), and hence they are outside the separation-detection range of Sect. \ref{sec_methodology} technique. \textbf{On the other end is WISPIT 2 b, which at $\sim57$ au (\citealp{2025ApJ...990L...8V}) is too separated for our detection range (Fig.~\ref{fig_LkCa15})}. MHO2 and RXJ1633.9-2442 have stellar companions reported at 7.3 and 3.3 au, respectively (\citealp{2011ApJ...731....8K, 2016MNRAS.463.3829R}), and thus they could have been detected with our methodology. This comparison with known companions sets a recovery fraction for Sect. \ref{sec_methodology} methodology of $85\%$ (11/13, excluding the three spectroscopic binaries \textbf{and WISPIT 2}).


\textbf{\citet{2025ApJ...990L...9C} propose another nearby planetary mass companion to WISPIT 2 at 15 au with $9\pm4$ M$\rm{_{J}}$. As shown in our non-detection plot (Fig.~\ref{fig_LkCa15}), a companion with that mass and separation lies at the upper mass limit consistent with our non-detection and could therefore have escaped detection with the methodology of this work. Our results align best with the lower end of the mass range proposed by \citet{2025ApJ...990L...9C} and disfavour substantially higher masses.} 


We also highlight the case of PDS 70, \textbf{an intensively studied system because of} the broad consensus regarding the planetary nature of the companions identified in its disk (e.g. \citealp{2018A&A...617L...2M, 2021AJ....161..148W, 2024A&A...685L...1C,2025MNRAS.539.1613H}). Notably, among the \textbf{98} disks with inner dust cavities analysed homogeneously in this study, PDS 70 stands out. We detect a significant proper motion anomaly in this system that is consistent with the presence of a planetary-mass companion. However, along with the hierarchical triple system GG Tau (Sect. \ref{s_significant}), PDS 70 is one of only two systems exhibiting astrometric behaviour that is difficult to model using a simple two-body interaction. Very few of our two-body simulations reproduce the observed astrometric signature of PDS 70 (Fig. \ref{fig:all_detect}). We attribute this to the system's multiple-planet configuration, where both PDS 70 b and c are of high mass and contribute significantly to the overall astrometric signal. Our results therefore support the idea that PDS 70 is a rare planetary system. From the perspective of this study, the planets in PDS 70 are unusually massive (4.9 M$\rm{_{J}}$ for b and 13.6 M$\rm{_{J}}$ for c, \citealp{2025A&A...698A..19T}), which may have facilitated their detection, whereas planets in other systems may still lie below current detection thresholds. In addition, because PDS 70 c is at $\sim$ 33 au, at the limit of our technique separation sensitivity range, it is possible we are seeing the effect of other massive bodies in the system which are closer to the star (e.g. PDS 70 b and d, \citealp{2024A&A...685L...1C}).

The separations and masses we obtain for the companion of HD 142527 (Fig. \ref{fig_HD142527}) are in agreement with the independently characterised separation and mass of the known companion in this system. The semi-major axis has been measured to $10.80\pm0.22$ au (\citealp{2024A&A...683A...6N}). Reported companion mass estimates range from $\sim$100 to 400 M$\rm{_{J}}$ ($136\pm31$ M$\rm{_{J}}$ from SED, \citealp{2016A&A...590A..90L}, $356\pm63$ M$\rm{_{J}}$ and $115\pm63$ M$\rm{_{J}}$ from spectral fitting, \citealp{2018A&A...617A..37C} and \citealp{2019A&A...622A..96C}, and $270^{+170}_{-150}$ M$\rm{_{J}}$ from dynamical considerations, \citealp{2019A&A...622A..96C}). Our results indicate that companion masses above $300$ M$\rm{_{J}}$ are more likely. However, we note that we obtain a higher stellar mass for HD 142527 than the one reported in \citet[2.2 instead of 2.0 M$_{\odot}$, both consistent within uncertainties]{2014ApJ...790...21M}, which has been used as a benchmark in the aforementioned studies. A lower stellar mass for HD 142527 would result in lower companion masses in our astrometric analysis.

Some other sources with companion detections in this work deserve special mention:
\begin{itemize}
    \item MWC 758 and CQ Tau. We detect a significant proper motion anomaly in these systems, consistent with a brown dwarf or stellar companion (Fig. \ref{fig:all_detect}). For MWC 758, \citet{2018A&A...611A..74R} reported a companion at 20 au, which could be the one we are tracing (although it was questioned by \citealp{2019ApJ...882...20W}). In CQ Tau a nearby massive embedded companion has been proposed (\citealp{2021A&A...648A..19W}), which is confirmed here. Both systems can be reconciled with no previous detections with direct imaging (\citealp{2025A&A...698A.102R}).\vspace{2pt}
    
    \item MP Mus. A gas giant was detected using the methodology of this work. This detection is presented in \citet{2025NatAs...9.1176R}.\vspace{2pt}
    
    \item RY Tau stands out among the detections of this work because the predicted companion has a very high median mass (although it is also compatible with lower masses, Fig. \ref{fig:all_detect}), yet RY Tau has no known stellar companion. It also shows a significant proper motion anomaly in the independent work of \citet{2022A&A...657A...7K}. \citet{2019A&A...628A..68G} and \citet{2021MNRAS.504..871P} found indirect evidence of an unseen planetary or substellar companion at sub-au to few‑au scales, which might be the one detected here.
\end{itemize}

We also report new companion detections with a high mass ratio in the systems Sz 100, CIDA 9 A, and SR 24 S. The remaining 13 YSOs with inner dust cavities for which we find companions are: AA Tau, HD 100453, IRAS 16072-2057, J04343128+1722201, J16102955-3922144, MHO6, PDS 111, RXJ1852.3-3700, RY Lup, Sz 118, Sz 76, Sz 84, and T Cha. All detections are presented in Table \ref{Table_1} and Fig. \ref{fig:all_detect}. 


\begin{figure*}[ht!]
    \centering
    \includegraphics[width=0.38\textwidth]{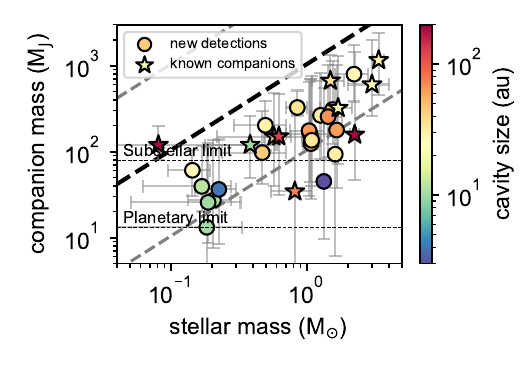}  
    \includegraphics[width=0.305\textwidth]{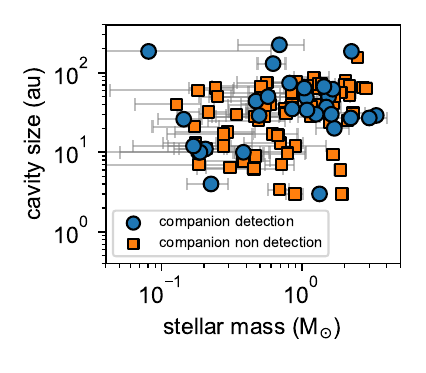}     
    \includegraphics[width=0.305\textwidth]{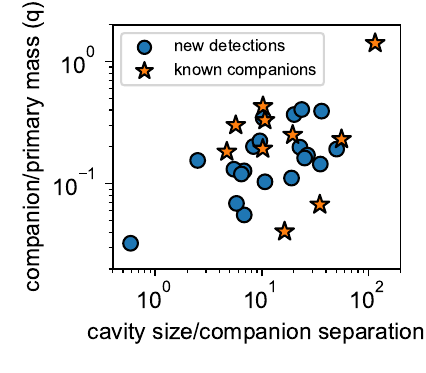} 
    \caption{Left panel: Stellar mass of the primary star vs companion mass. One-to-one line is shown in black and $\pm1$ dex lines are shown in grey. Centre panel: Stellar mass of the primary star vs cavity size at millimetre wavelengths. Right panel: Mass ratio ($q$) vs cavity size to semi-major axis ratio. We note only median values (50th percentiles) are shown.}
    \label{fig_population_detections}
\end{figure*}

In contrast, there are sources with proposed companions for which we obtained non-detections. These are AB Aur (\citealp{2022NatAs...6..751C}), HD 100546 (\citealp{2013ApJ...766L...1Q,2025AJ....169..152B}), HD 135344 B (\citealp{2025A&A...699L..10M}), HD 163296 (\citealp{2018ApJ...860L..13P}), HD 169142 (\citealp{2023MNRAS.522L..51H}), HD 97048 (\citealp{2019NatAs...3.1109P}), LkCa 15 (\citealp{2012ApJ...745....5K,2015Natur.527..342S}, Fig. \ref{fig_LkCa15}), and 2MASS J16120668-3010270 (\citealp{2024ApJ...974..102S,2025A&A...699A.237G,2025ApJ...990L..70L}). However, some of these companions are debated and have unclear status, and we are only sensitive to a particular range of companion separations and masses. We refer the reader to Fig. \ref{fig:all_non_detect} where the companion separations and masses that can be discarded for these systems with our methodology are presented. 



\section{Transition disk population analysis}\label{S_analysis}

Of the \textbf{98} transition disks considered, 31 have significant proper motion anomalies (\textbf{32\%}, Sect. \ref{s_significant}) and \textbf{67} do not (\textbf{68\%}, Sect. \ref{s_nondetections}). In this section we analyse this sample of transition disks from a population perspective.

\begin{figure*}[ht!]
    \centering \includegraphics[width=1\textwidth]{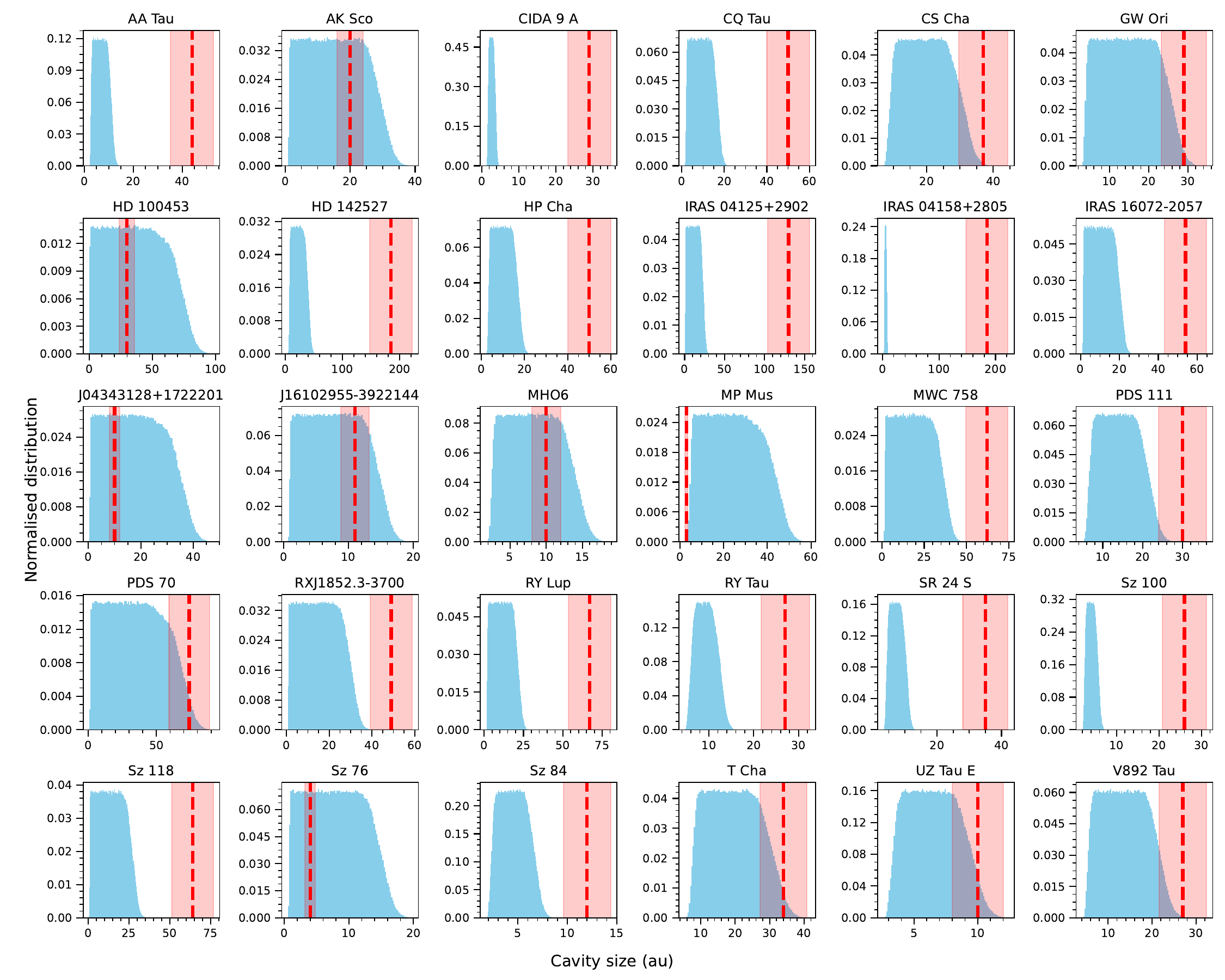} 
    \caption{Dust-disk cavity sizes that could be produced by the detected companions. For each source with a significant proper motion anomaly, blue distributions show the cavity sizes each Sect. \ref{s_significant} (Fig \ref{fig:all_detect}) companion \textbf{could} produce in their disks (from bootstrapping uncertainties). These distributions consider the mass ratio of each source and the theoretical prescription presented in \citet{2025A&A...698A.102R}. Uncertainties in stellar mass, companion mass, and semi-major axis have been propagated consistently. \textbf{We consider all possible} companion inclinations and eccentricities. Vertical red lines show the observed dust cavity sizes at millimetre wavelengths (Fig. \ref{big_mosaic_detections}, with a generic 20\% uncertainty). }\label{Fig_big_mosaic_of_ratios}
\end{figure*}

\subsection{Companions in the transition disk population}\label{S_analysis_1}

\textbf{A two-sided Kolmogorov-Smirnov (KS) test shows no significant difference in stellar mass between the population of transition disks with companion detections and that with non-detections (considering both $|\Delta \mu|/\sigma_{|\Delta \mu|}\geq3\sigma$ and $\geq5\sigma$ thresholds for companion detection). Indeed, we do not recover the expected trend that more massive stars have a higher probability of hosting a companion (\citealp{2023ASPC..534..275O}). However, the methodology described in Sect.~\ref{sec_methodology} loses sensitivity to low mass companions as the mass of the primary star increases (because we trace the companion’s gravitational influence on the system’s photocentre), which biases the inferred companion probability as low mass companions fall below our detection threshold and are removed from the statistics. This effect can be seen in the left panel of Fig. \ref{fig_population_detections}, which shows we detect the lowest mass companions around the lower mass stars, and mostly find stellar mass companions for stars $\gtrsim0.4$ M$_{\odot}$ (the exceptions being MP Mus and PDS 70, with sub-stellar mass companions). Hence, with the aid of the right panel of Fig.~\ref{fig_population_detections}, we place a lower limit on the mass ratio to which our method is sensitive, $q \gtrsim 0.01$. By construction, we do not detect companions more massive than the primary star (as we would have considered them the primary star, note we are working under the assumption of bright companions, $l = q^{3.5}$). IRAS 04158+2805 is an exception to this, but it is indeed an equal mass binary (\citealp{2021MNRAS.507.1157R}).}

The centre panel of Fig. \ref{fig_population_detections} shows the distribution of sources with detected companions with the methodology of this work as a function of stellar mass and dust cavity size. From this plot we note that we detect companions only in a specific region of the cavity-size versus stellar-mass plane. In particular, we do not detect companions in high mass stars ($>0.4$ M$_{\odot}$) with small cavities ($\lesssim20$ au, the exception to this being MP Mus), and in low mass stars ($<0.4$ M$_{\odot}$) with large cavities ($\gtrsim20$ au, the exception to this being IRAS 04158+2805). We speculate that this is a consequence of selection effects arising from the method’s detection limits in mass ratio and separation (Sect.~\ref{sec_methodology}). In particular, we theorise that small cavities may host low mass bodies for which our sensitivity decreases as the primary mass increases, explaining the lack of detected companions in small cavities around high mass stars. Similarly, the lack of companion detections in large cavities around low mass stars can be explained if these systems host companions at separations larger than our detection range ($0.1$--$30$ au, sensitivity peaks at $1$--$10$ au).

\textbf{We find that the fraction of companions in transition disks identified with \textit{Gaia} DR2-DR3 proper motion anomalies ($32\%$) is similar to the fraction found in random populations of main-sequence and YSO stars. In particular, we compare to all \textit{Gaia} stars within 100 to 200 pc (where $92\%$ of the considered transition disks are), to the Sco-Cen YSO catalogues of \citet{2022AJ....163...25L} and \citet{2023A&A...678A..71R}, and to the Taurus catalogue of YSOs from \citet{2019AJ....158...54E} and \citet{2023AJ....165...37L}. In all those catalogues we find a fraction of companions compatible within uncertainties with the fraction found for transition disks (accounting for the dependence of proper motion anomalies with distance and brightness). These fractions are smaller than, and inconsistent with, the companion fractions found in populations of known binary stars. For example, we recover a $\sim$$75\%$ companion fraction in the \textit{nss\_two\_body\_orbit} \textit{Gaia} DR3 table of sources compatible with an orbital two-body solution (\citealp{2023A&A...674A..34G,2023A&A...674A...9H,2023A&A...674A..10H}), a $\sim$$53\%$ companion fraction in the Washington Visual Double Star Catalog (\citealp{2001AJ....122.3466M}, 24-Feb-2025 version) and a $\sim$$73\%$ companion fraction in the circumbinary catalogue of \citet{2025Symm...17..344C}.}

Hence, the population of transition disks cannot be fully described as a circumbinary population, and transition disks host as many binaries as do randomly sampled groups of YSOs and main-sequence stars \textbf{within the separations and mass ratios probed by the methodology of this work (Sect.~\ref{sec_methodology})}. This is remarkable given that our sample of transition disks has 14 sources in common with the circumbinary sample \textbf{of \citet[Sect. \ref{S_individual_sources}]{2025Symm...17..344C}}. In addition, many of the considered transition disks are on the high end of the stellar mass distribution (Herbig stars, \citealp{2018A&A...620A.128V, 2022ApJ...930...39V}). Hence, if normalised by stellar mass the fraction of companion detections in transition disks could be even smaller. 




\subsection{About the detected companions carving the dust cavities}\label{S_cavities}

\begin{figure*}[ht!]
    \centering
    \includegraphics[width=1.0\textwidth]{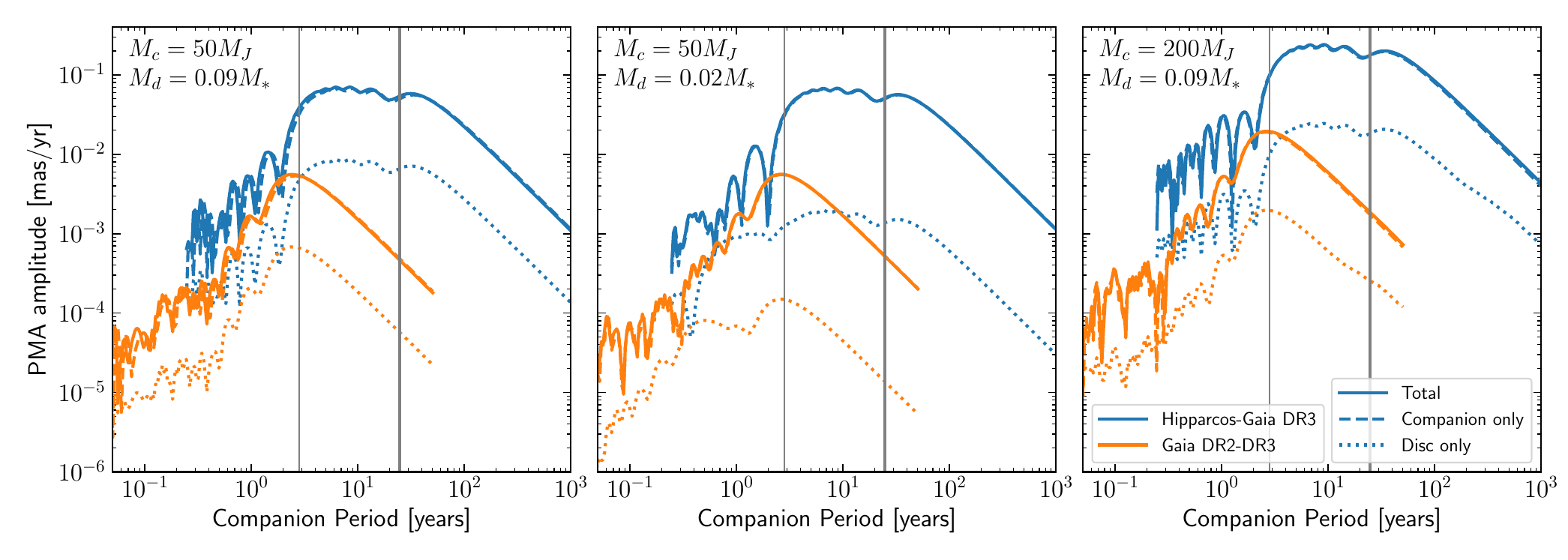}
    \caption{Measured proper motion anomalies ($|\Delta \mu|$) from hydrodynamical simulations describing the effect of protoplanetary disk gravity on \textit{Gaia} DR2-DR3 and \textit{Hipparcos}-\textit{Gaia} DR3 $|\Delta \mu|$ (see \citealp{2021ApJS..254...42B,2022A&A...657A...7K} for the latter). We consider a 1 M$_{\odot}$ central star at 150 pc. Mass of the simulated companion and disk are shown in the legend of each plot. Vertical grey lines are \textit{Hipparcos}-\textit{Gaia} and \textit{Gaia} DR2-\textit{Gaia} DR3 time baselines. The effect of the disk on the measured proper motion anomaly is at most $\sim10\%$ for very massive disks (M$\rm{_{d}}\sim0.1$M$_{\star}$) across different mass ratios ($q$).}
    \label{Richard_plot}
\end{figure*}

\textbf{The formation of transition disk dust cavities has often been attributed to unseen companions (e.g. \citealp{2017MNRAS.464.1449R,2018ApJ...854..177V,2018ApJ...859...32P,2020ApJ...892..111F,2023A&A...671A.140G,2025ApJ...989....9V}). In this section, we evaluate whether the companions detected with our method are consistent with carving the observed cavities (Fig. \ref{big_mosaic_detections}).}

Companions able to carve a dust cavity are expected to have a semi-major axis 2-3 times smaller than the size of the cavity (e.g. \citealp{2017MNRAS.466.1170M, 2020MNRAS.498.2936H, 2020MNRAS.499.3362R,2024ApJ...967...12D,2024MNRAS.532.3166P}). In particular, \citet{2025A&A...698A.102R} report from theoretical considerations an average ratio between cavity size and semi-major axis of $\sim$3.5 for stellar companions and  $\sim$1.7 for planetary companions. Typical cavities do not exceed 7 times the companion’s semi-major axis (\citealp{Sudarshan2022, 2025MNRAS.537.2422P}).

We find a preferred locus for the companions \textbf{detected in this work} in the mass ratio vs cavity radius-to-semimajor axis ratio plot (Fig. \ref{fig_population_detections}, \textbf{right} panel). Most companions have $q=0.1-0.5$ and \textbf{their cavities} have sizes 4 to 60 times larger than their semi-major axes. We find that the four sources with the more massive companions ($>400$ M$\rm{_{J}}$) are all located at $\sim1/10$ of their cavity size. This trend breaks for lower mass companions, although only two sources (MP Mus and Sz 76) have companions closer than one-fourth of their cavity size. However, we note these numbers only consider the median values of the distribution for companion mass and separation. To evaluate if the detected companions can be responsible for carving the observed cavities, we need to propagate uncertainties consistently in stellar mass, companion mass, and companion location. To do this, we apply the theoretical prescription described in \citet{2025A&A...698A.102R} to derive the cavity size that each companion could carve. To propagate uncertainties, we bootstrap the stellar mass of the primary star and the companion’s mass and separation from the results of Sect.~\ref{s_significant}. \textbf{We consider all possible companion inclinations and eccentricities over the full range of companion mass and separation.} \textbf{This provides an assessment of which companions are entirely inconsistent with having carved the inner dust cavity in their respective protoplanetary disks under current theoretical considerations (\citealp{2025A&A...698A.102R}).}

Our results are shown in Fig. \ref{Fig_big_mosaic_of_ratios}. \textbf{We find that in the following 14 sources the companion detected with our method could be responsible for carving the cavity} (considering a 20\% cavity size uncertainty): AK Sco, CS Cha, GW Ori, HD 100453, J04343128+1722201, J16102955-3922144, MHO6, MP Mus, PDS 111, PDS 70, Sz 76, T Cha, UZ Tau E, and V892 Tau. \textbf{We remark that, although the cavities in these systems are consistent with tidal truncation from the detected companions, this consistency alone is not sufficient to claim such companions are responsible for carving the cavity.} \textbf{In contrast, even considering large companion eccentricities (expected to produce larger cavities)}, we find that the companions in the following 16 sources are unable to carve the dust cavity of their protoplanetary disks: AA Tau, CIDA 9 A, CQ Tau, HD 142527 (in agreement with \citealp{2024A&A...683A...6N}), HP Cha, IRAS 04125+2902, IRAS 04158+2805, IRAS 16072-2057, MWC 758, RXJ1852.3-3700, RY Lup, RY Tau, SR 24 S, Sz 100, Sz 118, and Sz 84. Our results agree with the dust modelling predictions of \citet[for the sources in common for which we detect companions, CQ Tau and SR 24 S]{2024ApJ...974..306S}.




In conclusion, our analysis suggests that tidal truncation from the detected companions cannot explain the dust cavities in 53\% of the protoplanetary disks with companion detections, as their orbital semi-major axes are too small. However, as discussed in Sect.~\ref{S_analysis_1}, a non-detection does not preclude unseen companions at larger separations that could be responsible for carving these cavities. If present, such putative, unseen companions must be sufficiently low-mass to remain undetected by direct imaging campaigns (they should be below $\sim$10-15$\,{\rm M_J}$, e.g. \citealp{2021AJ....161...33V}, \citealp{2023A&A...680A.114R}, \citealp{2024A&A...682A.101S}, \citealp{2025A&A...700A.190R}, \citealp{2025A&A...698A.102R}). This is consistent with the limited sensitivity of our method to companions with mass ratios $q \lesssim 0.01$, as discussed in Sect.~\ref{S_analysis_1}.

\begin{figure}[h!]
    \centering
    \includegraphics[width=1.0\columnwidth]{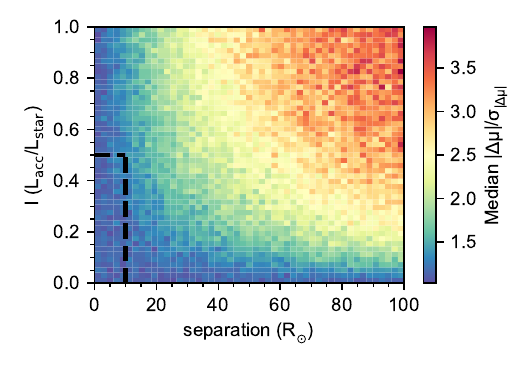}
    \includegraphics[width=1.0\columnwidth]{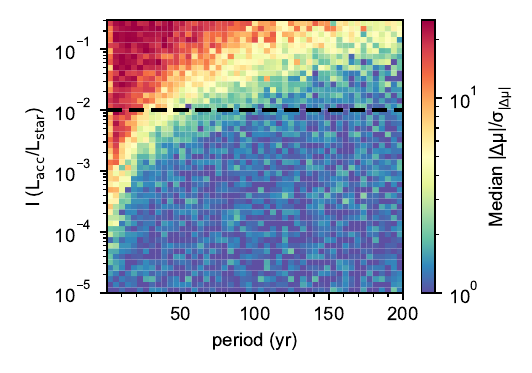}
    \caption{Simulations of a point-like accretion element (top) and scattered light element (bottom) and the significance of the \textbf{\textit{Gaia} DR2-DR3} proper motion anomaly they induce ($|\Delta \mu|/\sigma_{|\Delta \mu|}$). \textbf{Black} dashed lines indicate the conservative range where YSO accretion ($l<0.5$, separation $<$10 R$_{\odot}$) and scattered light (any period, $l<0.01$) \textbf{are} observed.}
    \label{Fig_Acc_sim}
\end{figure}

\section{Other possible sources of astrometric signal}\label{S_other_sources}

\textbf{YSOs have additional sources of astrometric noise than main-sequence stars.} In this section we evaluate the impact of disk gravity, accretion, disk-scattered light, dippers (mostly caused by variable inner extinction or misaligned inner disks), \textbf{starspots}, jets, and outflows, in the measured proper motion anomaly ($|\Delta \mu|$), and whether any of these effects can significantly cause false positive companion detections in YSOs. 


The disk mass can contribute a proper motion anomaly signature via its gravitational influence on the system if the disk is not axisymmetric. This could be important for transition disks, as many are Herbig stars known to host massive disks (\citealp{2024A&A...682A.149S,2025A&A...693A.286S}). While many processes can break the disk's symmetry, we focus here on cases where a companion is present and breaks the symmetry by making the disk eccentric. Since the disk eccentricities produced by a massive companion can reach 0.5 \citep[e.g.][]{2024MNRAS.532.3166P}, this represents one of the largest expected \textbf{sources of astrometric noise}. The process likely to produce the largest signature in the absence of a companion would be the large-scale spiral arms generated by gravitational instability. However, transition disks do not typically exhibit large-scale spirals in the continuum at millimetre wavelengths and are therefore unlikely to be significantly affected by gravitational instability.

To test the effect of disk mass on $|\Delta \mu|$, we model companions on protoplanetary disks of different masses using hydrodynamical simulations. This effect mainly depends on the companion to stellar mass ratio ($q$) and the disk to stellar mass ratio (see Appendix \ref{AppendixC} for details). Disk masses are notoriously challenging to measure (\citealp{2023ASPC..534..501M,2025ApJ...984L...6T,2025ApJ...989....1Z}), but from the work of \citet{2024A&A...686A...9M} and \citet{2025ApJ...984L..17L} we constrain them to be at most 10\% of the stellar mass for the vast majority of YSOs. Hence we explore a range of mass ratios from $q=0.005$ to 0.2, and a range of disk masses from 1\% to 10\% the stellar mass. Higher $q$ mass ratios result in a $|\Delta \mu|$ entirely dominated by the binary system. Fig.~\ref{Richard_plot} shows the impact of the disk's gravity on $|\Delta \mu|$ for some representative cases. We also evaluate the impact of disk gravity on the \textit{Hipparcos}-\textit{Gaia} DR3 $|\Delta \mu|$ (see \citealp{2021ApJS..254...42B,2022A&A...657A...7K}). The main conclusions are that while the effect of the disk increases with companion mass from 5~$M_{\rm J}$ to 200~$M_{\rm J}$, it remains a small fraction of the signal due to the companion. While increasing the disk mass increases the importance of the disk's contribution to $|\Delta \mu|$ (to at most $\sim10$\%), the disk mass would have to be implausibly high (and also gravitationally unstable) before its contribution to $|\Delta \mu|$ is significant. While other disk parameters such as the viscosity, the locally-isothermal assumption, or the 2D nature of the simulations will affect the precise amplitude of the disk's contribution \citep[see, e.g.][]{Sudarshan2022}, the companion should always dominate the measured $|\Delta \mu|$. \textbf{Empirically,} in our sample we find no correlation between $|\Delta \mu|$ and the reported disk masses, nor a higher fraction of detected companions in the more massive disks. 



Independently, the accretion luminosity in YSOs can account for a significant amount of the total system luminosity (\citealp{2025A&A...699A.145D}). In the case of the transition disks considered in this work, 26 have direct measurements of their accretion luminosity (\citealp{2020MNRAS.493..234W}, \citealp{2021A&A...650A.182G}, \citealp{2023ASPC..534..539M}, and references therein).  We find that the median accretion luminosity for this sample is around $7.4\%$ of the stellar luminosity, with some sources having values as high as $28\%$ (SZ Cha). To evaluate whether the accretion luminosity can have an effect on $|\Delta \mu|$ or $|\Delta \mu|/\sigma_{|\Delta \mu|}$, we simulate the effect of a point-like accretion element in rotation at a range of separations, periods, and luminosities. As accretion is often more axisymmetrical and stochastic than a single rotating accretion element, any real stellar accretion phase would produce a net effect in $|\Delta \mu|$ and $|\Delta \mu|/\sigma_{|\Delta \mu|}$ smaller than these simulations. 

To recreate the point-like accretion element, we simulate a single star with a companion of no mass but a range of light-ratios $l=0.0$ to $l=1.0$ (i.e. from no accretion luminosity to half the luminosity coming from the accretion phase). Because accretion luminosity originates on scales near the stellar radius (\citealp{2016ARA&A..54..135H}), we simulate separations between $0.1$ to $100$ R$_{\odot}$. Stellar masses between $0.1$ and $1.5$ M$_{\odot}$ are considered. We run $1\,000\,000$ simulations uniformly covering these ranges of $l$, separation, and stellar mass. The rest of the parameters, including magnitude (within G=$6.5$ to $19$  mag), parallax (within $5$-$10$ mas, or $100$-$200$ pc), inclination, position angle, eccentricity, and coordinates (which determine the \textit{Gaia} observational coverage) are taken at random in each simulation. The accretion periodicity is randomly sampled between periods of 0.1 and 100 days at each simulation (Keplerian rotation is not assumed), a range that encompasses the rotation period of most stars. 


The result of these simulations is shown in Fig. \ref{Fig_Acc_sim}. We find no dependence of the accretion element producing the detection of companions (i.e. $|\Delta \mu|/\sigma_{|\Delta \mu|}>3$) with any parameter except for separation and light-ratio. Only 1\% of simulations in a conservative range where accretion luminosity is expected to originate (red box, $l<0.5$ and separation $<10$ R$_{\odot}$) produce false positive companions. Hence, we conclude that fewer than 1\% of false positives can be attributed to accretion. Fig. \ref{Fig_Acc_sim} shows that only accretion elements at separations larger than $20$ R$_{\odot}$ with $l>0.1$ have the potential to cause a significant proper motion anomaly.



Similarly, we simulate the effect of a single point of disk-scattered light on the proper motion anomaly. Again, scattered light is often extended and axisymmetrical, so any real scattered light would produce a net effect in $|\Delta \mu|$ and $|\Delta \mu|/\sigma_{|\Delta \mu|}$ smaller than these simulations. For the scattered light element we also assume no Keplerian rotation, and sample at random periods from $0.1$ days to 1370 years and separations from 0.1 to 1000 au. We find no dependence of the scattered-light element producing the detection of companions (i.e. $|\Delta \mu|/\sigma_{|\Delta \mu|}>3$) with any parameter except for period and light-ratio. The sources with the brightest scattered light in the sample have on the order of 1--2\% of their stellar luminosity at optical wavelengths, although \textbf{this light-ratio} is much below 1\% for most sources (\citealp{2022A&A...658A.137G}, \citealp{2023A&A...680A.114R}, \citealp{2026arXiv260301703G}). Considering this, Fig. \ref{Fig_Acc_sim} shows that scattered light can lead to false positives if localised in periods under $\sim50$ years. However, we find that the scattered light elements of Fig. \ref{Fig_Acc_sim} with $|\Delta \mu|/\sigma_{|\Delta \mu|}>3$ have periods at least 100 times shorter than Keplerian. Hence, only very non-axisymmetric, bright, and fast-moving scattered light elements can contribute significantly to $|\Delta \mu|$ \textbf{and produce false positive companion detections}.

We note that the analyses of Fig. \ref{Fig_Acc_sim} can be extrapolated to evaluate the effect of dippers \textbf{or photospheric cold spots}. Dippers are irregularly variable YSOs whose variability is believed to be caused by misaligned inner disks and irregular extinction events (\citealp{2025arXiv251026449E}). For example, the accretion element interpretation can be replaced by one in which the visible photosphere appears to move due to occultation by a misaligned inner disk or \textbf{starspots}. This argument also leads to the conclusion that dippers \textbf{or starspots} cannot contribute significantly to $|\Delta \mu|$. \textbf{Indeed, we find a similar companion fraction ($\sim 30\%$) in the dipper catalogue of \citet{2022ApJS..263...14C} to that in randomly sampled groups of YSOs and main-sequence stars (see Sect.~\ref{S_analysis_1}).} Likewise, the analyses of Fig. \ref{Fig_Acc_sim} can be extrapolated to jet and outflow emission. Some sources can have very bright jets (e.g. \citealp{2023A&A...670A.126F}), which could potentially shift their photocentre. As shown in Fig. \ref{Fig_Acc_sim}, if the outflow emission is far from to the central star, or if it rotates rapidly, it could indeed lead to false positive companion detections.


To provide additional support for the conclusions of this section, we note that in our sample there are disks with very high accretion rates (e.g. SZ Cha), bright scattered light (e.g. AT Pyx and J160421.7-213028: \citealp{2022A&A...662A..74G,2018ApJ...868...85P}), and misaligned inner disks (e.g. DoAr 44, HD 139614), all of which show no significant $|\Delta \mu|$. 

The analyses in this section illustrate that astrometric techniques, such as the one employed in this work, can be applied to YSOs to infer the presence of companions, and the different caveats involved. We conclude that only disk-scattered light or jet emission could have a substantial impact \textbf{in the astrometry and the proper motion anomaly}, and only if they are highly non-axisymmetric, bright, and fast-moving.









\section{Conclusions}\label{S_conclusions}

In this work, our focus has been on the search for planetary and stellar companions in a sample of \textbf{98} young stellar objects (YSOs) with inner dust cavities in their protoplanetary disks (or ``transition disks''). We re-derived stellar masses in a homogeneous manner for all sources and compile disk geometries and cavity sizes from the literature (Table \ref{Table_1}). We then computed \textit{Gaia} proper motion anomalies for the entire sample, which \textbf{together with the RUWE} can reveal the presence of companions with mass ratios $q = M_{\rm{comp}} / M_{\rm{primary}} \gtrsim 0.01$ at separations between 0.1 and 30 au. \textbf{Our main conclusions are the following:}

\begin{itemize}
    \item Of the \textbf{98} transition disks considered, 31 have significant proper motion anomalies (\textbf{32\%}) indicative of companions in the system, and \textbf{67} do not (\textbf{68\%}). These are presented in Table \ref{Table_1}, and Figs. \ref{big_mosaic_detections} and \ref{big_mosaic_no_detections}. Twenty of these systems have either no previously reported companions or only indirect detections that are confirmed here (Sect. \ref{S_individual_sources}). We recover 85\% of the known companions within the sample at separations between 0.1 and 30 au.\vspace{5pt}
    
    \item \textbf{For the 31 detections}, assuming that the \textbf{astrometry of} each system is dominated by a single companion, we simulate the \textbf{companion} mass and semi-major axis required to reproduce the observed astrometric signals (Fig. \ref{fig:all_detect}). We present the 10th, 50th, and 90th percentiles of the distribution of possible companion masses and semi-major axes in Table \ref{Table_1}. Most recovered companions have $M>30$ M$\rm_{J}$, placing many of them within or near the stellar mass regime. Eight sources have median companion mass estimations in the brown-dwarf regime. Seven sources are compatible with hosting a planetary mass companion (HD 100453, J04343128+1722201, J16102955-3922144, MHO6, MP Mus, PDS 70, and Sz 76).\vspace{5pt} 
    
    
    \item \textbf{For the 67} non-detections, we estimate the semi-major axis and mass that a companion would need to produce a significant proper motion anomaly \textbf{or a high RUWE}, thereby defining the region of the separation-mass parameter space that can be excluded in future searches (Fig. \ref{fig:all_non_detect}).\vspace{5pt}

    \item Several sources with previously proposed companions yielded non-detections in our analysis. These include AB Aur, HD 100546, HD 135344 B, HD 163296, HD 169142, HD 97048, LkCa 15, 2MASS J16120668-3010270\textbf{, and WISPIT 2}. We emphasise that these non-detections do not imply the absence of companions in these systems. Rather, we exclude only a narrow range of companion masses and separations (typically masses $> 40$ M$\rm_{J}$ at separations of 0.3 to 10 au, though the exact limits vary from source to source). We encourage the reader to consult Fig. \ref{fig:all_non_detect} for the detailed constraints for each source.\vspace{5pt}

    \item We find that 53\% of detected companions cannot be reconciled with having carved the inner dust cavities (Sect. \ref{S_cavities}). If such cavities are indeed shaped by companions, they must reside at larger orbital separations than those of the companions detected here. Provided they have not been identified in other surveys sensitive to these wider separations, we predict that, if these companions exist, they are likely to be of planetary mass. This hypothesis that dust cavities are caused by planetary-mass companions located relatively close to the cavity edge is also consistent with the large fraction of non-detections observed in sources with both small and large inner dust cavities (Sect. \ref{S_analysis_1}).\vspace{5pt}

    \item The population of transition disks cannot be fully described as a circumbinary population, and transition disks host as many binaries \textbf{within the sensitivity range of this work} as do randomly sampled groups of YSOs and main-sequence stars. We test this using the Sco-Cen and Taurus regions, together with known binary samples (Sect. \ref{S_analysis_1}). This further supports the idea that inner dust cavities are not a consequence of stellar binary companions.\vspace{5pt}


    \item We assess the impact of disk gravity, accretion, disk-scattered light, dippers (typically caused by variable inner extinction or misaligned inner disks), \textbf{starspots}, jets, and outflows on the measured proper motion anomaly, evaluating whether any of these effects could significantly contribute to the astrometric signal. We conclude that only disk-scattered light or jet emission could have a substantial impact, and only if such emission is highly non-axisymmetric, bright, and fast-moving (with periods shorter than 100 times the Keplerian period). This demonstrates that astrometric analyses such as the one presented here can be reliably applied to forming stars.
    
\end{itemize}

In this work, we have gathered indirect evidence of the presence of companions in a large sample of transition disks. Our results lay the groundwork for applying this and similar astrometric techniques to other populations of forming stars for planetary and stellar companion searches. 



\section{Data availability}\label{sec_data_availability}
Table \ref{Table_1} is available at the CDS via \href{https://cdsarc.cds.unistra.fr/viz-bin/cat/J/A+A/705/A238}{https://cdsarc.cds.unistra.fr/viz-bin/cat/J/A+A/705/A238}.

\begin{acknowledgements}
We thank the Star and Planet Formation group at ESO and Nienke van der Marel for insightful discussions which improved this work. This work has made use of data from the European Space Agency (ESA) mission Gaia (https://www.cosmos.esa.int/gaia), processed by the Gaia Data Processing and Analysis Consortium (DPAC, https://www.cosmos.esa.int/web/gaia/dpac/consortium). Funding for the DPAC has been provided by national institutions, in particular the institutions participating in the Gaia Multilateral Agreement. This paper makes use of the following ALMA data: ADS/JAO.ALMA\#2012.1.00631.S, ADS/JAO.ALMA\#2012.1.00698.S, ADS/JAO.ALMA\#2013.1.00291.S, ADS/JAO.ALMA\#2013.1.00437.S, ADS/JAO.ALMA\#2015.1.00979.S, ADS/JAO.ALMA\#2015.1.01301.S, ADS/JAO.ALMA\#2015.A.00005.S, ADS/JAO.ALMA\#2016.1.00344.S, ADS/JAO.ALMA\#2016.1.00484.L, ADS/JAO.ALMA\#2016.1.00715.S, ADS/JAO.ALMA\#2016.1.01164.S, ADS/JAO.ALMA\#2016.1.01286.S, ADS/JAO.ALMA\#2016.1.01344.S,
ADS/JAO.ALMA\#2016.1.01511.S,
ADS/JAO.ALMA\#2017.1.00388.S,
ADS/JAO.ALMA\#2017.1.00449.S,
ADS/JAO.ALMA\#2017.1.00492.S,
ADS/JAO.ALMA\#2017.1.00969.S,
ADS/JAO.ALMA\#2017.1.01151.S,
ADS/JAO.ALMA\#2017.1.01404.S,
ADS/JAO.ALMA\#2017.1.01460.S,
ADS/JAO.ALMA\#2017.1.01578.S,
ADS/JAO.ALMA\#2017.1.01631.S,
ADS/JAO.ALMA\#2017.1.01678.S,
ADS/JAO.ALMA\#2018.1.00028.S,
ADS/JAO.ALMA\#2018.1.00310.S,
ADS/JAO.ALMA\#2018.1.00350.S,
ADS/JAO.ALMA\#2018.1.00532.S,
ADS/JAO.ALMA\#2018.1.00536.S,
ADS/JAO.ALMA\#2018.1.00689.S,
ADS/JAO.ALMA\#2018.1.01020.S,
ADS/JAO.ALMA\#2018.1.01054.S,
ADS/JAO.ALMA\#2018.1.01055.L,
ADS/JAO.ALMA\#2018.1.01066.S,
ADS/JAO.ALMA\#2018.1.01255.S,
ADS/JAO.ALMA\#2018.1.01302.S,
ADS/JAO.ALMA\#2018.1.01309.S,
ADS/JAO.ALMA\#2018.1.01458.S,
ADS/JAO.ALMA\#2018.1.01755.S,
ADS/JAO.ALMA\#2018.1.01829.S,
ADS/JAO.ALMA\#2018.A.00030.S,
ADS/JAO.ALMA\#2019.1.00566.S,
ADS/JAO.ALMA\#2019.1.00607.S,
ADS/JAO.ALMA\#2019.1.00847.S,
ADS/JAO.ALMA\#2019.1.01059.S,
ADS/JAO.ALMA\#2019.1.01091.S,
ADS/JAO.ALMA\#2019.1.01210.S,
ADS/JAO.ALMA\#2019.1.01270.S,
ADS/JAO.ALMA\#2021.1.00128.L,
ADS/JAO.ALMA\#2021.1.00378.S,
ADS/JAO.ALMA\#2021.1.00709.S,
ADS/JAO.ALMA\#2021.1.00994.S,
ADS/JAO.ALMA\#2021.1.01123.L,
ADS/JAO.ALMA\#2021.1.01137.S,
ADS/JAO.ALMA\#2021.1.01661.S,
ADS/JAO.ALMA\#2021.1.01705.S,
ADS/JAO.ALMA\#2022.1.00154.S,
ADS/JAO.ALMA\#2022.1.00313.S,
ADS/JAO.ALMA\#2022.1.00340.S,
ADS/JAO.ALMA\#2022.1.00646.S,
ADS/JAO.ALMA\#2022.1.00742.S,
ADS/JAO.ALMA\#2022.1.00760.S,
ADS/JAO.ALMA\#2022.1.00908.S,
ADS/JAO.ALMA\#2022.1.01302.S.
ALMA is a partnership of ESO (representing its member states), NSF (USA), and NINS (Japan), together with NRC (Canada), MOST and ASIAA (Taiwan), and KASI (Republic of Korea), in cooperation with the Republic of Chile. The Joint ALMA Observatory is operated by ESO, AUI/NRAO, and
NAOJ. The National Radio Astronomy Observatory is a
facility of the National Science Foundation operated under
cooperative agreement by Associated Universities, Inc. This research has made use of the Washington Double Star Catalog maintained at the U.S. Naval Observatory. RAB thanks the Royal Society for their support through a University Research Fellowship. ER acknowledges financial support from the European Union's Horizon Europe research and innovation programme under the Marie Sk\l{}odowska-Curie grant agreement No. 101102964 (ORBIT-D). A.R. has been supported by the UK Science and Technology Facilities Council (STFC) via the consolidated grant ST/W000997/1 and by the European Union’s Horizon 2020 research and innovation program under the Marie Sklodowska-Curie grant agreement No. 823823 (RISE DUSTBUSTERS project). GR and ER acknowledge support from the European Union (ERC Starting Grant DiscEvol, project number 101039651) and from Fondazione Cariplo, grant No. 2022-1217. Views and opinions expressed are, however, those of the author(s) only and do not necessarily reflect those of the European Union or the European Research Council. Neither the European Union nor the granting authority can be held responsible for them. SF acknowledges financial contribution from the European Union (ERC, UNVEIL, 101076613) and from PRIN-MUR 2022YP5ACE. MB has received funding from the European Research Council (ERC) under the European Union’s Horizon 2020 research and innovation programme (PROTOPLANETS, grant agreement No. 101002188). NH is funded by the Spanish grant MCIN/AEI/10.13039/501100011033 PID2023-150468NB-I00. This project has received funding from the European Research Council (ERC) under the European Union Horizon Europe programme (grant agreement No. 101042275, project Stellar-MADE). This work was partly funded by ANID -- Millennium Science Initiative Program -- Center Code NCN2024\_001. S.P. acknowledges support from FONDECYT 1231663. Funded by the European Union (ERC, WANDA, 101039452). Views and opinions expressed are however those of the author(s) only and do not necessarily reflect those of the European Union or the European Research Council Executive Agency. Neither the European Union nor the granting authority can be held responsible for them. IM’s research is funded by grants PID2022-138366NA-I00, by the Spanish Ministry of Science and Innovation/State Agency of Research MCIN/AEI/10.13039/501100011033 and by the European Union, and by a Ramón y Cajal fellowship RyC2019-026992-I.
\end{acknowledgements}

\bibliographystyle{aa} 
\bibliography{MyBib} 

@ARTICLE{2018A&A...620A.128V,
       author = {{Vioque}, M. and {Oudmaijer}, R.~D. and {Baines}, D. and
         {Mendigut{\'\i}a}, I. and {P{\'e}rez-Mart{\'\i}nez}, R.},
        title = "{Gaia DR2 study of Herbig Ae/Be stars}",
      journal = {\aap},
         year = "2018",
        month = "Dec",
       volume = {620},
          eid = {A128},
        pages = {A128},
          doi = {10.1051/0004-6361/201832870},
archivePrefix = {arXiv},
       eprint = {1808.00476},
 primaryClass = {astro-ph.SR},
       adsurl = {https://ui.adsabs.harvard.edu/\#abs/2018A&A...620A.128V}
}

@ARTICLE{2026arXiv260301703G,
       author = {{Garufi}, Antonio and {Ginski}, Christian and {Benisty}, Myriam and {Vioque}, Miguel and {Winter}, Andrew and {Huang}, Jane and {Manara}, Carlo Felice and {Dominik}, Carsten},
        title = "{Planet-forming disks and their environment across regions and time from the full NIR census}",
      journal = {arXiv e-prints},
         year = 2026,
        month = mar,
          eid = {arXiv:2603.01703},
        pages = {arXiv:2603.01703},
          doi = {10.48550/arXiv.2603.01703},
archivePrefix = {arXiv},
       eprint = {2603.01703},
 primaryClass = {astro-ph.SR},
       adsurl = {https://ui.adsabs.harvard.edu/abs/2026arXiv260301703G}
}

@ARTICLE{2026arXiv260303422Z,
       author = {{Zallio}, Luigi and {Vioque}, Miguel and {Andrews}, Sean M. and {Empey}, Aaron and {Rosotti}, Giovanni P. and {Miotello}, Anna and {Manara}, Carlo F. and {Carpenter}, John M. and {Deng}, Dingshan and {Kurtovic}, Nicol{\'a}s T. and {Law}, Charles J. and {Longarini}, Cristiano and {Paneque-Carreno}, Teresa and {Teague}, Richard and {Villenave}, Marion and {Yen}, Hsi-Wei and {Zagaria}, Francesco},
        title = "{Benchmarking pre-main sequence stellar evolutionary tracks using disk-based dynamical stellar masses}",
      journal = {arXiv e-prints},
         year = 2026,
        month = mar,
          eid = {arXiv:2603.03422},
        pages = {arXiv:2603.03422},
          doi = {10.48550/arXiv.2603.03422},
archivePrefix = {arXiv},
       eprint = {2603.03422},
 primaryClass = {astro-ph.SR},
       adsurl = {https://ui.adsabs.harvard.edu/abs/2026arXiv260303422Z}
}

@ARTICLE{2025ApJ...990L...8V,
       author = {{van Capelleveen}, Richelle F. and {Ginski}, Christian and {Kenworthy}, Matthew A. and {Byrne}, Jake and {Lawlor}, Chloe and {McLachlan}, Dan and {Mamajek}, Eric E. and {Stolker}, Tomas and {Benisty}, Myriam and {Bohn}, Alexander J. and {Close}, Laird M. and {Dominik}, Carsten and {Haffert}, Sebastiaan and {Landman}, Rico and {Ma}, Jie and {Snellen}, Ignas and {Tazaki}, Ryo and {van der Marel}, Nienke and {Welzel}, Lukas and {Zhang}, Yapeng},
        title = "{WIde Separation Planets In Time (WISPIT): A Gap-clearing Planet in a Multi-ringed Disk around the Young Solar-type Star WISPIT 2}",
      journal = {\apjl},
         year = 2025,
        month = sep,
       volume = {990},
       number = {1},
          eid = {L8},
        pages = {L8},
          doi = {10.3847/2041-8213/adf721},
archivePrefix = {arXiv},
       eprint = {2508.19053},
 primaryClass = {astro-ph.EP},
       adsurl = {https://ui.adsabs.harvard.edu/abs/2025ApJ...990L...8V}
}

@ARTICLE{2017A&A...600A..20A,
       author = {{Alcal{\'a}}, J.~M. and {Manara}, C.~F. and {Natta}, A. and {Frasca}, A. and {Testi}, L. and {Nisini}, B. and {Stelzer}, B. and {Williams}, J.~P. and {Antoniucci}, S. and {Biazzo}, K. and {Covino}, E. and {Esposito}, M. and {Getman}, F. and {Rigliaco}, E.},
        title = "{X-shooter spectroscopy of young stellar objects in Lupus. Accretion properties of class II and transitional objects}",
      journal = {\aap},
         year = 2017,
        month = apr,
       volume = {600},
          eid = {A20},
        pages = {A20},
          doi = {10.1051/0004-6361/201629929},
archivePrefix = {arXiv},
       eprint = {1612.07054},
 primaryClass = {astro-ph.SR},
       adsurl = {https://ui.adsabs.harvard.edu/abs/2017A&A...600A..20A}
}

@ARTICLE{2025A&A...696A.232G,
       author = {{Guerra-Alvarado}, Osmar M. and {van der Marel}, Nienke and {Williams}, Jonathan P. and {Pinilla}, Paola and {Mulders}, Gijs D. and {Lambrechts}, Michiel and {Sanchez}, Mariana},
        title = "{A high-resolution survey of protoplanetary disks in Lupus and the nature of compact disks}",
      journal = {\aap},
         year = 2025,
        month = apr,
       volume = {696},
          eid = {A232},
        pages = {A232},
          doi = {10.1051/0004-6361/202453338},
archivePrefix = {arXiv},
       eprint = {2503.19504},
 primaryClass = {astro-ph.EP},
       adsurl = {https://ui.adsabs.harvard.edu/abs/2025A&A...696A.232G}
}

@ARTICLE{2024A&A...692A.155K,
       author = {{Kurtovic}, N.~T. and {Facchini}, S. and {Benisty}, M. and {Pinilla}, P. and {Cabrit}, S. and {Jensen}, E.~L.~N. and {Dougados}, C. and {Booth}, R. and {Kimmig}, C.~N. and {Manara}, C.~F. and {Rodriguez}, J.~E.},
        title = "{Binary orbit and disks properties of the RW Aur system using ALMA observations}",
      journal = {\aap},
         year = 2024,
        month = dec,
       volume = {692},
          eid = {A155},
        pages = {A155},
          doi = {10.1051/0004-6361/202347583},
archivePrefix = {arXiv},
       eprint = {2407.18828},
 primaryClass = {astro-ph.EP},
       adsurl = {https://ui.adsabs.harvard.edu/abs/2024A&A...692A.155K}
}

@ARTICLE{1999MNRAS.307..909W,
       author = {{Wichmann}, R. and {Covino}, E. and {Alcal{\'a}}, J.~M. and {Krautter}, J. and {Allain}, S. and {Hauschildt}, P.~H.},
        title = "{High-resolution spectroscopy of ROSAT-discovered weak-line T Tauri stars near Lupus}",
      journal = {\mnras},
         year = 1999,
        month = aug,
       volume = {307},
       number = {4},
        pages = {909-918},
          doi = {10.1046/j.1365-8711.1999.02666.x},
       adsurl = {https://ui.adsabs.harvard.edu/abs/1999MNRAS.307..909W}
}

@ARTICLE{2022arXiv220408225V,
       author = {{van der Marel}, Nienke and {Williams}, Jonathan P. and {Picogna}, Giovanni and {van Terwisga}, Sierk and {Facchini}, Stefano and {Manara}, Carlo F. and {Zormpas}, Apostolos and {Ansdell}, Megan and {.}},
        title = "{High-resolution ALMA observations of transition disk candidates in Lupus}",
      journal = {arXiv e-prints},
         year = 2022,
        month = apr,
          eid = {arXiv:2204.08225},
        pages = {arXiv:2204.08225},
          doi = {10.48550/arXiv.2204.08225},
archivePrefix = {arXiv},
       eprint = {2204.08225},
 primaryClass = {astro-ph.EP},
       adsurl = {https://ui.adsabs.harvard.edu/abs/2022arXiv220408225V}
}

@ARTICLE{2024ApJ...966...59S,
       author = {{Shi}, Yangfan and {Long}, Feng and {Herczeg}, Gregory J. and {Harsono}, Daniel and {Liu}, Yao and {Pinilla}, Paola and {Ragusa}, Enrico and {Johnstone}, Doug and {Bai}, Xue-Ning and {Pascucci}, Ilaria and {Manara}, Carlo F. and {Mulders}, Gijs D. and {Cieza}, Lucas A.},
        title = "{Small and Large Dust Cavities in Disks around Mid-M Stars in Taurus}",
      journal = {\apj},
         year = 2024,
        month = may,
       volume = {966},
       number = {1},
          eid = {59},
        pages = {59},
          doi = {10.3847/1538-4357/ad2e94},
archivePrefix = {arXiv},
       eprint = {2402.18720},
 primaryClass = {astro-ph.EP},
       adsurl = {https://ui.adsabs.harvard.edu/abs/2024ApJ...966...59S}
}

@ARTICLE{2023MNRAS.522.2611A,
       author = {{Antilen}, Juanita and {Casassus}, Simon and {Cieza}, Lucas A. and {Gonz{\'a}lez-Ruilova}, Camilo},
        title = "{Gas distribution in ODISEA sources from ALMA long-baseline observations in $^{12}$CO(2-1)}",
      journal = {\mnras},
         year = 2023,
        month = jun,
       volume = {522},
       number = {2},
        pages = {2611-2627},
          doi = {10.1093/mnras/stad975},
archivePrefix = {arXiv},
       eprint = {2304.15002},
 primaryClass = {astro-ph.EP},
       adsurl = {https://ui.adsabs.harvard.edu/abs/2023MNRAS.522.2611A}
}

@ARTICLE{2018ApJ...859...32P,
       author = {{Pinilla}, P. and {Tazzari}, M. and {Pascucci}, I. and {Youdin}, A.~N. and {Garufi}, A. and {Manara}, C.~F. and {Testi}, L. and {van der Plas}, G. and {Barenfeld}, S.~A. and {Canovas}, H. and {Cox}, E.~G. and {Hendler}, N.~P. and {P{\'e}rez}, L.~M. and {van der Marel}, N.},
        title = "{Homogeneous Analysis of the Dust Morphology of Transition Disks Observed with ALMA: Investigating Dust Trapping and the Origin of the Cavities}",
      journal = {\apj},
         year = 2018,
        month = may,
       volume = {859},
       number = {1},
          eid = {32},
        pages = {32},
          doi = {10.3847/1538-4357/aabf94},
archivePrefix = {arXiv},
       eprint = {1804.07301},
 primaryClass = {astro-ph.EP},
       adsurl = {https://ui.adsabs.harvard.edu/abs/2018ApJ...859...32P}
}

@ARTICLE{2020A&A...642A.164V,
       author = {{Villenave}, M. and {M{\'e}nard}, F. and {Dent}, W.~R.~F. and {Duch{\^e}ne}, G. and {Stapelfeldt}, K.~R. and {Benisty}, M. and {Boehler}, Y. and {van der Plas}, G. and {Pinte}, C. and {Telkamp}, Z. and {Wolff}, S. and {Flores}, C. and {Lesur}, G. and {Louvet}, F. and {Riols}, A. and {Dougados}, C. and {Williams}, H. and {Padgett}, D.},
        title = "{Observations of edge-on protoplanetary disks with ALMA. I. Results from continuum data}",
      journal = {\aap},
         year = 2020,
        month = oct,
       volume = {642},
          eid = {A164},
        pages = {A164},
          doi = {10.1051/0004-6361/202038087},
archivePrefix = {arXiv},
       eprint = {2008.06518},
 primaryClass = {astro-ph.SR},
       adsurl = {https://ui.adsabs.harvard.edu/abs/2020A&A...642A.164V}
}

@ARTICLE{2021MNRAS.501.2934C,
       author = {{Cieza}, Lucas A. and {Gonz{\'a}lez-Ruilova}, Camilo and {Hales}, Antonio S. and {Pinilla}, Paola and {Ru{\'\i}z-Rodr{\'\i}guez}, Dary and {Zurlo}, Alice and {Casassus}, Sim{\'o}n and {P{\'e}rez}, Sebasti{\'a}n and {C{\'a}novas}, Hector and {Arce-Tord}, Carla and {Flock}, Mario and {Kurtovic}, Nicolas and {Marino}, Sebastian and {Nogueira}, Pedro H. and {Perez}, Laura and {Price}, Daniel J. and {Principe}, David A. and {Williams}, Jonathan P.},
        title = "{The Ophiuchus DIsc Survey Employing ALMA (ODISEA) - III. The evolution of substructures in massive discs at 3-5 au resolution}",
      journal = {\mnras},
         year = 2021,
        month = feb,
       volume = {501},
       number = {2},
        pages = {2934-2953},
          doi = {10.1093/mnras/staa3787},
archivePrefix = {arXiv},
       eprint = {2012.00189},
 primaryClass = {astro-ph.EP},
       adsurl = {https://ui.adsabs.harvard.edu/abs/2021MNRAS.501.2934C}
}

@ARTICLE{2006A&A...460..695T,
       author = {{Torres}, C.~A.~O. and {Quast}, G.~R. and {da Silva}, L. and {de La Reza}, R. and {Melo}, C.~H.~F. and {Sterzik}, M.},
        title = "{Search for associations containing young stars (SACY). I. Sample and searching method}",
      journal = {\aap},
         year = 2006,
        month = dec,
       volume = {460},
       number = {3},
        pages = {695-708},
          doi = {10.1051/0004-6361:20065602},
archivePrefix = {arXiv},
       eprint = {astro-ph/0609258},
 primaryClass = {astro-ph},
       adsurl = {https://ui.adsabs.harvard.edu/abs/2006A&A...460..695T}
}

@ARTICLE{2016MNRAS.461..794P,
       author = {{Pecaut}, Mark J. and {Mamajek}, Eric E.},
        title = "{The star formation history and accretion-disc fraction among the K-type members of the Scorpius-Centaurus OB association}",
      journal = {\mnras},
         year = 2016,
        month = sep,
       volume = {461},
       number = {1},
        pages = {794-815},
          doi = {10.1093/mnras/stw1300},
archivePrefix = {arXiv},
       eprint = {1605.08789},
 primaryClass = {astro-ph.SR},
       adsurl = {https://ui.adsabs.harvard.edu/abs/2016MNRAS.461..794P}
}

@ARTICLE{2024A&A...688A.149D,
       author = {{Derkink}, Annelotte and {Ginski}, Christian and {Pinilla}, Paola and {Kurtovic}, Nicolas and {Kaper}, Lex and {de Koter}, Alex and {Valeg{\r{a}}rd}, Per-Gunnar and {Mamajek}, Eric and {Backs}, Frank and {Benisty}, Myriam and {Birnstiel}, Til and {Columba}, Gabriele and {Dominik}, Carsten and {Garufi}, Antonio and {Hogerheijde}, Michiel and {van Holstein}, Rob and {Huang}, Jane and {M{\'e}nard}, Fran{\c{c}}ois and {Rab}, Christian and {Ram{\'\i}rez-Tannus}, Mar{\'\i}a Claudia and {Ribas}, {\'A}lvaro and {Williams}, Jonathan P. and {Zurlo}, Alice},
        title = "{Disk Evolution Study Through Imaging of Nearby Young Stars (DESTINYS): PDS 111, an old T Tauri star with a young-looking disk}",
      journal = {\aap},
         year = 2024,
        month = aug,
       volume = {688},
          eid = {A149},
        pages = {A149},
          doi = {10.1051/0004-6361/202348555},
archivePrefix = {arXiv},
       eprint = {2406.04160},
 primaryClass = {astro-ph.EP},
       adsurl = {https://ui.adsabs.harvard.edu/abs/2024A&A...688A.149D}
}

@ARTICLE{2002AJ....124.1670M,
       author = {{Mamajek}, Eric E. and {Meyer}, Michael R. and {Liebert}, James},
        title = "{Post-T Tauri Stars in the Nearest OB Association}",
      journal = {\aj},
         year = 2002,
        month = sep,
       volume = {124},
       number = {3},
        pages = {1670-1694},
          doi = {10.1086/341952},
archivePrefix = {arXiv},
       eprint = {astro-ph/0205417},
 primaryClass = {astro-ph},
       adsurl = {https://ui.adsabs.harvard.edu/abs/2002AJ....124.1670M}
}

@ARTICLE{2010ApJS..186..111L,
       author = {{Luhman}, K.~L. and {Allen}, P.~R. and {Espaillat}, C. and {Hartmann}, L. and {Calvet}, N.},
        title = "{The Disk Population of the Taurus Star-Forming Region}",
      journal = {\apjs},
         year = 2010,
        month = jan,
       volume = {186},
       number = {1},
        pages = {111-174},
          doi = {10.1088/0067-0049/186/1/111},
archivePrefix = {arXiv},
       eprint = {0911.5457},
 primaryClass = {astro-ph.GA},
       adsurl = {https://ui.adsabs.harvard.edu/abs/2010ApJS..186..111L}
}

@ARTICLE{2022AJ....163...24L,
       author = {{Luhman}, K.~L.},
        title = "{A Census of the Stellar Populations in the Sco-Cen Complex}",
      journal = {\aj},
         year = 2022,
        month = jan,
       volume = {163},
       number = {1},
          eid = {24},
        pages = {24},
          doi = {10.3847/1538-3881/ac35e2},
archivePrefix = {arXiv},
       eprint = {2111.13946},
 primaryClass = {astro-ph.GA},
       adsurl = {https://ui.adsabs.harvard.edu/abs/2022AJ....163...24L}
}

@ARTICLE{2023ApJ...945..112F,
       author = {{Fang}, Min and {Pascucci}, Ilaria and {Edwards}, Suzan and {Gorti}, Uma and {Hillenbrand}, Lynne A. and {Carpenter}, John M.},
        title = "{A High-resolution Optical Survey of Upper Sco: Evidence for Coevolution of Accretion and Disk Winds}",
      journal = {\apj},
         year = 2023,
        month = mar,
       volume = {945},
       number = {2},
          eid = {112},
        pages = {112},
          doi = {10.3847/1538-4357/acb2c9},
archivePrefix = {arXiv},
       eprint = {2301.09240},
 primaryClass = {astro-ph.SR},
       adsurl = {https://ui.adsabs.harvard.edu/abs/2023ApJ...945..112F}
}

@ARTICLE{2018AJ....156..271L,
       author = {{Luhman}, K.~L.},
        title = "{The Stellar Membership of the Taurus Star-forming Region}",
      journal = {\aj},
         year = 2018,
        month = dec,
       volume = {156},
       number = {6},
          eid = {271},
        pages = {271},
          doi = {10.3847/1538-3881/aae831},
archivePrefix = {arXiv},
       eprint = {1811.01359},
 primaryClass = {astro-ph.SR},
       adsurl = {https://ui.adsabs.harvard.edu/abs/2018AJ....156..271L}
}

@ARTICLE{2023ApJ...949...27L,
       author = {{Long}, Feng and {Ren}, Bin B. and {Wallack}, Nicole L. and {Harsono}, Daniel and {Herczeg}, Gregory J. and {Pinilla}, Paola and {Mawet}, Dimitri and {Liu}, Michael C. and {Andrews}, Sean M. and {Bai}, Xue-Ning and {Cabrit}, Sylvie and {Cieza}, Lucas A. and {Johnstone}, Doug and {Leisenring}, Jarron M. and {Lodato}, Giuseppe and {Liu}, Yao and {Manara}, Carlo F. and {Mulders}, Gijs D. and {Ragusa}, Enrico and {Sallum}, Steph and {Shi}, Yangfan and {Tazzari}, Marco and {Uyama}, Taichi and {Wagner}, Kevin and {Wilner}, David J. and {Xuan}, Jerry W.},
        title = "{A Large Double-ring Disk Around the Taurus M Dwarf J04124068+2438157}",
      journal = {\apj},
         year = 2023,
        month = may,
       volume = {949},
       number = {1},
          eid = {27},
        pages = {27},
          doi = {10.3847/1538-4357/acc843},
archivePrefix = {arXiv},
       eprint = {2303.14586},
 primaryClass = {astro-ph.EP},
       adsurl = {https://ui.adsabs.harvard.edu/abs/2023ApJ...949...27L}
}

@ARTICLE{2006A&A...460..547G,
       author = {{Gatti}, T. and {Testi}, L. and {Natta}, A. and {Randich}, S. and {Muzerolle}, J.},
        title = "{Accretion in {\ensuremath{\rho}} Ophiuchus brown dwarfs: infrared hydrogen line ratios}",
      journal = {\aap},
         year = 2006,
        month = dec,
       volume = {460},
       number = {2},
        pages = {547-553},
          doi = {10.1051/0004-6361:20066095},
archivePrefix = {arXiv},
       eprint = {astro-ph/0609291},
 primaryClass = {astro-ph},
       adsurl = {https://ui.adsabs.harvard.edu/abs/2006A&A...460..547G}
}

@ARTICLE{2012ApJ...744..116B,
       author = {{Brown}, J.~M. and {Herczeg}, G.~J. and {Pontoppidan}, K.~M. and {van Dishoeck}, E.~F.},
        title = "{A 30 AU Radius CO Gas Hole in the Disk around the Herbig Ae Star Oph IRS 48}",
      journal = {\apj},
         year = 2012,
        month = jan,
       volume = {744},
       number = {2},
          eid = {116},
        pages = {116},
          doi = {10.1088/0004-637X/744/2/116},
archivePrefix = {arXiv},
       eprint = {1110.2095},
 primaryClass = {astro-ph.GA},
       adsurl = {https://ui.adsabs.harvard.edu/abs/2012ApJ...744..116B}
}

@ARTICLE{2018ApJ...865..157A,
       author = {{Andrews}, Sean M. and {Terrell}, Marie and {Tripathi}, Anjali and {Ansdell}, Megan and {Williams}, Jonathan P. and {Wilner}, David J.},
        title = "{Scaling Relations Associated with Millimeter Continuum Sizes in Protoplanetary Disks}",
      journal = {\apj},
         year = 2018,
        month = oct,
       volume = {865},
       number = {2},
          eid = {157},
        pages = {157},
          doi = {10.3847/1538-4357/aadd9f},
archivePrefix = {arXiv},
       eprint = {1808.10510},
 primaryClass = {astro-ph.EP},
       adsurl = {https://ui.adsabs.harvard.edu/abs/2018ApJ...865..157A}
}

@ARTICLE{2025ApJ...978..117C,
       author = {{Carpenter}, John M. and {Esplin}, Taran L. and {Luhman}, Kevin L. and {Mamajek}, Eric E. and {Andrews}, Sean M.},
        title = "{Extending the ALMA Census of Circumstellar Disks in the Upper Scorpius OB Association}",
      journal = {\apj},
         year = 2025,
        month = jan,
       volume = {978},
       number = {1},
          eid = {117},
        pages = {117},
          doi = {10.3847/1538-4357/ad8ebc},
archivePrefix = {arXiv},
       eprint = {2410.21598},
 primaryClass = {astro-ph.SR},
       adsurl = {https://ui.adsabs.harvard.edu/abs/2025ApJ...978..117C}
}

@ARTICLE{2004ApJ...616..998W,
       author = {{White}, Russel J. and {Hillenbrand}, Lynne A.},
        title = "{On the Evolutionary Status of Class I Stars and Herbig-Haro Energy Sources in Taurus-Auriga}",
      journal = {\apj},
         year = 2004,
        month = dec,
       volume = {616},
       number = {2},
        pages = {998-1032},
          doi = {10.1086/425115},
archivePrefix = {arXiv},
       eprint = {astro-ph/0408244},
 primaryClass = {astro-ph},
       adsurl = {https://ui.adsabs.harvard.edu/abs/2004ApJ...616..998W}
}

@ARTICLE{2021A&A...649A.122P,
       author = {{Pinilla}, P. and {Kurtovic}, N.~T. and {Benisty}, M. and {Manara}, C.~F. and {Natta}, A. and {Sanchis}, E. and {Tazzari}, M. and {Stammler}, S.~M. and {Ricci}, L. and {Testi}, L.},
        title = "{A bright inner disk and structures in the transition disk around the very low-mass star CIDA 1}",
      journal = {\aap},
         year = 2021,
        month = may,
       volume = {649},
          eid = {A122},
        pages = {A122},
          doi = {10.1051/0004-6361/202140371},
archivePrefix = {arXiv},
       eprint = {2103.10465},
 primaryClass = {astro-ph.EP},
       adsurl = {https://ui.adsabs.harvard.edu/abs/2021A&A...649A.122P}
}

@ARTICLE{2024Natur.633...58S,
       author = {{Speedie}, Jessica and {Dong}, Ruobing and {Hall}, Cassandra and {Longarini}, Cristiano and {Veronesi}, Benedetta and {Paneque-Carre{\~n}o}, Teresa and {Lodato}, Giuseppe and {Tang}, Ya-Wen and {Teague}, Richard and {Hashimoto}, Jun},
        title = "{Gravitational instability in a planet-forming disk}",
      journal = {\nat},
         year = 2024,
        month = sep,
       volume = {633},
       number = {8028},
        pages = {58-62},
          doi = {10.1038/s41586-024-07877-0},
archivePrefix = {arXiv},
       eprint = {2409.02196},
 primaryClass = {astro-ph.EP},
       adsurl = {https://ui.adsabs.harvard.edu/abs/2024Natur.633...58S}
}

@ARTICLE{2020ApJ...891...48H,
       author = {{Huang}, Jane and {Andrews}, Sean M. and {Dullemond}, Cornelis P. and {{\"O}berg}, Karin I. and {Qi}, Chunhua and {Zhu}, Zhaohuan and {Birnstiel}, Tilman and {Carpenter}, John M. and {Isella}, Andrea and {Mac{\'\i}as}, Enrique and {McClure}, Melissa K. and {P{\'e}rez}, Laura M. and {Teague}, Richard and {Wilner}, David J. and {Zhang}, Shangjia},
        title = "{A Multifrequency ALMA Characterization of Substructures in the GM Aur Protoplanetary Disk}",
      journal = {\apj},
         year = 2020,
        month = mar,
       volume = {891},
       number = {1},
          eid = {48},
        pages = {48},
          doi = {10.3847/1538-4357/ab711e},
archivePrefix = {arXiv},
       eprint = {2001.11040},
 primaryClass = {astro-ph.EP},
       adsurl = {https://ui.adsabs.harvard.edu/abs/2020ApJ...891...48H}
}

@ARTICLE{2024A&A...681A..19C,
       author = {{Columba}, G. and {Rigliaco}, E. and {Gratton}, R. and {Mesa}, D. and {D'Orazi}, V. and {Ginski}, C. and {Engler}, N. and {Williams}, J.~P. and {Bae}, J. and {Benisty}, M. and {Birnstiel}, T. and {Delorme}, P. and {Dominik}, C. and {Facchini}, S. and {Menard}, F. and {Pinilla}, P. and {Rab}, C. and {Ribas}, {\'A}. and {Squicciarini}, V. and {van Holstein}, R.~G. and {Zurlo}, A.},
        title = "{Disk Evolution Study Through Imaging of Nearby Young Stars (DESTINYS): HD 34700 A unveils an inner ring}",
      journal = {\aap},
         year = 2024,
        month = jan,
       volume = {681},
          eid = {A19},
        pages = {A19},
          doi = {10.1051/0004-6361/202347109},
archivePrefix = {arXiv},
       eprint = {2310.16873},
 primaryClass = {astro-ph.EP},
       adsurl = {https://ui.adsabs.harvard.edu/abs/2024A&A...681A..19C}
}

@ARTICLE{2018ApJ...859..111H,
       author = {{Hales}, A.~S. and {P{\'e}rez}, S. and {Saito}, M. and {Pinte}, C. and {Knee}, L.~B.~G. and {de Gregorio-Monsalvo}, I. and {Dent}, B. and {L{\'o}pez}, C. and {Plunkett}, A. and {Cort{\'e}s}, P. and {Corder}, S. and {Cieza}, L.},
        title = "{The Circumstellar Disk and Asymmetric Outflow of the EX Lup Outburst System}",
      journal = {\apj},
         year = 2018,
        month = jun,
       volume = {859},
       number = {2},
          eid = {111},
        pages = {111},
          doi = {10.3847/1538-4357/aac018},
archivePrefix = {arXiv},
       eprint = {1804.10340},
 primaryClass = {astro-ph.SR},
       adsurl = {https://ui.adsabs.harvard.edu/abs/2018ApJ...859..111H}
}

@ARTICLE{2022A&A...664A.151K,
       author = {{Kurtovic}, N.~T. and {Pinilla}, P. and {Penzlin}, Anna B.~T. and {Benisty}, M. and {P{\'e}rez}, L. and {Ginski}, C. and {Isella}, A. and {Kley}, W. and {Menard}, F. and {P{\'e}rez}, S. and {Bayo}, A.},
        title = "{The morphology of CS Cha circumbinary disk suggesting the existence of a Saturn-mass planet}",
      journal = {\aap},
         year = 2022,
        month = aug,
       volume = {664},
          eid = {A151},
        pages = {A151},
          doi = {10.1051/0004-6361/202243505},
archivePrefix = {arXiv},
       eprint = {2206.04427},
 primaryClass = {astro-ph.EP},
       adsurl = {https://ui.adsabs.harvard.edu/abs/2022A&A...664A.151K}
}

@ARTICLE{2021A&A...645A.139K,
       author = {{Kurtovic}, N.~T. and {Pinilla}, P. and {Long}, F. and {Benisty}, M. and {Manara}, C.~F. and {Natta}, A. and {Pascucci}, I. and {Ricci}, L. and {Scholz}, A. and {Testi}, L.},
        title = "{Size and structures of disks around very low mass stars in the Taurus star-forming region}",
      journal = {\aap},
         year = 2021,
        month = jan,
       volume = {645},
          eid = {A139},
        pages = {A139},
          doi = {10.1051/0004-6361/202038983},
archivePrefix = {arXiv},
       eprint = {2012.02225},
 primaryClass = {astro-ph.EP},
       adsurl = {https://ui.adsabs.harvard.edu/abs/2021A&A...645A.139K}
}

@ARTICLE{2025A&A...694A.147G,
       author = {{Gasman}, Danny and {Temmink}, Milou and {van Dishoeck}, Ewine F. and {Kurtovic}, Nicolas T. and {Grant}, Sierra L. and {Sellek}, Andrew and {Tabone}, Beno{\^\i}t and {Henning}, Thomas and {Kamp}, Inga and {G{\"u}del}, Manuel and {Barrado}, David and {Caratti o Garatti}, Alessio and {Glauser}, Adrian M. and {Waters}, Laurens B.~F.~M. and {Arabhavi}, Aditya M. and {Jang}, Hyerin and {Kanwar}, Jayatee and {Lienert}, Julia L. and {Perotti}, Giulia and {Schwarz}, Kamber and {Vlasblom}, Marissa},
        title = "{MINDS: The influence of outer dust disc structure on the volatile delivery to the inner disc}",
      journal = {\aap},
         year = 2025,
        month = feb,
       volume = {694},
          eid = {A147},
        pages = {A147},
          doi = {10.1051/0004-6361/202452152},
archivePrefix = {arXiv},
       eprint = {2501.04587},
 primaryClass = {astro-ph.EP},
       adsurl = {https://ui.adsabs.harvard.edu/abs/2025A&A...694A.147G}
}

@ARTICLE{2018ApJ...869L..42H,
       author = {{Huang}, Jane and {Andrews}, Sean M. and {Dullemond}, Cornelis P. and {Isella}, Andrea and {P{\'e}rez}, Laura M. and {Guzm{\'a}n}, Viviana V. and {{\"O}berg}, Karin I. and {Zhu}, Zhaohuan and {Zhang}, Shangjia and {Bai}, Xue-Ning and {Benisty}, Myriam and {Birnstiel}, Tilman and {Carpenter}, John M. and {Hughes}, A. Meredith and {Ricci}, Luca and {Weaver}, Erik and {Wilner}, David J.},
        title = "{The Disk Substructures at High Angular Resolution Project (DSHARP). II. Characteristics of Annular Substructures}",
      journal = {\apjl},
         year = 2018,
        month = dec,
       volume = {869},
       number = {2},
          eid = {L42},
        pages = {L42},
          doi = {10.3847/2041-8213/aaf740},
archivePrefix = {arXiv},
       eprint = {1812.04041},
 primaryClass = {astro-ph.EP},
       adsurl = {https://ui.adsabs.harvard.edu/abs/2018ApJ...869L..42H}
}

@ARTICLE{2018ApJ...869L..44K,
       author = {{Kurtovic}, Nicol{\'a}s T. and {P{\'e}rez}, Laura M. and {Benisty}, Myriam and {Zhu}, Zhaohuan and {Zhang}, Shangjia and {Huang}, Jane and {Andrews}, Sean M. and {Dullemond}, Cornelis P. and {Isella}, Andrea and {Bai}, Xue-Ning and {Carpenter}, John M. and {Guzm{\'a}n}, Viviana V. and {Ricci}, Luca and {Wilner}, David J.},
        title = "{The Disk Substructures at High Angular Resolution Project (DSHARP). IV. Characterizing Substructures and Interactions in Disks around Multiple Star Systems}",
      journal = {\apjl},
         year = 2018,
        month = dec,
       volume = {869},
       number = {2},
          eid = {L44},
        pages = {L44},
          doi = {10.3847/2041-8213/aaf746},
archivePrefix = {arXiv},
       eprint = {1812.04536},
 primaryClass = {astro-ph.SR},
       adsurl = {https://ui.adsabs.harvard.edu/abs/2018ApJ...869L..44K}
}

@ARTICLE{2025ApJ...984L...9C,
       author = {{Curone}, Pietro and {Facchini}, Stefano and {Andrews}, Sean M. and {Testi}, Leonardo and {Benisty}, Myriam and {Czekala}, Ian and {Huang}, Jane and {Ilee}, John D. and {Isella}, Andrea and {Lodato}, Giuseppe and {Loomis}, Ryan A. and {Stadler}, Jochen and {Winter}, Andrew J. and {Bae}, Jaehan and {Barraza-Alfaro}, Marcelo and {Cataldi}, Gianni and {Cuello}, Nicol{\'a}s and {Fasano}, Daniele and {Flock}, Mario and {Fukagawa}, Misato and {Galloway-Sprietsma}, Maria and {Garg}, Himanshi and {Hall}, Cassandra and {Izquierdo}, Andr{\'e}s F. and {Kanagawa}, Kazuhiro and {Lesur}, Geoffroy and {Longarini}, Cristiano and {Menard}, Francois and {Orihara}, Ryuta and {Pinte}, Christophe and {Price}, Daniel J. and {Rosotti}, Giovanni and {Teague}, Richard and {Wafflard-Fernandez}, Gaylor and {Wilner}, David J. and {W{\"o}lfer}, Lisa and {Yen}, Hsi-Wei and {Yoshida}, Tomohiro C. and {Zawadzki}, Brianna},
        title = "{exoALMA. IV. Substructures, Asymmetries, and the Faint Outer Disk in Continuum Emission}",
      journal = {\apjl},
         year = 2025,
        month = may,
       volume = {984},
       number = {1},
          eid = {L9},
        pages = {L9},
          doi = {10.3847/2041-8213/adc438},
archivePrefix = {arXiv},
       eprint = {2504.18725},
 primaryClass = {astro-ph.EP},
       adsurl = {https://ui.adsabs.harvard.edu/abs/2025ApJ...984L...9C}
}

@ARTICLE{1999ApJ...520..811W,
       author = {{White}, Russel J. and {Ghez}, A.~M. and {Reid}, I. Neill and {Schultz}, Greg},
        title = "{A Test of Pre-Main-Sequence Evolutionary Models across the Stellar/Substellar Boundary Based on Spectra of the Young Quadruple GG Tauri}",
      journal = {\apj},
         year = 1999,
        month = aug,
       volume = {520},
       number = {2},
        pages = {811-821},
          doi = {10.1086/307494},
archivePrefix = {arXiv},
       eprint = {astro-ph/9902318},
 primaryClass = {astro-ph},
       adsurl = {https://ui.adsabs.harvard.edu/abs/1999ApJ...520..811W}
}

@ARTICLE{2014A&A...568A..18M,
       author = {{Manara}, C.~F. and {Testi}, L. and {Natta}, A. and {Rosotti}, G. and {Benisty}, M. and {Ercolano}, B. and {Ricci}, L.},
        title = "{Gas content of transitional disks: a VLT/X-Shooter study of accretion and winds}",
      journal = {\aap},
         year = 2014,
        month = aug,
       volume = {568},
          eid = {A18},
        pages = {A18},
          doi = {10.1051/0004-6361/201323318},
archivePrefix = {arXiv},
       eprint = {1406.1428},
 primaryClass = {astro-ph.SR},
       adsurl = {https://ui.adsabs.harvard.edu/abs/2014A&A...568A..18M}
}

@ARTICLE{2005ApJ...623..952E,
       author = {{Eisner}, J.~A. and {Hillenbrand}, L.~A. and {White}, R.~J. and {Akeson}, R.~L. and {Sargent}, A.~I.},
        title = "{Observations of T Tauri Disks at Sub-AU Radii: Implications for Magnetospheric Accretion and Planet Formation}",
      journal = {\apj},
         year = 2005,
        month = apr,
       volume = {623},
       number = {2},
        pages = {952-966},
          doi = {10.1086/428828},
archivePrefix = {arXiv},
       eprint = {astro-ph/0501308},
 primaryClass = {astro-ph},
       adsurl = {https://ui.adsabs.harvard.edu/abs/2005ApJ...623..952E}
}

@ARTICLE{2014ApJ...786...97H,
       author = {{Herczeg}, Gregory J. and {Hillenbrand}, Lynne A.},
        title = "{An Optical Spectroscopic Study of T Tauri Stars. I. Photospheric Properties}",
      journal = {\apj},
         year = 2014,
        month = may,
       volume = {786},
       number = {2},
          eid = {97},
        pages = {97},
          doi = {10.1088/0004-637X/786/2/97},
archivePrefix = {arXiv},
       eprint = {1403.1675},
 primaryClass = {astro-ph.SR},
       adsurl = {https://ui.adsabs.harvard.edu/abs/2014ApJ...786...97H}
}

@ARTICLE{2025arXiv251026449E,
       author = {{Empey}, A. and {Garcia Lopez}, R. and {Natta}, A. and {Manara}, C.~F. and {Benisty}, M. and {Claes}, R. and {McGinnis}, P.},
        title = "{The dance of dust: Investigating young stellar object dipper variability}",
      journal = {arXiv e-prints},
         year = 2025,
        month = oct,
          eid = {arXiv:2510.26449},
        pages = {arXiv:2510.26449},
archivePrefix = {arXiv},
       eprint = {2510.26449},
 primaryClass = {astro-ph.SR},
       adsurl = {https://ui.adsabs.harvard.edu/abs/2025arXiv251026449E}
}

@ARTICLE{2025ApJ...989....9V,
       author = {{Vioque}, Miguel and {Kurtovic}, Nicol{\'a}s T. and {Trapman}, Leon and {Sierra}, Anibal and {P{\'e}rez}, Laura M. and {Zhang}, Ke and {Curone}, Pietro and {Rosotti}, Giovanni P. and {Carpenter}, John and {Tabone}, Beno{\^\i}t and {Pinilla}, Paola and {Deng}, Dingshan and {Pascucci}, Ilaria and {Miley}, James and {Agurto-Gangas}, Carolina and {Cieza}, Lucas A. and {Anania}, Rossella and {Ruiz-Rodriguez}, Dary A. and {Gonz{\'a}lez-Ruilova}, Camilo and {TorresVillanueva}, Estephani E. and {Kuznetsova}, Aleksandra},
        title = "{The ALMA Survey of Gas Evolution of PROtoplanetary Disks (AGE-PRO). X. Dust Substructures, Disk Geometries, and Dust-disk Radii}",
      journal = {\apj},
         year = 2025,
        month = aug,
       volume = {989},
       number = {1},
          eid = {9},
        pages = {9},
          doi = {10.3847/1538-4357/adc7b0},
archivePrefix = {arXiv},
       eprint = {2506.10746},
 primaryClass = {astro-ph.EP},
       adsurl = {https://ui.adsabs.harvard.edu/abs/2025ApJ...989....9V}
}

@ARTICLE{2021A&A...649A.124C,
       author = {{Cantat-Gaudin}, Tristan and {Brandt}, Timothy D.},
        title = "{Characterizing and correcting the proper motion bias of the bright Gaia EDR3 sources}",
      journal = {\aap},
         year = 2021,
        month = may,
       volume = {649},
          eid = {A124},
        pages = {A124},
          doi = {10.1051/0004-6361/202140807},
archivePrefix = {arXiv},
       eprint = {2103.07432},
 primaryClass = {astro-ph.GA},
       adsurl = {https://ui.adsabs.harvard.edu/abs/2021A&A...649A.124C}
}

@ARTICLE{2025ApJ...990L...9C,
       author = {{Close}, Laird M. and {van Capelleveen}, Richelle F. and {Weible}, Gabriel and {Wagner}, Kevin and {Haffert}, Sebastiaan Y. and {Males}, Jared R. and {Ilyin}, Ilya and {Kenworthy}, Matthew A. and {Li}, Jialin and {Long}, Joseph D. and {Ertel}, Steve and {Ginski}, Christian and {Weinberger}, Alycia J. and {Follette}, Kate and {Liberman}, Joshua and {Twitchell}, Katie and {Johnson}, Parker and {Kueny}, Jay and {Apai}, Daniel and {Doyon}, Rene and {Foster}, Warren and {Gasho}, Victor and {Van Gorkom}, Kyle and {Guyon}, Olivier and {Kautz}, Maggie Y. and {McLeod}, Avalon and {McEwen}, Eden and {Pearce}, Logan and {Schatz}, Lauren and {Hedglen}, Alexander D. and {Wu}, Ya-Lin and {Isbell}, Jacob and {Power}, Jenny and {Carlson}, Jared and {Close}, Emmeline and {Tonucci}, Elena and {Mars}, Matthijs},
        title = "{Wide Separation Planets in Time (WISPIT): Discovery of a Gap H{\ensuremath{\alpha}} Protoplanet WISPIT 2b with MagAO-X}",
      journal = {\apjl},
         year = 2025,
        month = sep,
       volume = {990},
       number = {1},
          eid = {L9},
        pages = {L9},
          doi = {10.3847/2041-8213/adf7a5},
archivePrefix = {arXiv},
       eprint = {2508.19046},
 primaryClass = {astro-ph.EP},
       adsurl = {https://ui.adsabs.harvard.edu/abs/2025ApJ...990L...9C}
}

@ARTICLE{2012A&A...538A..78L,
       author = {{Lindegren}, L. and {Lammers}, U. and {Hobbs}, D. and {O'Mullane}, W. and {Bastian}, U. and {Hern{\'a}ndez}, J.},
        title = "{The astrometric core solution for the Gaia mission. Overview of models, algorithms, and software implementation}",
      journal = {\aap},
         year = 2012,
        month = feb,
       volume = {538},
          eid = {A78},
        pages = {A78},
          doi = {10.1051/0004-6361/201117905},
archivePrefix = {arXiv},
       eprint = {1112.4139},
 primaryClass = {astro-ph.IM},
       adsurl = {https://ui.adsabs.harvard.edu/abs/2012A&A...538A..78L}
}

@ARTICLE{2023A&A...674A..10H,
       author = {{Holl}, B. and {Sozzetti}, A. and {Sahlmann}, J. and {Giacobbe}, P. and {S{\'e}gransan}, D. and {Unger}, N. and {Delisle}, J. -B. and {Barbato}, D. and {Lattanzi}, M.~G. and {Morbidelli}, R. and {Sosnowska}, D.},
        title = "{Gaia Data Release 3. Astrometric orbit determination with Markov chain Monte Carlo and genetic algorithms: Systems with stellar, sub-stellar, and planetary mass companions}",
      journal = {\aap},
         year = 2023,
        month = jun,
       volume = {674},
          eid = {A10},
        pages = {A10},
          doi = {10.1051/0004-6361/202244161},
archivePrefix = {arXiv},
       eprint = {2206.05439},
 primaryClass = {astro-ph.EP},
       adsurl = {https://ui.adsabs.harvard.edu/abs/2023A&A...674A..10H}
}

@ARTICLE{2023A&A...674A...9H,
       author = {{Halbwachs}, Jean-Louis and {Pourbaix}, Dimitri and {Arenou}, Fr{\'e}d{\'e}ric and {Galluccio}, Laurent and {Guillout}, Patrick and {Bauchet}, Nathalie and {Marchal}, Olivier and {Sadowski}, Gilles and {Teyssier}, David},
        title = "{Gaia Data Release 3. Astrometric binary star processing}",
      journal = {\aap},
         year = 2023,
        month = jun,
       volume = {674},
          eid = {A9},
        pages = {A9},
          doi = {10.1051/0004-6361/202243969},
archivePrefix = {arXiv},
       eprint = {2206.05726},
 primaryClass = {astro-ph.SR},
       adsurl = {https://ui.adsabs.harvard.edu/abs/2023A&A...674A...9H}
}

@ARTICLE{2023A&A...674A..34G,
       author = {{Gaia Collaboration} and {Arenou}, F. and {Babusiaux}, C. and {Barstow}, M.~A. and {Faigler}, S. and {Jorissen}, A. and {Kervella}, P. and {Mazeh}, T. and {Mowlavi}, N. and {Panuzzo}, P. and {Sahlmann}, J. and {Shahaf}, S. and {Sozzetti}, A. and {Bauchet}, N. and {Damerdji}, Y. and {Gavras}, P. and {Giacobbe}, P. and {Gosset}, E. and {Halbwachs}, J. -L. and {Holl}, B. and {Lattanzi}, M.~G. and {Leclerc}, N. and {Morel}, T. and {Pourbaix}, D. and {Re Fiorentin}, P. and {Sadowski}, G. and {S{\'e}gransan}, D. and {Siopis}, C. and {Teyssier}, D. and {Zwitter}, T. and {Planquart}, L. and {Brown}, A.~G.~A. and {Vallenari}, A. and {Prusti}, T. and {de Bruijne}, J.~H.~J. and {Biermann}, M. and {Creevey}, O.~L. and {Ducourant}, C. and {Evans}, D.~W. and {Eyer}, L. and {Guerra}, R. and {Hutton}, A. and {Jordi}, C. and {Klioner}, S.~A. and {Lammers}, U.~L. and {Lindegren}, L. and {Luri}, X. and {Mignard}, F. and {Panem}, C. and {Randich}, S. and {Sartoretti}, P. and {Soubiran}, C. and {Tanga}, P. and {Walton}, N.~A. and {Bailer-Jones}, C.~A.~L. and {Bastian}, U. and {Drimmel}, R. and {Jansen}, F. and {Katz}, D. and {van Leeuwen}, F. and {Bakker}, J. and {Cacciari}, C. and {Casta{\~n}eda}, J. and {De Angeli}, F. and {Fabricius}, C. and {Fouesneau}, M. and {Fr{\'e}mat}, Y. and {Galluccio}, L. and {Guerrier}, A. and {Heiter}, U. and {Masana}, E. and {Messineo}, R. and {Nicolas}, C. and {Nienartowicz}, K. and {Pailler}, F. and {Riclet}, F. and {Roux}, W. and {Seabroke}, G.~M. and {Sordo}, R. and {Th{\'e}venin}, F. and {Gracia-Abril}, G. and {Portell}, J. and {Altmann}, M. and {Andrae}, R. and {Audard}, M. and {Bellas-Velidis}, I. and {Benson}, K. and {Berthier}, J. and {Blomme}, R. and {Burgess}, P.~W. and {Busonero}, D. and {Busso}, G. and {C{\'a}novas}, H. and {Carry}, B. and {Cellino}, A. and {Cheek}, N. and {Clementini}, G. and {Davidson}, M. and {de Teodoro}, P. and {Nu{\~n}ez Campos}, M. and {Delchambre}, L. and {Dell'Oro}, A. and {Esquej}, P. and {Fern{\'a}ndez-Hern{\'a}ndez}, J. and {Fraile}, E. and {Garabato}, D. and {Garc{\'\i}a-Lario}, P. and {Haigron}, R. and {Hambly}, N.~C. and {Harrison}, D.~L. and {Hern{\'a}ndez}, J. and {Hestroffer}, D. and {Hodgkin}, S.~T. and {Jan{\ss}en}, K. and {Jevardat de Fombelle}, G. and {Jordan}, S. and {Krone-Martins}, A. and {Lanzafame}, A.~C. and {L{\"o}ffler}, W. and {Marchal}, O. and {Marrese}, P.~M. and {Moitinho}, A. and {Muinonen}, K. and {Osborne}, P. and {Pancino}, E. and {Pauwels}, T. and {Recio-Blanco}, A. and {Reyl{\'e}}, C. and {Riello}, M. and {Rimoldini}, L. and {Roegiers}, T. and {Rybizki}, J. and {Sarro}, L.~M. and {Smith}, M. and {Utrilla}, E. and {van Leeuwen}, M. and {Abbas}, U. and {{\'A}brah{\'a}m}, P. and {Abreu Aramburu}, A. and {Aerts}, C. and {Aguado}, J.~J. and {Ajaj}, M. and {Aldea-Montero}, F. and {Altavilla}, G. and {{\'A}lvarez}, M.~A. and {Alves}, J. and {Anders}, F. and {Anderson}, R.~I. and {Anglada Varela}, E. and {Antoja}, T. and {Baines}, D. and {Baker}, S.~G. and {Balaguer-N{\'u}{\~n}ez}, L. and {Balbinot}, E. and {Balog}, Z. and {Barache}, C. and {Barbato}, D. and {Barros}, M. and {Bartolom{\'e}}, S. and {Bassilana}, J. -L. and {Becciani}, U. and {Bellazzini}, M. and {Berihuete}, A. and {Bernet}, M. and {Bertone}, S. and {Bianchi}, L. and {Binnenfeld}, A. and {Blanco-Cuaresma}, S. and {Blazere}, A. and {Boch}, T. and {Bombrun}, A. and {Bossini}, D. and {Bouquillon}, S. and {Bragaglia}, A. and {Bramante}, L. and {Breedt}, E. and {Bressan}, A. and {Brouillet}, N. and {Brugaletta}, E. and {Bucciarelli}, B. and {Burlacu}, A. and {Butkevich}, A.~G. and {Buzzi}, R. and {Caffau}, E. and {Cancelliere}, R. and {Cantat-Gaudin}, T. and {Carballo}, R. and {Carlucci}, T. and {Carnerero}, M.~I. and {Carrasco}, J.~M. and {Casamiquela}, L. and {Castellani}, M. and {Castro-Ginard}, A. and {Chaoul}, L. and {Charlot}, P. and {Chemin}, L. and {Chiaramida}, V. and {Chiavassa}, A. and {Chornay}, N. and {Comoretto}, G. and {Contursi}, G. and {Cooper}, W.~J. and {Cornez}, T. and {Cowell}, S. and {Crifo}, F. and {Cropper}, M. and {Crosta}, M. and {Crowley}, C. and {Dafonte}, C. and {Dapergolas}, A. and {David}, P. and {de Laverny}, P. and {De Luise}, F. and {De March}, R. and {De Ridder}, J. and {de Souza}, R. and {de Torres}, A. and {del Peloso}, E.~F. and {del Pozo}, E. and {Delbo}, M. and {Delgado}, A. and {Delisle}, J. -B. and {Demouchy}, C. and {Dharmawardena}, T.~E. and {Diakite}, S. and {Diener}, C. and {Distefano}, E. and {Dolding}, C. and {Enke}, H. and {Fabre}, C. and {Fabrizio}, M. and {Fedorets}, G. and {Fernique}, P. and {Figueras}, F. and {Fournier}, Y. and {Fouron}, C. and {Fragkoudi}, F. and {Gai}, M. and {Garcia-Gutierrez}, A. and {Garcia-Reinaldos}, M. and {Garc{\'\i}a-Torres}, M. and {Garofalo}, A. and {Gavel}, A. and {Gerlach}, E. and {Geyer}, R. and {Gilmore}, G. and {Girona}, S. and {Giuffrida}, G. and {Gomel}, R. and {Gomez}, A. and {Gonz{\'a}lez-N{\'u}{\~n}ez}, J. and {Gonz{\'a}lez-Santamar{\'\i}a}, I. and {Gonz{\'a}lez-Vidal}, J.~J. and {Granvik}, M. and {Guillout}, P. and {Guiraud}, J. and {Guti{\'e}rrez-S{\'a}nchez}, R. and {Guy}, L.~P. and {Hatzidimitriou}, D. and {Hauser}, M. and {Haywood}, M. and {Helmer}, A. and {Helmi}, A. and {Sarmiento}, M.~H. and {Hidalgo}, S.~L. and {Hilger}, T. and {H{\l}adczuk}, N. and {Hobbs}, D. and {Holland}, G. and {Huckle}, H.~E. and {Jardine}, K. and {Jasniewicz}, G. and {Jean-Antoine Piccolo}, A. and {Jim{\'e}nez-Arranz}, {\'O}. and {Juaristi Campillo}, J. and {Julbe}, F. and {Karbevska}, L. and {Khanna}, S. and {Kordopatis}, G. and {Korn}, A.~J. and {K{\'o}sp{\'a}l}, {\'A}. and {Kostrzewa-Rutkowska}, Z. and {Kruszy{\'n}ska}, K. and {Kun}, M. and {Laizeau}, P. and {Lambert}, S. and {Lanza}, A.~F. and {Lasne}, Y. and {Le Campion}, J. -F. and {Lebreton}, Y. and {Lebzelter}, T. and {Leccia}, S. and {Lecoeur-Taibi}, I. and {Liao}, S. and {Licata}, E.~L. and {Lindstr{\o}m}, H.~E.~P. and {Lister}, T.~A. and {Livanou}, E. and {Lobel}, A. and {Lorca}, A. and {Loup}, C. and {Madrero Pardo}, P. and {Magdaleno Romeo}, A. and {Managau}, S. and {Mann}, R.~G. and {Manteiga}, M. and {Marchant}, J.~M. and {Marconi}, M. and {Marcos}, J. and {Marcos Santos}, M.~M.~S. and {Mar{\'\i}n Pina}, D. and {Marinoni}, S. and {Marocco}, F. and {Marshall}, D.~J. and {Martin Polo}, L. and {Mart{\'\i}n-Fleitas}, J.~M. and {Marton}, G. and {Mary}, N. and {Masip}, A. and {Massari}, D. and {Mastrobuono-Battisti}, A. and {McMillan}, P.~J. and {Messina}, S. and {Michalik}, D. and {Millar}, N.~R. and {Mints}, A. and {Molina}, D. and {Molinaro}, R. and {Moln{\'a}r}, L. and {Monari}, G. and {Mongui{\'o}}, M. and {Montegriffo}, P. and {Montero}, A. and {Mor}, R. and {Mora}, A. and {Morbidelli}, R. and {Morris}, D. and {Muraveva}, T. and {Murphy}, C.~P. and {Musella}, I. and {Nagy}, Z. and {Noval}, L. and {Oca{\~n}a}, F. and {Ogden}, A. and {Ordenovic}, C. and {Osinde}, J.~O. and {Pagani}, C. and {Pagano}, I. and {Palaversa}, L. and {Palicio}, P.~A. and {Pallas-Quintela}, L. and {Panahi}, A. and {Payne-Wardenaar}, S. and {Pe{\~n}alosa Esteller}, X. and {Penttil{\"a}}, A. and {Pichon}, B. and {Piersimoni}, A.~M. and {Pineau}, F. -X. and {Plachy}, E. and {Plum}, G. and {Poggio}, E. and {Pr{\v{s}}a}, A. and {Pulone}, L. and {Racero}, E. and {Ragaini}, S. and {Rainer}, M. and {Raiteri}, C.~M. and {Ramos}, P. and {Ramos-Lerate}, M. and {Regibo}, S. and {Richards}, P.~J. and {Rios Diaz}, C. and {Ripepi}, V. and {Riva}, A. and {Rix}, H. -W. and {Rixon}, G. and {Robichon}, N. and {Robin}, A.~C. and {Robin}, C. and {Roelens}, M. and {Rogues}, H.~R.~O. and {Rohrbasser}, L. and {Romero-G{\'o}mez}, M. and {Rowell}, N. and {Royer}, F. and {Ruz Mieres}, D. and {Rybicki}, K.~A. and {S{\'a}ez N{\'u}{\~n}ez}, A. and {Sagrist{\`a} Sell{\'e}s}, A. and {Salguero}, E. and {Samaras}, N. and {Sanchez Gimenez}, V. and {Sanna}, N. and {Santove{\~n}a}, R. and {Sarasso}, M. and {Schultheis}, M. and {Sciacca}, E. and {Segol}, M. and {Segovia}, J.~C. and {Semeux}, D. and {Siddiqui}, H.~I. and {Siebert}, A. and {Siltala}, L. and {Silvelo}, A. and {Slezak}, E. and {Slezak}, I. and {Smart}, R.~L. and {Snaith}, O.~N. and {Solano}, E. and {Solitro}, F. and {Souami}, D. and {Souchay}, J. and {Spagna}, A. and {Spina}, L. and {Spoto}, F. and {Steele}, I.~A. and {Steidelm{\"u}ller}, H. and {Stephenson}, C.~A. and {S{\"u}veges}, M. and {Surdej}, J. and {Szabados}, L. and {Szegedi-Elek}, E. and {Taris}, F. and {Taylor}, M.~B. and {Teixeira}, R. and {Tolomei}, L. and {Tonello}, N. and {Torra}, F. and {Torra}, J. and {Torralba Elipe}, G. and {Trabucchi}, M. and {Tsounis}, A.~T. and {Turon}, C. and {Ulla}, A. and {Unger}, N. and {Vaillant}, M.~V. and {van Dillen}, E. and {van Reeven}, W. and {Vanel}, O. and {Vecchiato}, A. and {Viala}, Y. and {Vicente}, D. and {Voutsinas}, S. and {Weiler}, M. and {Wevers}, T. and {Wyrzykowski}, {\L}. and {Yoldas}, A. and {Yvard}, P. and {Zhao}, H. and {Zorec}, J. and {Zucker}, S.},
        title = "{Gaia Data Release 3. Stellar multiplicity, a teaser for the hidden treasure}",
      journal = {\aap},
         year = 2023,
        month = jun,
       volume = {674},
          eid = {A34},
        pages = {A34},
          doi = {10.1051/0004-6361/202243782},
archivePrefix = {arXiv},
       eprint = {2206.05595},
 primaryClass = {astro-ph.SR},
       adsurl = {https://ui.adsabs.harvard.edu/abs/2023A&A...674A..34G}
}

@ARTICLE{2022MNRAS.513.5270P,
       author = {{Penoyre}, Zephyr and {Belokurov}, Vasily and {Evans}, N. Wyn},
        title = "{Astrometric identification of nearby binary stars - II. Astrometric binaries in the Gaia Catalogue of Nearby Stars}",
      journal = {\mnras},
         year = 2022,
        month = jul,
       volume = {513},
       number = {4},
        pages = {5270-5289},
          doi = {10.1093/mnras/stac1147},
archivePrefix = {arXiv},
       eprint = {2202.06963},
 primaryClass = {astro-ph.SR},
       adsurl = {https://ui.adsabs.harvard.edu/abs/2022MNRAS.513.5270P}
}

@ARTICLE{2022MNRAS.513.2437P,
       author = {{Penoyre}, Zephyr and {Belokurov}, Vasily and {Evans}, N. Wyn},
        title = "{Astrometric identification of nearby binary stars - I. Predicted astrometric signals}",
      journal = {\mnras},
         year = 2022,
        month = jun,
       volume = {513},
       number = {2},
        pages = {2437-2456},
          doi = {10.1093/mnras/stac959},
archivePrefix = {arXiv},
       eprint = {2111.10380},
 primaryClass = {astro-ph.SR},
       adsurl = {https://ui.adsabs.harvard.edu/abs/2022MNRAS.513.2437P}
}

@ARTICLE{2024MNRAS.527.3076D,
       author = {{Dodd}, Jonathan M. and {Oudmaijer}, Ren{\'e} D. and {Radley}, Isaac C. and {Vioque}, Miguel and {Frost}, Abigail J.},
        title = "{Gaia uncovers difference in B and Be star binarity at small scales: evidence for mass transfer causing the Be phenomenon}",
      journal = {\mnras},
         year = 2024,
        month = jan,
       volume = {527},
       number = {2},
        pages = {3076-3086},
          doi = {10.1093/mnras/stad3105},
archivePrefix = {arXiv},
       eprint = {2310.05653},
 primaryClass = {astro-ph.SR},
       adsurl = {https://ui.adsabs.harvard.edu/abs/2024MNRAS.527.3076D}
}

@INPROCEEDINGS{2023ASPC..534..539M,
       author = {{Manara}, C.~F. and {Ansdell}, M. and {Rosotti}, G.~P. and {Hughes}, A.~M. and {Armitage}, P.~J. and {Lodato}, G. and {Williams}, J.~P.},
        title = "{Demographics of Young Stars and their Protoplanetary Disks: Lessons Learned on Disk Evolution and its Connection to Planet Formation}",
    booktitle = {Protostars and Planets VII},
         year = 2023,
       editor = {{Inutsuka}, S. and {Aikawa}, Y. and {Muto}, T. and {Tomida}, K. and {Tamura}, M.},
       series = {Astronomical Society of the Pacific Conference Series},
       volume = {534},
        month = jul,
        pages = {539},
          doi = {10.48550/arXiv.2203.09930},
archivePrefix = {arXiv},
       eprint = {2203.09930},
 primaryClass = {astro-ph.SR},
       adsurl = {https://ui.adsabs.harvard.edu/abs/2023ASPC..534..539M}
}

@ARTICLE{2023AJ....165...37L,
       author = {{Luhman}, K.~L.},
        title = "{A Census of the Taurus Star-forming Region and Neighboring Associations with Gaia}",
      journal = {\aj},
         year = 2023,
        month = feb,
       volume = {165},
       number = {2},
          eid = {37},
        pages = {37},
          doi = {10.3847/1538-3881/ac9da3},
archivePrefix = {arXiv},
       eprint = {2211.09785},
 primaryClass = {astro-ph.GA},
       adsurl = {https://ui.adsabs.harvard.edu/abs/2023AJ....165...37L}
}

@ARTICLE{2019AJ....158...54E,
       author = {{Esplin}, T.~L. and {Luhman}, K.~L.},
        title = "{A Survey for New Members of Taurus from Stellar to Planetary Masses}",
      journal = {\aj},
         year = 2019,
        month = aug,
       volume = {158},
       number = {2},
          eid = {54},
        pages = {54},
          doi = {10.3847/1538-3881/ab2594},
archivePrefix = {arXiv},
       eprint = {1907.00055},
 primaryClass = {astro-ph.SR},
       adsurl = {https://ui.adsabs.harvard.edu/abs/2019AJ....158...54E}
}

@ARTICLE{2022ApJS..263...14C,
       author = {{Capistrant}, Benjamin K. and {Soares-Furtado}, Melinda and {Vanderburg}, Andrew and {Kounkel}, Marina and {Rappaport}, Saul A. and {Omohundro}, Mark and {Powell}, Brian P. and {Gagliano}, Robert and {Jacobs}, Thomas and {Kostov}, Veselin B. and {Kristiansen}, Martti H. and {LaCourse}, Daryll M. and {Schmitt}, Allan R. and {Schwengeler}, Hans Martin and {Terentev}, Ivan A.},
        title = "{A Population of Dipper Stars from the Transiting Exoplanet Survey Satellite Mission}",
      journal = {\apjs},
         year = 2022,
        month = nov,
       volume = {263},
       number = {1},
          eid = {14},
        pages = {14},
          doi = {10.3847/1538-4365/ac9125},
archivePrefix = {arXiv},
       eprint = {2209.03379},
 primaryClass = {astro-ph.SR},
       adsurl = {https://ui.adsabs.harvard.edu/abs/2022ApJS..263...14C}
}

@ARTICLE{2021ApJ...916L...2B,
       author = {{Benisty}, Myriam and {Bae}, Jaehan and {Facchini}, Stefano and {Keppler}, Miriam and {Teague}, Richard and {Isella}, Andrea and {Kurtovic}, Nicolas T. and {P{\'e}rez}, Laura M. and {Sierra}, Anibal and {Andrews}, Sean M. and {Carpenter}, John and {Czekala}, Ian and {Dominik}, Carsten and {Henning}, Thomas and {Menard}, Francois and {Pinilla}, Paola and {Zurlo}, Alice},
        title = "{A Circumplanetary Disk around PDS70c}",
      journal = {\apjl},
         year = 2021,
        month = jul,
       volume = {916},
       number = {1},
          eid = {L2},
        pages = {L2},
          doi = {10.3847/2041-8213/ac0f83},
archivePrefix = {arXiv},
       eprint = {2108.07123},
 primaryClass = {astro-ph.EP},
       adsurl = {https://ui.adsabs.harvard.edu/abs/2021ApJ...916L...2B}
}

@ARTICLE{2021AJ....161..148W,
       author = {{Wang}, J.~J. and {Vigan}, A. and {Lacour}, S. and {Nowak}, M. and {Stolker}, T. and {De Rosa}, R.~J. and {Ginzburg}, S. and {Gao}, P. and {Abuter}, R. and {Amorim}, A. and {Asensio-Torres}, R. and {Baub{\"o}ck}, M. and {Benisty}, M. and {Berger}, J.~P. and {Beust}, H. and {Beuzit}, J. -L. and {Blunt}, S. and {Boccaletti}, A. and {Bohn}, A. and {Bonnefoy}, M. and {Bonnet}, H. and {Brandner}, W. and {Cantalloube}, F. and {Caselli}, P. and {Charnay}, B. and {Chauvin}, G. and {Choquet}, E. and {Christiaens}, V. and {Cl{\'e}net}, Y. and {Coud{\'e} Du Foresto}, V. and {Cridland}, A. and {de Zeeuw}, P.~T. and {Dembet}, R. and {Dexter}, J. and {Drescher}, A. and {Duvert}, G. and {Eckart}, A. and {Eisenhauer}, F. and {Facchini}, S. and {Gao}, F. and {Garcia}, P. and {Garcia Lopez}, R. and {Gardner}, T. and {Gendron}, E. and {Genzel}, R. and {Gillessen}, S. and {Girard}, J. and {Haubois}, X. and {Hei{\ss}el}, G. and {Henning}, T. and {Hinkley}, S. and {Hippler}, S. and {Horrobin}, M. and {Houll{\'e}}, M. and {Hubert}, Z. and {Jim{\'e}nez-Rosales}, A. and {Jocou}, L. and {Kammerer}, J. and {Keppler}, M. and {Kervella}, P. and {Meyer}, M. and {Kreidberg}, L. and {Lagrange}, A. -M. and {Lapeyr{\`e}re}, V. and {Le Bouquin}, J. -B. and {L{\'e}na}, P. and {Lutz}, D. and {Maire}, A. -L. and {M{\'e}nard}, F. and {M{\'e}rand}, A. and {Molli{\`e}re}, P. and {Monnier}, J.~D. and {Mouillet}, D. and {M{\"u}ller}, A. and {Nasedkin}, E. and {Ott}, T. and {Otten}, G.~P.~P.~L. and {Paladini}, C. and {Paumard}, T. and {Perraut}, K. and {Perrin}, G. and {Pfuhl}, O. and {Pueyo}, L. and {Rameau}, J. and {Rodet}, L. and {Rodr{\'\i}guez-Coira}, G. and {Rousset}, G. and {Scheithauer}, S. and {Shangguan}, J. and {Shimizu}, T. and {Stadler}, J. and {Straub}, O. and {Straubmeier}, C. and {Sturm}, E. and {Tacconi}, L.~J. and {van Dishoeck}, E.~F. and {Vincent}, F. and {von Fellenberg}, S.~D. and {Ward-Duong}, K. and {Widmann}, F. and {Wieprecht}, E. and {Wiezorrek}, E. and {Woillez}, J. and {Gravity Collaboration}},
        title = "{Constraining the Nature of the PDS 70 Protoplanets with VLTI/GRAVITY}",
      journal = {\aj},
         year = 2021,
        month = mar,
       volume = {161},
       number = {3},
          eid = {148},
        pages = {148},
          doi = {10.3847/1538-3881/abdb2d},
archivePrefix = {arXiv},
       eprint = {2101.04187},
 primaryClass = {astro-ph.EP},
       adsurl = {https://ui.adsabs.harvard.edu/abs/2021AJ....161..148W}
}

@ARTICLE{2018A&A...617L...2M,
       author = {{M{\"u}ller}, A. and {Keppler}, M. and {Henning}, Th. and {Samland}, M. and {Chauvin}, G. and {Beust}, H. and {Maire}, A. -L. and {Molaverdikhani}, K. and {van Boekel}, R. and {Benisty}, M. and {Boccaletti}, A. and {Bonnefoy}, M. and {Cantalloube}, F. and {Charnay}, B. and {Baudino}, J. -L. and {Gennaro}, M. and {Long}, Z.~C. and {Cheetham}, A. and {Desidera}, S. and {Feldt}, M. and {Fusco}, T. and {Girard}, J. and {Gratton}, R. and {Hagelberg}, J. and {Janson}, M. and {Lagrange}, A. -M. and {Langlois}, M. and {Lazzoni}, C. and {Ligi}, R. and {M{\'e}nard}, F. and {Mesa}, D. and {Meyer}, M. and {Molli{\`e}re}, P. and {Mordasini}, C. and {Moulin}, T. and {Pavlov}, A. and {Pawellek}, N. and {Quanz}, S.~P. and {Ramos}, J. and {Rouan}, D. and {Sissa}, E. and {Stadler}, E. and {Vigan}, A. and {Wahhaj}, Z. and {Weber}, L. and {Zurlo}, A.},
        title = "{Orbital and atmospheric characterization of the planet within the gap of the PDS 70 transition disk}",
      journal = {\aap},
         year = 2018,
        month = sep,
       volume = {617},
          eid = {L2},
        pages = {L2},
          doi = {10.1051/0004-6361/201833584},
archivePrefix = {arXiv},
       eprint = {1806.11567},
 primaryClass = {astro-ph.EP},
       adsurl = {https://ui.adsabs.harvard.edu/abs/2018A&A...617L...2M}
}

@ARTICLE{2025MNRAS.539.1613H,
       author = {{Hammond}, Iain and {Christiaens}, Valentin and {Price}, Daniel J. and {Blakely}, Dori and {Trevascus}, David and {Bonse}, Markus J. and {Cantalloube}, Faustine and {Marleau}, Gabriel-Dominique and {Pinte}, Christophe and {Juillard}, Sandrine and {Samland}, Matthias and {Thompson}, William and {Wallace}, Alex},
        title = "{Keplerian motion of a compact source orbiting the inner disc of PDS 70: a third protoplanet in resonance with b and c?}",
      journal = {\mnras},
         year = 2025,
        month = may,
       volume = {539},
       number = {2},
        pages = {1613-1627},
          doi = {10.1093/mnras/staf586},
archivePrefix = {arXiv},
       eprint = {2504.11127},
 primaryClass = {astro-ph.EP},
       adsurl = {https://ui.adsabs.harvard.edu/abs/2025MNRAS.539.1613H}
}

@ARTICLE{2024A&A...685L...1C,
       author = {{Christiaens}, V. and {Samland}, M. and {Henning}, Th. and {Portilla-Revelo}, B. and {Perotti}, G. and {Matthews}, E. and {Absil}, O. and {Decin}, L. and {Kamp}, I. and {Boccaletti}, A. and {Tabone}, B. and {Marleau}, G. -D. and {van Dishoeck}, E.~F. and {G{\"u}del}, M. and {Lagage}, P. -O. and {Barrado}, D. and {Caratti o Garatti}, A. and {Glauser}, A.~M. and {Olofsson}, G. and {Ray}, T.~P. and {Scheithauer}, S. and {Vandenbussche}, B. and {Waters}, L.~B.~F.~M. and {Arabhavi}, A.~M. and {Grant}, S.~L. and {Jang}, H. and {Kanwar}, J. and {Schreiber}, J. and {Schwarz}, K. and {Temmink}, M. and {{\"O}stlin}, G.},
        title = "{MINDS: JWST/NIRCam imaging of the protoplanetary disk PDS 70. A spiral accretion stream and a potential third protoplanet}",
      journal = {\aap},
         year = 2024,
        month = may,
       volume = {685},
          eid = {L1},
        pages = {L1},
          doi = {10.1051/0004-6361/202349089},
archivePrefix = {arXiv},
       eprint = {2403.04855},
 primaryClass = {astro-ph.EP},
       adsurl = {https://ui.adsabs.harvard.edu/abs/2024A&A...685L...1C}
}

@ARTICLE{2024A&A...691A.155H,
       author = {{Huang}, Shuo and {van der Marel}, Nienke and {Portegies Zwart}, Simon},
        title = "{Origin of transition disk cavities: Pebble-accreting protoplanets vs super-Jupiters}",
      journal = {\aap},
         year = 2024,
        month = nov,
       volume = {691},
          eid = {A155},
        pages = {A155},
          doi = {10.1051/0004-6361/202451511},
archivePrefix = {arXiv},
       eprint = {2410.02856},
 primaryClass = {astro-ph.EP},
       adsurl = {https://ui.adsabs.harvard.edu/abs/2024A&A...691A.155H}
}

@ARTICLE{2018A&A...611A..74R,
       author = {{Reggiani}, M. and {Christiaens}, V. and {Absil}, O. and {Mawet}, D. and {Huby}, E. and {Choquet}, E. and {Gomez Gonzalez}, C.~A. and {Ruane}, G. and {Femenia}, B. and {Serabyn}, E. and {Matthews}, K. and {Barraza}, M. and {Carlomagno}, B. and {Defr{\`e}re}, D. and {Delacroix}, C. and {Habraken}, S. and {Jolivet}, A. and {Karlsson}, M. and {Orban de Xivry}, G. and {Piron}, P. and {Surdej}, J. and {Vargas Catalan}, E. and {Wertz}, O.},
        title = "{Discovery of a point-like source and a third spiral arm in the transition disk around the Herbig Ae star MWC 758}",
      journal = {\aap},
         year = 2018,
        month = mar,
       volume = {611},
          eid = {A74},
        pages = {A74},
          doi = {10.1051/0004-6361/201732016},
archivePrefix = {arXiv},
       eprint = {1710.11393},
 primaryClass = {astro-ph.EP},
       adsurl = {https://ui.adsabs.harvard.edu/abs/2018A&A...611A..74R}
}

@ARTICLE{2025A&A...699A.145D,
       author = {{Delfini}, L. and {Vioque}, M. and {Ribas}, {\'A}. and {Hodgkin}, S.},
        title = "{Star formation and accretion rates within 500 pc as traced by Gaia DR3 XP spectra}",
      journal = {\aap},
         year = 2025,
        month = jul,
       volume = {699},
          eid = {A145},
        pages = {A145},
          doi = {10.1051/0004-6361/202453539},
archivePrefix = {arXiv},
       eprint = {2505.04699},
 primaryClass = {astro-ph.SR},
       adsurl = {https://ui.adsabs.harvard.edu/abs/2025A&A...699A.145D}
}

@ARTICLE{2016A&A...595A...1G,
       author = {{Gaia Collaboration} and {Prusti}, T. and {de Bruijne}, J.~H.~J. and {Brown}, A.~G.~A. and {Vallenari}, A. and {Babusiaux}, C. and {Bailer-Jones}, C.~A.~L. and {Bastian}, U. and {Biermann}, M. and {Evans}, D.~W. and {Eyer}, L. and {Jansen}, F. and {Jordi}, C. and {Klioner}, S.~A. and {Lammers}, U. and {Lindegren}, L. and {Luri}, X. and {Mignard}, F. and {Milligan}, D.~J. and {Panem}, C. and {Poinsignon}, V. and {Pourbaix}, D. and {Randich}, S. and {Sarri}, G. and {Sartoretti}, P. and {Siddiqui}, H.~I. and {Soubiran}, C. and {Valette}, V. and {van Leeuwen}, F. and {Walton}, N.~A. and {Aerts}, C. and {Arenou}, F. and {Cropper}, M. and {Drimmel}, R. and {H{\o}g}, E. and {Katz}, D. and {Lattanzi}, M.~G. and {O'Mullane}, W. and {Grebel}, E.~K. and {Holland}, A.~D. and {Huc}, C. and {Passot}, X. and {Bramante}, L. and {Cacciari}, C. and {Casta{\~n}eda}, J. and {Chaoul}, L. and {Cheek}, N. and {De Angeli}, F. and {Fabricius}, C. and {Guerra}, R. and {Hern{\'a}ndez}, J. and {Jean-Antoine-Piccolo}, A. and {Masana}, E. and {Messineo}, R. and {Mowlavi}, N. and {Nienartowicz}, K. and {Ord{\'o}{\~n}ez-Blanco}, D. and {Panuzzo}, P. and {Portell}, J. and {Richards}, P.~J. and {Riello}, M. and {Seabroke}, G.~M. and {Tanga}, P. and {Th{\'e}venin}, F. and {Torra}, J. and {Els}, S.~G. and {Gracia-Abril}, G. and {Comoretto}, G. and {Garcia-Reinaldos}, M. and {Lock}, T. and {Mercier}, E. and {Altmann}, M. and {Andrae}, R. and {Astraatmadja}, T.~L. and {Bellas-Velidis}, I. and {Benson}, K. and {Berthier}, J. and {Blomme}, R. and {Busso}, G. and {Carry}, B. and {Cellino}, A. and {Clementini}, G. and {Cowell}, S. and {Creevey}, O. and {Cuypers}, J. and {Davidson}, M. and {De Ridder}, J. and {de Torres}, A. and {Delchambre}, L. and {Dell'Oro}, A. and {Ducourant}, C. and {Fr{\'e}mat}, Y. and {Garc{\'\i}a-Torres}, M. and {Gosset}, E. and {Halbwachs}, J. -L. and {Hambly}, N.~C. and {Harrison}, D.~L. and {Hauser}, M. and {Hestroffer}, D. and {Hodgkin}, S.~T. and {Huckle}, H.~E. and {Hutton}, A. and {Jasniewicz}, G. and {Jordan}, S. and {Kontizas}, M. and {Korn}, A.~J. and {Lanzafame}, A.~C. and {Manteiga}, M. and {Moitinho}, A. and {Muinonen}, K. and {Osinde}, J. and {Pancino}, E. and {Pauwels}, T. and {Petit}, J. -M. and {Recio-Blanco}, A. and {Robin}, A.~C. and {Sarro}, L.~M. and {Siopis}, C. and {Smith}, M. and {Smith}, K.~W. and {Sozzetti}, A. and {Thuillot}, W. and {van Reeven}, W. and {Viala}, Y. and {Abbas}, U. and {Abreu Aramburu}, A. and {Accart}, S. and {Aguado}, J.~J. and {Allan}, P.~M. and {Allasia}, W. and {Altavilla}, G. and {{\'A}lvarez}, M.~A. and {Alves}, J. and {Anderson}, R.~I. and {Andrei}, A.~H. and {Anglada Varela}, E. and {Antiche}, E. and {Antoja}, T. and {Ant{\'o}n}, S. and {Arcay}, B. and {Atzei}, A. and {Ayache}, L. and {Bach}, N. and {Baker}, S.~G. and {Balaguer-N{\'u}{\~n}ez}, L. and {Barache}, C. and {Barata}, C. and {Barbier}, A. and {Barblan}, F. and {Baroni}, M. and {Barrado y Navascu{\'e}s}, D. and {Barros}, M. and {Barstow}, M.~A. and {Becciani}, U. and {Bellazzini}, M. and {Bellei}, G. and {Bello Garc{\'\i}a}, A. and {Belokurov}, V. and {Bendjoya}, P. and {Berihuete}, A. and {Bianchi}, L. and {Bienaym{\'e}}, O. and {Billebaud}, F. and {Blagorodnova}, N. and {Blanco-Cuaresma}, S. and {Boch}, T. and {Bombrun}, A. and {Borrachero}, R. and {Bouquillon}, S. and {Bourda}, G. and {Bouy}, H. and {Bragaglia}, A. and {Breddels}, M.~A. and {Brouillet}, N. and {Br{\"u}semeister}, T. and {Bucciarelli}, B. and {Budnik}, F. and {Burgess}, P. and {Burgon}, R. and {Burlacu}, A. and {Busonero}, D. and {Buzzi}, R. and {Caffau}, E. and {Cambras}, J. and {Campbell}, H. and {Cancelliere}, R. and {Cantat-Gaudin}, T. and {Carlucci}, T. and {Carrasco}, J.~M. and {Castellani}, M. and {Charlot}, P. and {Charnas}, J. and {Charvet}, P. and {Chassat}, F. and {Chiavassa}, A. and {Clotet}, M. and {Cocozza}, G. and {Collins}, R.~S. and {Collins}, P. and {Costigan}, G.},
        title = "{The Gaia mission}",
      journal = {\aap},
         year = 2016,
        month = nov,
       volume = {595},
          eid = {A1},
        pages = {A1},
          doi = {10.1051/0004-6361/201629272},
archivePrefix = {arXiv},
       eprint = {1609.04153},
 primaryClass = {astro-ph.IM},
       adsurl = {https://ui.adsabs.harvard.edu/abs/2016A&A...595A...1G}
}

@ARTICLE{2018A&A...616A...1G,
       author = {{Gaia Collaboration} and {Brown}, A.~G.~A. and {Vallenari}, A. and {Prusti}, T. and {de Bruijne}, J.~H.~J. and {Babusiaux}, C. and {Bailer-Jones}, C.~A.~L. and {Biermann}, M. and {Evans}, D.~W. and {Eyer}, L. and {Jansen}, F. and {Jordi}, C. and {Klioner}, S.~A. and {Lammers}, U. and {Lindegren}, L. and {Luri}, X. and {Mignard}, F. and {Panem}, C. and {Pourbaix}, D. and {Randich}, S. and {Sartoretti}, P. and {Siddiqui}, H.~I. and {Soubiran}, C. and {van Leeuwen}, F. and {Walton}, N.~A. and {Arenou}, F. and {Bastian}, U. and {Cropper}, M. and {Drimmel}, R. and {Katz}, D. and {Lattanzi}, M.~G. and {Bakker}, J. and {Cacciari}, C. and {Casta{\~n}eda}, J. and {Chaoul}, L. and {Cheek}, N. and {De Angeli}, F. and {Fabricius}, C. and {Guerra}, R. and {Holl}, B. and {Masana}, E. and {Messineo}, R. and {Mowlavi}, N. and {Nienartowicz}, K. and {Panuzzo}, P. and {Portell}, J. and {Riello}, M. and {Seabroke}, G.~M. and {Tanga}, P. and {Th{\'e}venin}, F. and {Gracia-Abril}, G. and {Comoretto}, G. and {Garcia-Reinaldos}, M. and {Teyssier}, D. and {Altmann}, M. and {Andrae}, R. and {Audard}, M. and {Bellas-Velidis}, I. and {Benson}, K. and {Berthier}, J. and {Blomme}, R. and {Burgess}, P. and {Busso}, G. and {Carry}, B. and {Cellino}, A. and {Clementini}, G. and {Clotet}, M. and {Creevey}, O. and {Davidson}, M. and {De Ridder}, J. and {Delchambre}, L. and {Dell'Oro}, A. and {Ducourant}, C. and {Fern{\'a}ndez-Hern{\'a}ndez}, J. and {Fouesneau}, M. and {Fr{\'e}mat}, Y. and {Galluccio}, L. and {Garc{\'\i}a-Torres}, M. and {Gonz{\'a}lez-N{\'u}{\~n}ez}, J. and {Gonz{\'a}lez-Vidal}, J.~J. and {Gosset}, E. and {Guy}, L.~P. and {Halbwachs}, J. -L. and {Hambly}, N.~C. and {Harrison}, D.~L. and {Hern{\'a}ndez}, J. and {Hestroffer}, D. and {Hodgkin}, S.~T. and {Hutton}, A. and {Jasniewicz}, G. and {Jean-Antoine-Piccolo}, A. and {Jordan}, S. and {Korn}, A.~J. and {Krone-Martins}, A. and {Lanzafame}, A.~C. and {Lebzelter}, T. and {L{\"o}ffler}, W. and {Manteiga}, M. and {Marrese}, P.~M. and {Mart{\'\i}n-Fleitas}, J.~M. and {Moitinho}, A. and {Mora}, A. and {Muinonen}, K. and {Osinde}, J. and {Pancino}, E. and {Pauwels}, T. and {Petit}, J. -M. and {Recio-Blanco}, A. and {Richards}, P.~J. and {Rimoldini}, L. and {Robin}, A.~C. and {Sarro}, L.~M. and {Siopis}, C. and {Smith}, M. and {Sozzetti}, A. and {S{\"u}veges}, M. and {Torra}, J. and {van Reeven}, W. and {Abbas}, U. and {Abreu Aramburu}, A. and {Accart}, S. and {Aerts}, C. and {Altavilla}, G. and {{\'A}lvarez}, M.~A. and {Alvarez}, R. and {Alves}, J. and {Anderson}, R.~I. and {Andrei}, A.~H. and {Anglada Varela}, E. and {Antiche}, E. and {Antoja}, T. and {Arcay}, B. and {Astraatmadja}, T.~L. and {Bach}, N. and {Baker}, S.~G. and {Balaguer-N{\'u}{\~n}ez}, L. and {Balm}, P. and {Barache}, C. and {Barata}, C. and {Barbato}, D. and {Barblan}, F. and {Barklem}, P.~S. and {Barrado}, D. and {Barros}, M. and {Barstow}, M.~A. and {Bartholom{\'e} Mu{\~n}oz}, S. and {Bassilana}, J. -L. and {Becciani}, U. and {Bellazzini}, M. and {Berihuete}, A. and {Bertone}, S. and {Bianchi}, L. and {Bienaym{\'e}}, O. and {Blanco-Cuaresma}, S. and {Boch}, T. and {Boeche}, C. and {Bombrun}, A. and {Borrachero}, R. and {Bossini}, D. and {Bouquillon}, S. and {Bourda}, G. and {Bragaglia}, A. and {Bramante}, L. and {Breddels}, M.~A. and {Bressan}, A. and {Brouillet}, N. and {Br{\"u}semeister}, T. and {Brugaletta}, E. and {Bucciarelli}, B. and {Burlacu}, A. and {Busonero}, D. and {Butkevich}, A.~G. and {Buzzi}, R. and {Caffau}, E. and {Cancelliere}, R. and {Cannizzaro}, G. and {Cantat-Gaudin}, T. and {Carballo}, R. and {Carlucci}, T. and {Carrasco}, J.~M. and {Casamiquela}, L. and {Castellani}, M. and {Castro-Ginard}, A. and {Charlot}, P. and {Chemin}, L. and {Chiavassa}, A. and {Cocozza}, G. and {Costigan}, G. and {Cowell}, S. and {Crifo}, F. and {Crosta}, M. and {Crowley}, C. and {Cuypers}, J. and {Dafonte}, C. and {Damerdji}, Y. and {Dapergolas}, A. and {David}, P. and {David}, M. and {de Laverny}, P. and {De Luise}, F.},
        title = "{Gaia Data Release 2. Summary of the contents and survey properties}",
      journal = {\aap},
         year = 2018,
        month = aug,
       volume = {616},
          eid = {A1},
        pages = {A1},
          doi = {10.1051/0004-6361/201833051},
archivePrefix = {arXiv},
       eprint = {1804.09365},
 primaryClass = {astro-ph.GA},
       adsurl = {https://ui.adsabs.harvard.edu/abs/2018A&A...616A...1G}
}

@ARTICLE{2021A&A...649A...1G,
       author = {{Gaia Collaboration} and {Brown}, A.~G.~A. and {Vallenari}, A. and {Prusti}, T. and {de Bruijne}, J.~H.~J. and {Babusiaux}, C. and {Biermann}, M. and {Creevey}, O.~L. and {Evans}, D.~W. and {Eyer}, L. and {Hutton}, A. and {Jansen}, F. and {Jordi}, C. and {Klioner}, S.~A. and {Lammers}, U. and {Lindegren}, L. and {Luri}, X. and {Mignard}, F. and {Panem}, C. and {Pourbaix}, D. and {Randich}, S. and {Sartoretti}, P. and {Soubiran}, C. and {Walton}, N.~A. and {Arenou}, F. and {Bailer-Jones}, C.~A.~L. and {Bastian}, U. and {Cropper}, M. and {Drimmel}, R. and {Katz}, D. and {Lattanzi}, M.~G. and {van Leeuwen}, F. and {Bakker}, J. and {Cacciari}, C. and {Casta{\~n}eda}, J. and {De Angeli}, F. and {Ducourant}, C. and {Fabricius}, C. and {Fouesneau}, M. and {Fr{\'e}mat}, Y. and {Guerra}, R. and {Guerrier}, A. and {Guiraud}, J. and {Jean-Antoine Piccolo}, A. and {Masana}, E. and {Messineo}, R. and {Mowlavi}, N. and {Nicolas}, C. and {Nienartowicz}, K. and {Pailler}, F. and {Panuzzo}, P. and {Riclet}, F. and {Roux}, W. and {Seabroke}, G.~M. and {Sordo}, R. and {Tanga}, P. and {Th{\'e}venin}, F. and {Gracia-Abril}, G. and {Portell}, J. and {Teyssier}, D. and {Altmann}, M. and {Andrae}, R. and {Bellas-Velidis}, I. and {Benson}, K. and {Berthier}, J. and {Blomme}, R. and {Brugaletta}, E. and {Burgess}, P.~W. and {Busso}, G. and {Carry}, B. and {Cellino}, A. and {Cheek}, N. and {Clementini}, G. and {Damerdji}, Y. and {Davidson}, M. and {Delchambre}, L. and {Dell'Oro}, A. and {Fern{\'a}ndez-Hern{\'a}ndez}, J. and {Galluccio}, L. and {Garc{\'\i}a-Lario}, P. and {Garcia-Reinaldos}, M. and {Gonz{\'a}lez-N{\'u}{\~n}ez}, J. and {Gosset}, E. and {Haigron}, R. and {Halbwachs}, J. -L. and {Hambly}, N.~C. and {Harrison}, D.~L. and {Hatzidimitriou}, D. and {Heiter}, U. and {Hern{\'a}ndez}, J. and {Hestroffer}, D. and {Hodgkin}, S.~T. and {Holl}, B. and {Jan{\ss}en}, K. and {Jevardat de Fombelle}, G. and {Jordan}, S. and {Krone-Martins}, A. and {Lanzafame}, A.~C. and {L{\"o}ffler}, W. and {Lorca}, A. and {Manteiga}, M. and {Marchal}, O. and {Marrese}, P.~M. and {Moitinho}, A. and {Mora}, A. and {Muinonen}, K. and {Osborne}, P. and {Pancino}, E. and {Pauwels}, T. and {Petit}, J. -M. and {Recio-Blanco}, A. and {Richards}, P.~J. and {Riello}, M. and {Rimoldini}, L. and {Robin}, A.~C. and {Roegiers}, T. and {Rybizki}, J. and {Sarro}, L.~M. and {Siopis}, C. and {Smith}, M. and {Sozzetti}, A. and {Ulla}, A. and {Utrilla}, E. and {van Leeuwen}, M. and {van Reeven}, W. and {Abbas}, U. and {Abreu Aramburu}, A. and {Accart}, S. and {Aerts}, C. and {Aguado}, J.~J. and {Ajaj}, M. and {Altavilla}, G. and {{\'A}lvarez}, M.~A. and {{\'A}lvarez Cid-Fuentes}, J. and {Alves}, J. and {Anderson}, R.~I. and {Anglada Varela}, E. and {Antoja}, T. and {Audard}, M. and {Baines}, D. and {Baker}, S.~G. and {Balaguer-N{\'u}{\~n}ez}, L. and {Balbinot}, E. and {Balog}, Z. and {Barache}, C. and {Barbato}, D. and {Barros}, M. and {Barstow}, M.~A. and {Bartolom{\'e}}, S. and {Bassilana}, J. -L. and {Bauchet}, N. and {Baudesson-Stella}, A. and {Becciani}, U. and {Bellazzini}, M. and {Bernet}, M. and {Bertone}, S. and {Bianchi}, L. and {Blanco-Cuaresma}, S. and {Boch}, T. and {Bombrun}, A. and {Bossini}, D. and {Bouquillon}, S. and {Bragaglia}, A. and {Bramante}, L. and {Breedt}, E. and {Bressan}, A. and {Brouillet}, N. and {Bucciarelli}, B. and {Burlacu}, A. and {Busonero}, D. and {Butkevich}, A.~G. and {Buzzi}, R. and {Caffau}, E. and {Cancelliere}, R. and {C{\'a}novas}, H. and {Cantat-Gaudin}, T. and {Carballo}, R. and {Carlucci}, T. and {Carnerero}, M.~I. and {Carrasco}, J.~M. and {Casamiquela}, L. and {Castellani}, M. and {Castro-Ginard}, A. and {Castro Sampol}, P. and {Chaoul}, L. and {Charlot}, P. and {Chemin}, L. and {Chiavassa}, A. and {Cioni}, M. -R.~L. and {Comoretto}, G. and {Cooper}, W.~J. and {Cornez}, T. and {Cowell}, S. and {Crifo}, F. and {Crosta}, M. and {Crowley}, C. and {Dafonte}, C. and {Dapergolas}, A. and {David}, M. and {David}, P.},
        title = "{Gaia Early Data Release 3. Summary of the contents and survey properties}",
      journal = {\aap},
         year = 2021,
        month = may,
       volume = {649},
          eid = {A1},
        pages = {A1},
          doi = {10.1051/0004-6361/202039657},
archivePrefix = {arXiv},
       eprint = {2012.01533},
 primaryClass = {astro-ph.GA},
       adsurl = {https://ui.adsabs.harvard.edu/abs/2021A&A...649A...1G}
}

@ARTICLE{2022AJ....163...25L,
       author = {{Luhman}, K.~L.},
        title = "{A Census of the Circumstellar Disk Populations in the Sco-Cen Complex}",
      journal = {\aj},
         year = 2022,
        month = jan,
       volume = {163},
       number = {1},
          eid = {25},
        pages = {25},
          doi = {10.3847/1538-3881/ac35e3},
archivePrefix = {arXiv},
       eprint = {2111.13939},
 primaryClass = {astro-ph.GA},
       adsurl = {https://ui.adsabs.harvard.edu/abs/2022AJ....163...25L}
}

@ARTICLE{2025NatAs...9.1176R,
       author = {{Ribas}, {\'A}lvaro and {Vioque}, Miguel and {Zagaria}, Francesco and {Longarini}, Cristiano and {Mac{\'\i}as}, Enrique and {Clarke}, Cathie J. and {P{\'e}rez}, Sebasti{\'a}n and {Carpenter}, John and {Cuello}, Nicol{\'a}s and {de Gregorio-Monsalvo}, Itziar},
        title = "{A young gas giant and hidden substructures in a protoplanetary disk}",
      journal = {Nature Astronomy},
         year = 2025,
        month = aug,
       volume = {9},
        pages = {1176-1183},
          doi = {10.1038/s41550-025-02576-w},
archivePrefix = {arXiv},
       eprint = {2507.11612},
 primaryClass = {astro-ph.EP},
       adsurl = {https://ui.adsabs.harvard.edu/abs/2025NatAs...9.1176R}
}

@ARTICLE{2018MNRAS.475L..62H,
       author = {{Hendler}, Nathanial P. and {Pinilla}, Paola and {Pascucci}, Ilaria and {Pohl}, Adriana and {Mulders}, Gijs and {Henning}, Thomas and {Dong}, Ruobing and {Clarke}, Cathie and {Owen}, James and {Hollenbach}, David},
        title = "{A likely planet-induced gap in the disc around T Cha}",
      journal = {\mnras},
         year = 2018,
        month = mar,
       volume = {475},
       number = {1},
        pages = {L62-L66},
          doi = {10.1093/mnrasl/slx184},
       adsurl = {https://ui.adsabs.harvard.edu/abs/2018MNRAS.475L..62H}
}

@ARTICLE{2022ApJ...941...66H,
       author = {{Hashimoto}, Jun and {Liu}, Hauyu Baobab and {Dong}, Ruobing and {Liu}, Beibei and {Muto}, Takayuki},
        title = "{Grain Growth in the Dust Ring with a Crescent around the Very Low-mass Star ZZ Tau IRS with JVLA}",
      journal = {\apj},
         year = 2022,
        month = dec,
       volume = {941},
       number = {1},
          eid = {66},
        pages = {66},
          doi = {10.3847/1538-4357/aca01d},
archivePrefix = {arXiv},
       eprint = {2211.01570},
 primaryClass = {astro-ph.EP},
       adsurl = {https://ui.adsabs.harvard.edu/abs/2022ApJ...941...66H}
}

@ARTICLE{2021ApJ...915..131L,
       author = {{Long}, Feng and {Andrews}, Sean M. and {Vega}, Justin and {Wilner}, David J. and {Chandler}, Claire J. and {Ragusa}, Enrico and {Teague}, Richard and {P{\'e}rez}, Laura M. and {Calvet}, Nuria and {Carpenter}, John M. and {Henning}, Thomas and {Kwon}, Woojin and {Linz}, Hendrik and {Ricci}, Luca},
        title = "{The Architecture of the V892 Tau System: The Binary and Its Circumbinary Disk}",
      journal = {\apj},
         year = 2021,
        month = jul,
       volume = {915},
       number = {2},
          eid = {131},
        pages = {131},
          doi = {10.3847/1538-4357/abff53},
archivePrefix = {arXiv},
       eprint = {2105.02918},
 primaryClass = {astro-ph.EP},
       adsurl = {https://ui.adsabs.harvard.edu/abs/2021ApJ...915..131L}
}

@ARTICLE{2018ApJ...852..122H,
       author = {{Huang}, Jane and {Andrews}, Sean M. and {Cleeves}, L. Ilsedore and {{\"O}berg}, Karin I. and {Wilner}, David J. and {Bai}, Xuening and {Birnstiel}, Til and {Carpenter}, John and {Hughes}, A. Meredith and {Isella}, Andrea and {P{\'e}rez}, Laura M. and {Ricci}, Luca and {Zhu}, Zhaohuan},
        title = "{CO and Dust Properties in the TW Hya Disk from High-resolution ALMA Observations}",
      journal = {\apj},
         year = 2018,
        month = jan,
       volume = {852},
       number = {2},
          eid = {122},
        pages = {122},
          doi = {10.3847/1538-4357/aaa1e7},
archivePrefix = {arXiv},
       eprint = {1801.03948},
 primaryClass = {astro-ph.SR},
       adsurl = {https://ui.adsabs.harvard.edu/abs/2018ApJ...852..122H}
}

@ARTICLE{2021ApJ...923..128M,
       author = {{Mauc{\'o}}, Karina and {Carrasco-Gonz{\'a}lez}, Carlos and {Schreiber}, Matthias R. and {Sierra}, Anibal and {Olofsson}, Johan and {Bayo}, Amelia and {Caceres}, Claudio and {Canovas}, Hector and {Palau}, Aina},
        title = "{The Characterization of the Dust Content in the Ring Around Sz 91: Indications of Planetesimal Formation?}",
      journal = {\apj},
         year = 2021,
        month = dec,
       volume = {923},
       number = {1},
          eid = {128},
        pages = {128},
          doi = {10.3847/1538-4357/ac21d0},
archivePrefix = {arXiv},
       eprint = {2108.12548},
 primaryClass = {astro-ph.EP},
       adsurl = {https://ui.adsabs.harvard.edu/abs/2021ApJ...923..128M}
}

@ARTICLE{2021ApJ...908..250H,
       author = {{Hashimoto}, Jun and {Muto}, Takayuki and {Dong}, Ruobing and {Hasegawa}, Yasuhiro and {Marel}, Nienke van der and {Tamura}, Motohide and {Takami}, Michihiro and {Momose}, Munetake},
        title = "{ALMA Observations of the Inner Cavity in the Protoplanetary Disk around Sz 84}",
      journal = {\apj},
         year = 2021,
        month = feb,
       volume = {908},
       number = {2},
          eid = {250},
        pages = {250},
          doi = {10.3847/1538-4357/abba76},
archivePrefix = {arXiv},
       eprint = {2009.09912},
 primaryClass = {astro-ph.EP},
       adsurl = {https://ui.adsabs.harvard.edu/abs/2021ApJ...908..250H}
}

@ARTICLE{2023PASJ...75..424O,
       author = {{Orihara}, Ryuta and {Momose}, Munetake and {Muto}, Takayuki and {Hashimoto}, Jun and {Liu}, Hauyu Baobab and {Tsukagoshi}, Takashi and {Kudo}, Tomoyuki and {Takahashi}, Sanemichi and {Yang}, Yi and {Hasegawa}, Yasuhiro and {Dong}, Ruobing and {Konishi}, Mihoko and {Akiyama}, Eiji},
        title = "{ALMA Band 6 high-resolution observations of the transitional disk around SY Chamaeleontis}",
      journal = {\pasj},
         year = 2023,
        month = apr,
       volume = {75},
       number = {2},
        pages = {424-445},
          doi = {10.1093/pasj/psad009},
archivePrefix = {arXiv},
       eprint = {2302.05659},
 primaryClass = {astro-ph.EP},
       adsurl = {https://ui.adsabs.harvard.edu/abs/2023PASJ...75..424O}
}

@ARTICLE{2024MNRAS.532.1752R,
       author = {{Ribas}, {\'A}. and {Clarke}, Cathie J. and {Zagaria}, Francesco},
        title = "{Inner walls or vortices? Crescent-shaped asymmetries in ALMA observations of protoplanetary discs}",
      journal = {\mnras},
         year = 2024,
        month = aug,
       volume = {532},
       number = {2},
        pages = {1752-1764},
          doi = {10.1093/mnras/stae1534},
archivePrefix = {arXiv},
       eprint = {2406.14626},
 primaryClass = {astro-ph.EP},
       adsurl = {https://ui.adsabs.harvard.edu/abs/2024MNRAS.532.1752R}
}

@ARTICLE{2018ApJ...860..124D,
       author = {{Dong}, Ruobing and {Liu}, Sheng-yuan and {Eisner}, Josh and {Andrews}, Sean and {Fung}, Jeffrey and {Zhu}, Zhaohuan and {Chiang}, Eugene and {Hashimoto}, Jun and {Liu}, Hauyu Baobab and {Casassus}, Simon and {Esposito}, Thomas and {Hasegawa}, Yasuhiro and {Muto}, Takayuki and {Pavlyuchenkov}, Yaroslav and {Wilner}, David and {Akiyama}, Eiji and {Tamura}, Motohide and {Wisniewski}, John},
        title = "{The Eccentric Cavity, Triple Rings, Two-armed Spirals, and Double Clumps of the MWC 758 Disk}",
      journal = {\apj},
         year = 2018,
        month = jun,
       volume = {860},
       number = {2},
          eid = {124},
        pages = {124},
          doi = {10.3847/1538-4357/aac6cb},
archivePrefix = {arXiv},
       eprint = {1805.12141},
 primaryClass = {astro-ph.SR},
       adsurl = {https://ui.adsabs.harvard.edu/abs/2018ApJ...860..124D}
}

@ARTICLE{2022A&A...665A.128P,
       author = {{Pinilla}, P. and {Benisty}, M. and {Kurtovic}, N.~T. and {Bae}, J. and {Dong}, R. and {Zhu}, Z. and {Andrews}, S. and {Carpenter}, J. and {Ginski}, C. and {Huang}, J. and {Isella}, A. and {P{\'e}rez}, L. and {Ricci}, L. and {Rosotti}, G. and {Villenave}, M. and {Wilner}, D.},
        title = "{Distributions of gas and small and large grains in the LkH{\ensuremath{\alpha}} 330 disk trace a young planetary system}",
      journal = {\aap},
         year = 2022,
        month = sep,
       volume = {665},
          eid = {A128},
        pages = {A128},
          doi = {10.1051/0004-6361/202243704},
archivePrefix = {arXiv},
       eprint = {2206.09975},
 primaryClass = {astro-ph.EP},
       adsurl = {https://ui.adsabs.harvard.edu/abs/2022A&A...665A.128P}
}

@ARTICLE{2022ApJ...937L...1L,
       author = {{Long}, Feng and {Andrews}, Sean M. and {Zhang}, Shangjia and {Qi}, Chunhua and {Benisty}, Myriam and {Facchini}, Stefano and {Isella}, Andrea and {Wilner}, David J. and {Bae}, Jaehan and {Huang}, Jane and {Loomis}, Ryan A. and {{\"O}berg}, Karin I. and {Zhu}, Zhaohuan},
        title = "{ALMA Detection of Dust Trapping around Lagrangian Points in the LkCa 15 Disk}",
      journal = {\apjl},
         year = 2022,
        month = sep,
       volume = {937},
       number = {1},
          eid = {L1},
        pages = {L1},
          doi = {10.3847/2041-8213/ac8b10},
archivePrefix = {arXiv},
       eprint = {2209.05535},
 primaryClass = {astro-ph.EP},
       adsurl = {https://ui.adsabs.harvard.edu/abs/2022ApJ...937L...1L}
}

@ARTICLE{2023A&A...670L...1S,
       author = {{Stadler}, J. and {Benisty}, M. and {Izquierdo}, A. and {Facchini}, S. and {Teague}, R. and {Kurtovic}, N. and {Pinilla}, P. and {Bae}, J. and {Ansdell}, M. and {Loomis}, R. and {Mayama}, S. and {Perez}, L.~M. and {Testi}, L.},
        title = "{A kinematically detected planet candidate in a transition disk}",
      journal = {\aap},
         year = 2023,
        month = feb,
       volume = {670},
          eid = {L1},
        pages = {L1},
          doi = {10.1051/0004-6361/202245381},
archivePrefix = {arXiv},
       eprint = {2301.01684},
 primaryClass = {astro-ph.EP},
       adsurl = {https://ui.adsabs.harvard.edu/abs/2023A&A...670L...1S}
}

@ARTICLE{2023ApJ...948L...2Y,
       author = {{Yang}, Haifeng and {Fern{\'a}ndez-L{\'o}pez}, Manuel and {Li}, Zhi-Yun and {Stephens}, Ian W. and {Looney}, Leslie W. and {Lin}, Zhe-Yu Daniel and {Harrison}, Rachel},
        title = "{Eccentric Dust Ring in the IRS 48 Transition Disk}",
      journal = {\apjl},
         year = 2023,
        month = may,
       volume = {948},
       number = {1},
          eid = {L2},
        pages = {L2},
          doi = {10.3847/2041-8213/acccf8},
archivePrefix = {arXiv},
       eprint = {2304.02937},
 primaryClass = {astro-ph.SR},
       adsurl = {https://ui.adsabs.harvard.edu/abs/2023ApJ...948L...2Y}
}

@ARTICLE{2025ApJ...981L...4D,
       author = {{Dasgupta}, Anuroop and {Cieza}, Lucas A. and {Gonz{\'a}lez-Ruilova}, Camilo and {Bhowmik}, Trisha and {Chavan}, Prachi and {Batalla-Falcon}, Grace and {Herczeg}, Gregory and {Ruiz-Rodriguez}, Dary and {Williams}, Jonathan P. and {Sierra}, Anibal and {Casassus}, Simon and {Guilera}, Octavio and {P{\'e}rez}, Sebastian and {Orcajo}, Santiago and {Nogueira}, P.~H. and {Hales}, A. S and {Miley}, J.~M. and {Rannou}, Fernando R. and {Zurlo}, Alice},
        title = "{The Ophiuchus DIsk Survey Employing ALMA (ODISEA): Complete Size Distributions for the 100 Brightest Disks across Multiplicity and Spectral Energy Distribution Classes}",
      journal = {\apjl},
         year = 2025,
        month = mar,
       volume = {981},
       number = {1},
          eid = {L4},
        pages = {L4},
          doi = {10.3847/2041-8213/adb03c},
archivePrefix = {arXiv},
       eprint = {2501.15789},
 primaryClass = {astro-ph.EP},
       adsurl = {https://ui.adsabs.harvard.edu/abs/2025ApJ...981L...4D}
}

@ARTICLE{2025ApJ...993...90S,
       author = {{Shoshi}, Ayumu and {Muto}, Takayuki and {Bosschaart}, Quincy and {van der Marel}, Nienke and {Mulders}, Gijs D. and {Omura}, Mitsuki and {Tokuda}, Kazuki and {Machida}, Masahiro N.},
        title = "{ALMA High-resolution Observation for the Transitional Disk Around IRAS 04125+2902}",
      journal = {\apj},
         year = 2025,
        month = nov,
       volume = {993},
       number = {1},
          eid = {90},
        pages = {90},
          doi = {10.3847/1538-4357/ae045a},
archivePrefix = {arXiv},
       eprint = {2509.01896},
 primaryClass = {astro-ph.EP},
       adsurl = {https://ui.adsabs.harvard.edu/abs/2025ApJ...993...90S}
}

@ARTICLE{2021MNRAS.502.5779N,
       author = {{Norfolk}, Brodie J. and {Maddison}, Sarah T. and {Pinte}, Christophe and {van der Marel}, Nienke and {Booth}, Richard A. and {Francis}, Logan and {Gonzalez}, Jean-Fran{\c{c}}ois and {M{\'e}nard}, Fran{\c{c}}ois and {Wright}, Chris M. and {van der Plas}, Gerrit and {Garg}, Himanshi},
        title = "{Dust traps and the formation of cavities in transition discs: a millimetre to sub-millimetre comparison survey}",
      journal = {\mnras},
         year = 2021,
        month = apr,
       volume = {502},
       number = {4},
        pages = {5779-5796},
          doi = {10.1093/mnras/stab313},
archivePrefix = {arXiv},
       eprint = {2102.02316},
 primaryClass = {astro-ph.EP},
       adsurl = {https://ui.adsabs.harvard.edu/abs/2021MNRAS.502.5779N}
}

@ARTICLE{2019AJ....158...15P,
       author = {{P{\'e}rez}, Sebasti{\'a}n and {Casassus}, Simon and {Baruteau}, Cl{\'e}ment and {Dong}, Ruobing and {Hales}, Antonio and {Cieza}, Lucas},
        title = "{Dust Unveils the Formation of a Mini-Neptune Planet in a Protoplanetary Ring}",
      journal = {\aj},
         year = 2019,
        month = jul,
       volume = {158},
       number = {1},
          eid = {15},
        pages = {15},
          doi = {10.3847/1538-3881/ab1f88},
archivePrefix = {arXiv},
       eprint = {1902.05143},
 primaryClass = {astro-ph.EP},
       adsurl = {https://ui.adsabs.harvard.edu/abs/2019AJ....158...15P}
}

@ARTICLE{2018ApJ...869L..50P,
       author = {{P{\'e}rez}, Laura M. and {Benisty}, Myriam and {Andrews}, Sean M. and {Isella}, Andrea and {Dullemond}, Cornelis P. and {Huang}, Jane and {Kurtovic}, Nicol{\'a}s T. and {Guzm{\'a}n}, Viviana V. and {Zhu}, Zhaohuan and {Birnstiel}, Tilman and {Zhang}, Shangjia and {Carpenter}, John M. and {Wilner}, David J. and {Ricci}, Luca and {Bai}, Xue-Ning and {Weaver}, Erik and {{\"O}berg}, Karin I.},
        title = "{The Disk Substructures at High Angular Resolution Project (DSHARP). X. Multiple Rings, a Misaligned Inner Disk, and a Bright Arc in the Disk around the T Tauri star HD 143006}",
      journal = {\apjl},
         year = 2018,
        month = dec,
       volume = {869},
       number = {2},
          eid = {L50},
        pages = {L50},
          doi = {10.3847/2041-8213/aaf745},
archivePrefix = {arXiv},
       eprint = {1812.04049},
 primaryClass = {astro-ph.SR},
       adsurl = {https://ui.adsabs.harvard.edu/abs/2018ApJ...869L..50P}
}

@ARTICLE{2021MNRAS.507.3789C,
       author = {{Casassus}, Simon and {Christiaens}, Valentin and {C{\'a}rcamo}, Miguel and {P{\'e}rez}, Sebasti{\'a}n and {Weber}, Philipp and {Ercolano}, Barbara and {van der Marel}, Nienke and {Pinte}, Christophe and {Dong}, Ruobing and {Baruteau}, Cl{\'e}ment and {Cieza}, Lucas and {van Dishoeck}, Ewine F. and {Jordan}, Andr{\'e}s and {Price}, Daniel J. and {Absil}, Olivier and {Arce-Tord}, Carla and {Faramaz}, Virginie and {Flores}, Christian and {Reggiani}, Maddalena},
        title = "{A dusty filament and turbulent CO spirals in HD 135344B - SAO 206462}",
      journal = {\mnras},
         year = 2021,
        month = nov,
       volume = {507},
       number = {3},
        pages = {3789-3809},
          doi = {10.1093/mnras/stab2359},
archivePrefix = {arXiv},
       eprint = {2104.08379},
 primaryClass = {astro-ph.EP},
       adsurl = {https://ui.adsabs.harvard.edu/abs/2021MNRAS.507.3789C}
}

@ARTICLE{2022ApJ...933L...4C,
       author = {{Casassus}, Simon and {C{\'a}rcamo}, Miguel and {Hales}, Antonio and {Weber}, Philipp and {Dent}, Bill},
        title = "{The Doppler Flip in HD 100546 as a Disk Eruption: The Elephant in the Room of Kinematic Protoplanet Searches}",
      journal = {\apjl},
         year = 2022,
        month = jul,
       volume = {933},
       number = {1},
          eid = {L4},
        pages = {L4},
          doi = {10.3847/2041-8213/ac75e8},
archivePrefix = {arXiv},
       eprint = {2206.03236},
 primaryClass = {astro-ph.EP},
       adsurl = {https://ui.adsabs.harvard.edu/abs/2022ApJ...933L...4C}
}

@ARTICLE{2020MNRAS.499.3837G,
       author = {{Gonzalez}, Jean-Fran{\c{c}}ois and {van der Plas}, Gerrit and {Pinte}, Christophe and {Cuello}, Nicol{\'a}s and {Nealon}, Rebecca and {M{\'e}nard}, Fran{\c{c}}ois and {Revol}, Alexandre and {Rodet}, Laetitia and {Langlois}, Maud and {Maire}, Anne-Lise},
        title = "{Spirals, shadows, and precession in HD 100453 - I. The orbit of the binary}",
      journal = {\mnras},
         year = 2020,
        month = dec,
       volume = {499},
       number = {3},
        pages = {3837-3856},
          doi = {10.1093/mnras/staa2938},
archivePrefix = {arXiv},
       eprint = {2009.10504},
 primaryClass = {astro-ph.EP},
       adsurl = {https://ui.adsabs.harvard.edu/abs/2020MNRAS.499.3837G}
}

@INPROCEEDINGS{2019ASPC..523..637Y,
       author = {{Yamaguchi}, Masayuki and {Akiyama}, Kazunori and {Kataoka}, Akimasa and {Tsukagoshi}, Takashi and {Muto}, Takayuki and {Ikeda}, Shrio and {Fukagawa}, Misato and {Honma}, Mareki and {Kawabe}, Ryohei},
        title = "{Super-resolution Imaging of the Protoplanetary Disk HD 142527 using Sparse Modeling}",
    booktitle = {Astronomical Data Analysis Software and Systems XXVII},
         year = 2019,
       editor = {{Teuben}, Peter J. and {Pound}, Marc W. and {Thomas}, Brian A. and {Warner}, Elizabeth M.},
       series = {Astronomical Society of the Pacific Conference Series},
       volume = {523},
        month = oct,
        pages = {637},
       adsurl = {https://ui.adsabs.harvard.edu/abs/2019ASPC..523..637Y}
}

@ARTICLE{2024A&A...684A.134R,
       author = {{Rota}, A.~A. and {Meijerhof}, J.~D. and {van der Marel}, N. and {Francis}, L. and {van der Tak}, F.~F.~S. and {Sellek}, A.~D.},
        title = "{Correlation between accretion rate and free-free emission in protoplanetary disks. A multiwavelength analysis of central mm/cm emission in transition disks}",
      journal = {\aap},
         year = 2024,
        month = apr,
       volume = {684},
          eid = {A134},
        pages = {A134},
          doi = {10.1051/0004-6361/202348387},
archivePrefix = {arXiv},
       eprint = {2401.05798},
 primaryClass = {astro-ph.EP},
       adsurl = {https://ui.adsabs.harvard.edu/abs/2024A&A...684A.134R}
}

@ARTICLE{2021ApJ...911....5H,
       author = {{Hashimoto}, Jun and {Muto}, Takayuki and {Dong}, Ruobing and {Liu}, Hauyu Baobab and {van der Marel}, Nienke and {Francis}, Logan and {Hasegawa}, Yasuhiro and {Tsukagoshi}, Takashi},
        title = "{ALMA Observations of the Asymmetric Dust Disk around DM Tau}",
      journal = {\apj},
         year = 2021,
        month = apr,
       volume = {911},
       number = {1},
          eid = {5},
        pages = {5},
          doi = {10.3847/1538-4357/abe59f},
archivePrefix = {arXiv},
       eprint = {2102.05905},
 primaryClass = {astro-ph.EP},
       adsurl = {https://ui.adsabs.harvard.edu/abs/2021ApJ...911....5H}
}

@ARTICLE{2024ApJ...961...28H,
       author = {{Harsono}, Daniel and {Long}, Feng and {Pinilla}, Paola and {Rota}, Alessia A. and {Manara}, Carlo F. and {Herczeg}, Gregory J. and {Johnstone}, Doug and {Rosotti}, Giovanni and {Lodato}, Giuseppe and {Menard}, Francois and {Tazzari}, Marco and {Shi}, Yangfan},
        title = "{Dual-band Observations of the Asymmetric Ring around CIDA 9A: Dead or Alive?}",
      journal = {\apj},
         year = 2024,
        month = jan,
       volume = {961},
       number = {1},
          eid = {28},
        pages = {28},
          doi = {10.3847/1538-4357/ad0835},
archivePrefix = {arXiv},
       eprint = {2310.11007},
 primaryClass = {astro-ph.SR},
       adsurl = {https://ui.adsabs.harvard.edu/abs/2024ApJ...961...28H}
}

@ARTICLE{2021AJ....161..264H,
       author = {{Hashimoto}, Jun and {Dong}, Ruobing and {Muto}, Takayuki},
        title = "{An Asymmetric Dust Ring around a Very Low Mass Star ZZ Tau IRS}",
      journal = {\aj},
         year = 2021,
        month = jun,
       volume = {161},
       number = {6},
          eid = {264},
        pages = {264},
          doi = {10.3847/1538-3881/abf431},
archivePrefix = {arXiv},
       eprint = {2103.16731},
 primaryClass = {astro-ph.EP},
       adsurl = {https://ui.adsabs.harvard.edu/abs/2021AJ....161..264H}
}

@ARTICLE{2024A&A...687A.311A,
       author = {{Alaguero}, Antoine and {Cuello}, Nicol{\'a}s and {M{\'e}nard}, Fran{\c{c}}ois and {Ceppi}, Simone and {Ribas}, {\'A}lvaro and {Nealon}, Rebecca and {Vioque}, Miguel and {Izquierdo}, Andr{\'e}s and {Miley}, James and {Mac{\'\i}as}, Enrique and {Price}, Daniel J.},
        title = "{V892 Tau: A tidally perturbed circumbinary disc in a triple stellar system}",
      journal = {\aap},
         year = 2024,
        month = jul,
       volume = {687},
          eid = {A311},
        pages = {A311},
          doi = {10.1051/0004-6361/202449683},
archivePrefix = {arXiv},
       eprint = {2405.12593},
 primaryClass = {astro-ph.EP},
       adsurl = {https://ui.adsabs.harvard.edu/abs/2024A&A...687A.311A}
}

@ARTICLE{2025A&A...703A.210A,
       author = {{Alaguero}, Antoine and {M{\'e}nard}, Fran{\c{c}}ois and {Cuello}, Nicol{\'a}s and {Ribas}, {\'A}lvaro and {Viscardi}, Elena and {Mac{\'\i}as}, Enrique and {Vioque}, Miguel and {Miley}, James},
        title = "{Probing dust and grain growth in the optically thick circumbinary ring of V892 Tau}",
      journal = {\aap},
         year = 2025,
        month = nov,
       volume = {703},
          eid = {A210},
        pages = {A210},
          doi = {10.1051/0004-6361/202556077},
archivePrefix = {arXiv},
       eprint = {2509.17424},
 primaryClass = {astro-ph.EP},
       adsurl = {https://ui.adsabs.harvard.edu/abs/2025A&A...703A.210A}
}

@ARTICLE{2004A&A...421.1159S,
       author = {{Stempels}, H.~C. and {Gahm}, G.~F.},
        title = "{The close T Tauri binary V 4046 Sagittarii}",
      journal = {\aap},
         year = 2004,
        month = jul,
       volume = {421},
        pages = {1159-1168},
          doi = {10.1051/0004-6361:20034502},
       adsurl = {https://ui.adsabs.harvard.edu/abs/2004A&A...421.1159S}
}

@ARTICLE{2019ApJ...882...49L,
       author = {{Long}, Feng and {Herczeg}, Gregory J. and {Harsono}, Daniel and {Pinilla}, Paola and {Tazzari}, Marco and {Manara}, Carlo F. and {Pascucci}, Ilaria and {Cabrit}, Sylvie and {Nisini}, Brunella and {Johnstone}, Doug and {Edwards}, Suzan and {Salyk}, Colette and {Menard}, Francois and {Lodato}, Giuseppe and {Boehler}, Yann and {Mace}, Gregory N. and {Liu}, Yao and {Mulders}, Gijs D. and {Hendler}, Nathanial and {Ragusa}, Enrico and {Fischer}, William J. and {Banzatti}, Andrea and {Rigliaco}, Elisabetta and {van de Plas}, Gerrit and {Dipierro}, Giovanni and {Gully-Santiago}, Michael and {Lopez-Valdivia}, Ricardo},
        title = "{Compact Disks in a High-resolution ALMA Survey of Dust Structures in the Taurus Molecular Cloud}",
      journal = {\apj},
         year = 2019,
        month = sep,
       volume = {882},
       number = {1},
          eid = {49},
        pages = {49},
          doi = {10.3847/1538-4357/ab2d2d},
archivePrefix = {arXiv},
       eprint = {1906.10809},
 primaryClass = {astro-ph.SR},
       adsurl = {https://ui.adsabs.harvard.edu/abs/2019ApJ...882...49L}
}

@ARTICLE{2009A&A...501.1013S,
       author = {{Schisano}, E. and {Covino}, E. and {Alcal{\'a}}, J.~M. and {Esposito}, M. and {Gandolfi}, D. and {Guenther}, E.~W.},
        title = "{Variability of the transitional T Tauri star T Chamaeleontis}",
      journal = {\aap},
         year = 2009,
        month = jul,
       volume = {501},
       number = {3},
        pages = {1013-1030},
          doi = {10.1051/0004-6361/200811073},
archivePrefix = {arXiv},
       eprint = {0904.0101},
 primaryClass = {astro-ph.SR},
       adsurl = {https://ui.adsabs.harvard.edu/abs/2009A&A...501.1013S}
}

@ARTICLE{2006A&A...452..245N,
       author = {{Natta}, A. and {Testi}, L. and {Randich}, S.},
        title = "{Accretion in the {\ensuremath{\rho}}-Ophiuchi pre-main sequence stars}",
      journal = {\aap},
         year = 2006,
        month = jun,
       volume = {452},
       number = {1},
        pages = {245-252},
          doi = {10.1051/0004-6361:20054706},
archivePrefix = {arXiv},
       eprint = {astro-ph/0602618},
 primaryClass = {astro-ph},
       adsurl = {https://ui.adsabs.harvard.edu/abs/2006A&A...452..245N}
}

@ARTICLE{2022RNAAS...6...18F,
       author = {{Fitton}, Shannon and {Tofflemire}, Benjamin M. and {Kraus}, Adam L.},
        title = "{Disk Material Inflates Gaia RUWE Values in Single Stars}",
      journal = {Research Notes of the American Astronomical Society},
         year = 2022,
        month = jan,
       volume = {6},
       number = {1},
          eid = {18},
        pages = {18},
          doi = {10.3847/2515-5172/ac4bb7},
archivePrefix = {arXiv},
       eprint = {2206.02695},
 primaryClass = {astro-ph.SR},
       adsurl = {https://ui.adsabs.harvard.edu/abs/2022RNAAS...6...18F}
}

@ARTICLE{2022A&A...659A...6B,
       author = {{Bergez-Casalou}, C. and {Bitsch}, B. and {Kurtovic}, N.~T. and {Pinilla}, P.},
        title = "{Constraining giant planet formation with synthetic ALMA images of the Solar System's natal protoplanetary disk}",
      journal = {\aap},
         year = 2022,
        month = mar,
       volume = {659},
          eid = {A6},
        pages = {A6},
          doi = {10.1051/0004-6361/202142490},
archivePrefix = {arXiv},
       eprint = {2201.03383},
 primaryClass = {astro-ph.EP},
       adsurl = {https://ui.adsabs.harvard.edu/abs/2022A&A...659A...6B}
}

@ARTICLE{2016MNRAS.463.3829R,
       author = {{Ru{\'\i}z-Rodr{\'\i}guez}, D. and {Ireland}, M. and {Cieza}, L. and {Kraus}, A.},
        title = "{The frequency of binary star interlopers amongst transitional discs}",
      journal = {\mnras},
         year = 2016,
        month = dec,
       volume = {463},
       number = {4},
        pages = {3829-3847},
          doi = {10.1093/mnras/stw2297},
archivePrefix = {arXiv},
       eprint = {1609.02979},
 primaryClass = {astro-ph.SR},
       adsurl = {https://ui.adsabs.harvard.edu/abs/2016MNRAS.463.3829R}
}

@ARTICLE{2011ApJ...731....8K,
       author = {{Kraus}, Adam L. and {Ireland}, Michael J. and {Martinache}, Frantz and {Hillenbrand}, Lynne A.},
        title = "{Mapping the Shores of the Brown Dwarf Desert. II. Multiple Star Formation in Taurus-Auriga}",
      journal = {\apj},
         year = 2011,
        month = apr,
       volume = {731},
       number = {1},
          eid = {8},
        pages = {8},
          doi = {10.1088/0004-637X/731/1/8},
archivePrefix = {arXiv},
       eprint = {1101.4016},
 primaryClass = {astro-ph.SR},
       adsurl = {https://ui.adsabs.harvard.edu/abs/2011ApJ...731....8K}
}

@ARTICLE{2004AJ....127.1187T,
       author = {{Torres}, Guillermo},
        title = "{A Double-lined Spectroscopic Orbit for the Young Star HD 34700}",
      journal = {\aj},
         year = 2004,
        month = feb,
       volume = {127},
       number = {2},
        pages = {1187-1193},
          doi = {10.1086/381066},
archivePrefix = {arXiv},
       eprint = {astro-ph/0311183},
 primaryClass = {astro-ph},
       adsurl = {https://ui.adsabs.harvard.edu/abs/2004AJ....127.1187T}
}

@ARTICLE{2024A&A...688A.102T,
       author = {{Toci}, Claudia and {Ceppi}, Simone and {Cuello}, Nicol{\'a}s and {Duch{\^e}ne}, Gaspard and {Ragusa}, Enrico and {Lodato}, Giuseppe and {Farina}, Francesca and {M{\'e}nard}, Fran{\c{c}}ois and {Aly}, Hossam},
        title = "{Orbital dynamics in the GG Tau A system: Investigating its enigmatic disc}",
      journal = {\aap},
         year = 2024,
        month = aug,
       volume = {688},
          eid = {A102},
        pages = {A102},
          doi = {10.1051/0004-6361/202348470},
archivePrefix = {arXiv},
       eprint = {2404.07565},
 primaryClass = {astro-ph.EP},
       adsurl = {https://ui.adsabs.harvard.edu/abs/2024A&A...688A.102T}
}

@ARTICLE{2025A&A...702A..76K,
       author = {{Kiefer}, F. and {Lagrange}, A.-M. and {Rubini}, P. and {Philipot}, F.},
        title = "{Searching for substellar companion candidates with Gaia: I. Introducing the GaiaPMEX tool}",
      journal = {\aap},
         year = 2025,
        month = oct,
       volume = {702},
          eid = {A76},
        pages = {A76},
          doi = {10.1051/0004-6361/202449335},
archivePrefix = {arXiv},
       eprint = {2409.16992},
 primaryClass = {astro-ph.EP},
       adsurl = {https://ui.adsabs.harvard.edu/abs/2025A&A...702A..76K}
}

@ARTICLE{2025ApJ...984L...6T,
       author = {{Teague}, Richard and {Benisty}, Myriam and {Facchini}, Stefano and {Fukagawa}, Misato and {Pinte}, Christophe and {Andrews}, Sean M. and {Bae}, Jaehan and {Barraza-Alfaro}, Marcelo and {Cataldi}, Gianni and {Cuello}, Nicol{\'a}s and {Curone}, Pietro and {Czekala}, Ian and {Fasano}, Daniele and {Flock}, Mario and {Galloway-Sprietsma}, Maria and {Garg}, Himanshi and {Hall}, Cassandra and {Hammond}, Iain and {Hilder}, Thomas and {Huang}, Jane and {Ilee}, John D. and {Izquierdo}, Andr{\'e}s F. and {Kanagawa}, Kazuhiro and {Lesur}, Geoffroy and {Lodato}, Giuseppe and {Longarini}, Cristiano and {Loomis}, Ryan A. and {Masset}, Fr{\'e}d{\'e}ric and {Menard}, Francois and {Orihara}, Ryuta and {Price}, Daniel J. and {Rosotti}, Giovanni and {Stadler}, Jochen and {Testi}, Leonardo and {Yen}, Hsi-Wei and {Wafflard-Fernandez}, Gaylor and {Wilner}, David J. and {Winter}, Andrew J. and {W{\"o}lfer}, Lisa and {Yoshida}, Tomohiro C. and {Zawadzki}, Brianna},
        title = "{exoALMA. I. Science Goals, Project Design, and Data Products}",
      journal = {\apjl},
         year = 2025,
        month = may,
       volume = {984},
       number = {1},
          eid = {L6},
        pages = {L6},
          doi = {10.3847/2041-8213/adc43b},
archivePrefix = {arXiv},
       eprint = {2504.18688},
 primaryClass = {astro-ph.EP},
       adsurl = {https://ui.adsabs.harvard.edu/abs/2025ApJ...984L...6T}
}

@ARTICLE{2025ApJ...984L..17L,
       author = {{Longarini}, Cristiano and {Lodato}, Giuseppe and {Rosotti}, Giovanni and {Andrews}, Sean and {Winter}, Andrew and {Stadler}, Jochen and {Izquierdo}, Andr{\'e}s and {Galloway-Sprietsma}, Maria and {Facchini}, Stefano and {Curone}, Pietro and {Benisty}, Myriam and {Teague}, Richard and {Bae}, Jaehan and {Barraza-Alfaro}, Marcelo and {Cataldi}, Gianni and {Czekala}, Ian and {Cuello}, Nicol{\'a}s and {Fasano}, Daniele and {Flock}, Mario and {Fukagawa}, Misato and {Garg}, Himanshi and {Hall}, Cassandra and {Hammond}, Iain and {Hardiman}, Caitlyn and {Hilder}, Thomas and {Huang}, Jane and {Ilee}, John D. and {Isella}, Andrea and {Kanagawa}, Kazuhiro and {Lesur}, Geoffroy and {Loomis}, Ryan A. and {M{\'e}nard}, Francois and {Orihara}, Ryuta and {Pinte}, Christophe and {Price}, Daniel and {Testi}, Leonardo and {Fernandez}, Gaylor Wafflard- and {W{\"o}lfer}, Lisa and {Yen}, Hsi-Wei and {Yoshida}, Tomohiro C. and {Zawadzki}, Brianna},
        title = "{exoALMA. XII. Weighing and Sizing exoALMA Disks with Rotation Curve Modelling}",
      journal = {\apjl},
         year = 2025,
        month = may,
       volume = {984},
       number = {1},
          eid = {L17},
        pages = {L17},
          doi = {10.3847/2041-8213/adc431},
archivePrefix = {arXiv},
       eprint = {2504.18726},
 primaryClass = {astro-ph.EP},
       adsurl = {https://ui.adsabs.harvard.edu/abs/2025ApJ...984L..17L}
}

@ARTICLE{2022A&A...665A.126N,
       author = {{Nguyen}, C.~T. and {Costa}, G. and {Girardi}, L. and {Volpato}, G. and {Bressan}, A. and {Chen}, Y. and {Marigo}, P. and {Fu}, X. and {Goudfrooij}, P.},
        title = "{PARSEC V2.0: Stellar tracks and isochrones of low- and intermediate-mass stars with rotation}",
      journal = {\aap},
         year = 2022,
        month = sep,
       volume = {665},
          eid = {A126},
        pages = {A126},
          doi = {10.1051/0004-6361/202244166},
archivePrefix = {arXiv},
       eprint = {2207.08642},
 primaryClass = {astro-ph.SR},
       adsurl = {https://ui.adsabs.harvard.edu/abs/2022A&A...665A.126N}
}

@ARTICLE{2015A&A...577A..42B,
       author = {{Baraffe}, Isabelle and {Homeier}, Derek and {Allard}, France and {Chabrier}, Gilles},
        title = "{New evolutionary models for pre-main sequence and main sequence low-mass stars down to the hydrogen-burning limit}",
      journal = {\aap},
         year = 2015,
        month = may,
       volume = {577},
          eid = {A42},
        pages = {A42},
          doi = {10.1051/0004-6361/201425481},
archivePrefix = {arXiv},
       eprint = {1503.04107},
 primaryClass = {astro-ph.SR},
       adsurl = {https://ui.adsabs.harvard.edu/abs/2015A&A...577A..42B}
}

@ARTICLE{2021ApJ...907L..33S,
       author = {{Stassun}, Keivan G. and {Torres}, Guillermo},
        title = "{Parallax Systematics and Photocenter Motions of Benchmark Eclipsing Binaries in Gaia EDR3}",
      journal = {\apjl},
         year = 2021,
        month = feb,
       volume = {907},
       number = {2},
          eid = {L33},
        pages = {L33},
          doi = {10.3847/2041-8213/abdaad},
archivePrefix = {arXiv},
       eprint = {2101.03425},
 primaryClass = {astro-ph.SR},
       adsurl = {https://ui.adsabs.harvard.edu/abs/2021ApJ...907L..33S}
}

@ARTICLE{2020MNRAS.499.3362R,
       author = {{Ragusa}, Enrico and {Alexander}, Richard and {Calcino}, Josh and {Hirsh}, Kieran and {Price}, Daniel J.},
        title = "{The evolution of large cavities and disc eccentricity in circumbinary discs}",
      journal = {\mnras},
         year = 2020,
        month = dec,
       volume = {499},
       number = {3},
        pages = {3362-3380},
          doi = {10.1093/mnras/staa2954},
archivePrefix = {arXiv},
       eprint = {2009.10738},
 primaryClass = {astro-ph.EP},
       adsurl = {https://ui.adsabs.harvard.edu/abs/2020MNRAS.499.3362R}
}

@ARTICLE{2024ApJ...974..102S,
       author = {{Sierra}, Anibal and {P{\'e}rez}, Laura M. and {Agurto-Gangas}, Carolina and {Miley}, James and {Zhang}, Ke and {Pinilla}, Paola and {Pascucci}, Ilaria and {Trapman}, Leon and {Kurtovic}, Nicolas and {Vioque}, Miguel and {Deng}, Dingshan and {Anania}, Rossella and {Carpenter}, John and {Cieza}, Lucas A. and {Gonz{\'a}lez-Ruilova}, Camilo and {Hogerheijde}, Michiel and {Kuznetsova}, Aleksandra and {Rosotti}, Giovanni P. and {Ruiz-Rodriguez}, Dary A. and {Schwarz}, Kamber and {Tabone}, Beno{\^\i}t and {TorresVillanueva}, Estephani E.},
        title = "{Hints of Planet Formation Signatures in a Large-cavity Disk Studied in the AGE-PRO ALMA Large Program}",
      journal = {\apj},
         year = 2024,
        month = oct,
       volume = {974},
       number = {1},
          eid = {102},
        pages = {102},
          doi = {10.3847/1538-4357/ad6e73},
archivePrefix = {arXiv},
       eprint = {2407.16651},
 primaryClass = {astro-ph.EP},
       adsurl = {https://ui.adsabs.harvard.edu/abs/2024ApJ...974..102S}
}

@ARTICLE{2014ApJ...797...14P,
       author = {{Perryman}, Michael and {Hartman}, Joel and {Bakos}, G{\'a}sp{\'a}r {\'A}. and {Lindegren}, Lennart},
        title = "{Astrometric Exoplanet Detection with Gaia}",
      journal = {\apj},
         year = 2014,
        month = dec,
       volume = {797},
       number = {1},
          eid = {14},
        pages = {14},
          doi = {10.1088/0004-637X/797/1/14},
archivePrefix = {arXiv},
       eprint = {1411.1173},
 primaryClass = {astro-ph.EP},
       adsurl = {https://ui.adsabs.harvard.edu/abs/2014ApJ...797...14P}
}

@ARTICLE{2018ApJS..239...31B,
       author = {{Brandt}, Timothy D.},
        title = "{The Hipparcos-Gaia Catalog of Accelerations}",
      journal = {\apjs},
         year = 2018,
        month = dec,
       volume = {239},
       number = {2},
          eid = {31},
        pages = {31},
          doi = {10.3847/1538-4365/aaec06},
archivePrefix = {arXiv},
       eprint = {1811.07283},
 primaryClass = {astro-ph.SR},
       adsurl = {https://ui.adsabs.harvard.edu/abs/2018ApJS..239...31B}
}

@ARTICLE{2025A&A...702A..77K,
       author = {{Kiefer}, F. and {Lagrange}, A.-M. and {Rubini}, P. and {Philipot}, F.},
        title = "{Searching for substellar companion candidates with Gaia: II. A catalog of 9698 planet candidate solar-type hosts}",
      journal = {\aap},
         year = 2025,
        month = oct,
       volume = {702},
          eid = {A77},
        pages = {A77},
          doi = {10.1051/0004-6361/202451745},
archivePrefix = {arXiv},
       eprint = {2409.16993},
 primaryClass = {astro-ph.EP},
       adsurl = {https://ui.adsabs.harvard.edu/abs/2025A&A...702A..77K}
}

@ARTICLE{2024A&A...688A...1C,
       author = {{Castro-Ginard}, Alfred and {Penoyre}, Zephyr and {Casey}, Andrew R. and {Brown}, Anthony G.~A. and {Belokurov}, Vasily and {Cantat-Gaudin}, Tristan and {Drimmel}, Ronald and {Fouesneau}, Morgan and {Khanna}, Shourya and {Kurbatov}, Evgeny P. and {Price-Whelan}, Adrian M. and {Rix}, Hans-Walter and {Smart}, Richard L.},
        title = "{Gaia DR3 detectability of unresolved binary systems}",
      journal = {\aap},
         year = 2024,
        month = aug,
       volume = {688},
          eid = {A1},
        pages = {A1},
          doi = {10.1051/0004-6361/202450172},
archivePrefix = {arXiv},
       eprint = {2404.14127},
 primaryClass = {astro-ph.GA},
       adsurl = {https://ui.adsabs.harvard.edu/abs/2024A&A...688A...1C}
}

@INPROCEEDINGS{2014prpl.conf..267R,
       author = {{Reipurth}, B. and {Clarke}, C.~J. and {Boss}, A.~P. and {Goodwin}, S.~P. and {Rodr{\'\i}guez}, L.~F. and {Stassun}, K.~G. and {Tokovinin}, A. and {Zinnecker}, H.},
        title = "{Multiplicity in Early Stellar Evolution}",
    booktitle = {Protostars and Planets VI},
         year = 2014,
       editor = {{Beuther}, Henrik and {Klessen}, Ralf S. and {Dullemond}, Cornelis P. and {Henning}, Thomas},
        month = jan,
        pages = {267-290},
          doi = {10.2458/azu_uapress_9780816531240-ch012},
archivePrefix = {arXiv},
       eprint = {1403.1907},
 primaryClass = {astro-ph.SR},
       adsurl = {https://ui.adsabs.harvard.edu/abs/2014prpl.conf..267R}
}

@INPROCEEDINGS{2023ASPC..534..605B,
       author = {{Benisty}, M. and {Dominik}, C. and {Follette}, K. and {Garufi}, A. and {Ginski}, C. and {Hashimoto}, J. and {Keppler}, M. and {Kley}, W. and {Monnier}, J.},
        title = "{Optical and Near-infrared View of Planet-forming Disks and Protoplanets}",
    booktitle = {Protostars and Planets VII},
         year = 2023,
       editor = {{Inutsuka}, S. and {Aikawa}, Y. and {Muto}, T. and {Tomida}, K. and {Tamura}, M.},
       series = {Astronomical Society of the Pacific Conference Series},
       volume = {534},
        month = jul,
        pages = {605},
          doi = {10.48550/arXiv.2203.09991},
archivePrefix = {arXiv},
       eprint = {2203.09991},
 primaryClass = {astro-ph.EP},
       adsurl = {https://ui.adsabs.harvard.edu/abs/2023ASPC..534..605B}
}

@ARTICLE{2010ApJS..190....1R,
       author = {{Raghavan}, Deepak and {McAlister}, Harold A. and {Henry}, Todd J. and {Latham}, David W. and {Marcy}, Geoffrey W. and {Mason}, Brian D. and {Gies}, Douglas R. and {White}, Russel J. and {ten Brummelaar}, Theo A.},
        title = "{A Survey of Stellar Families: Multiplicity of Solar-type Stars}",
      journal = {\apjs},
         year = 2010,
        month = sep,
       volume = {190},
       number = {1},
        pages = {1-42},
          doi = {10.1088/0067-0049/190/1/1},
archivePrefix = {arXiv},
       eprint = {1007.0414},
 primaryClass = {astro-ph.SR},
       adsurl = {https://ui.adsabs.harvard.edu/abs/2010ApJS..190....1R}
}

@ARTICLE{2021ApJS..254...42B,
       author = {{Brandt}, Timothy D.},
        title = "{The Hipparcos-Gaia Catalog of Accelerations: Gaia EDR3 Edition}",
      journal = {\apjs},
         year = 2021,
        month = jun,
       volume = {254},
       number = {2},
          eid = {42},
        pages = {42},
          doi = {10.3847/1538-4365/abf93c},
archivePrefix = {arXiv},
       eprint = {2105.11662},
 primaryClass = {astro-ph.GA},
       adsurl = {https://ui.adsabs.harvard.edu/abs/2021ApJS..254...42B}
}

@ARTICLE{2023A&A...680A.114R,
       author = {{Ren}, Bin B. and {Benisty}, Myriam and {Ginski}, Christian and {Tazaki}, Ryo and {Wallack}, Nicole L. and {Milli}, Julien and {Garufi}, Antonio and {Bae}, Jaehan and {Facchini}, Stefano and {M{\'e}nard}, Fran{\c{c}}ois and {Pinilla}, Paola and {Swastik}, C. and {Teague}, Richard and {Wahhaj}, Zahed},
        title = "{Protoplanetary disks in K$_{s}$-band total intensity and polarized light}",
      journal = {\aap},
         year = 2023,
        month = dec,
       volume = {680},
          eid = {A114},
        pages = {A114},
          doi = {10.1051/0004-6361/202347353},
archivePrefix = {arXiv},
       eprint = {2310.08589},
 primaryClass = {astro-ph.EP},
       adsurl = {https://ui.adsabs.harvard.edu/abs/2023A&A...680A.114R}
}

@ARTICLE{2014A&A...565L...2D,
       author = {{Di Folco}, E. and {Dutrey}, A. and {Le Bouquin}, J. -B. and {Lacour}, S. and {Berger}, J. -P. and {K{\"o}hler}, R. and {Guilloteau}, S. and {Pi{\'e}tu}, V. and {Bary}, J. and {Beck}, T. and {Beust}, H. and {Pantin}, E.},
        title = "{GG Tauri: the fifth element}",
      journal = {\aap},
         year = 2014,
        month = may,
       volume = {565},
          eid = {L2},
        pages = {L2},
          doi = {10.1051/0004-6361/201423675},
archivePrefix = {arXiv},
       eprint = {1404.2205},
 primaryClass = {astro-ph.SR},
       adsurl = {https://ui.adsabs.harvard.edu/abs/2014A&A...565L...2D}
}

@INPROCEEDINGS{2023ASPC..534..799C,
       author = {{Currie}, T. and {Biller}, B. and {Lagrange}, A. and {Marois}, C. and {Guyon}, O. and {Nielsen}, E.~L. and {Bonnefoy}, M. and {De Rosa}, R.~J.},
        title = "{Direct Imaging and Spectroscopy of Extrasolar Planets}",
    booktitle = {Protostars and Planets VII},
         year = 2023,
       editor = {{Inutsuka}, S. and {Aikawa}, Y. and {Muto}, T. and {Tomida}, K. and {Tamura}, M.},
       series = {Astronomical Society of the Pacific Conference Series},
       volume = {534},
        month = jul,
        pages = {799},
          doi = {10.48550/arXiv.2205.05696},
archivePrefix = {arXiv},
       eprint = {2205.05696},
 primaryClass = {astro-ph.EP},
       adsurl = {https://ui.adsabs.harvard.edu/abs/2023ASPC..534..799C}
}

\begin{appendix}






\section{Detections and non-detections}\label{Sect_appen_B}

\textbf{Table~\ref{Table_1} lists the stellar (effective temperature, luminosity, and mass) and disk properties (cavity size, inclination, and position angle) of the 98 sources with inner dust cavities (``transition disks'') considered in this work, and whether we detect companions in them via proper motion anomalies ($|\Delta \mu|/\sigma_{|\Delta \mu|} \ge 3$). For the detected companions, we report the predicted companion mass and semi-major axis (Sect.~\ref{s_significant}). Figs.~\ref{fig:all_detect} and \ref{fig:all_non_detect} show the parameter space of semi-major axis versus companion mass for companion detections and non-detections (for the latter, Fig. \ref{fig:all_non_detect} shows the space of separations and masses where a companion dominating the astrometric signal can be excluded), as illustrated by Figs.~\ref{fig_HD142527} and \ref{fig_LkCa15}, respectively.}

\begin{figure*}[ht!]
    \centering
    \includegraphics[width=1\textwidth]{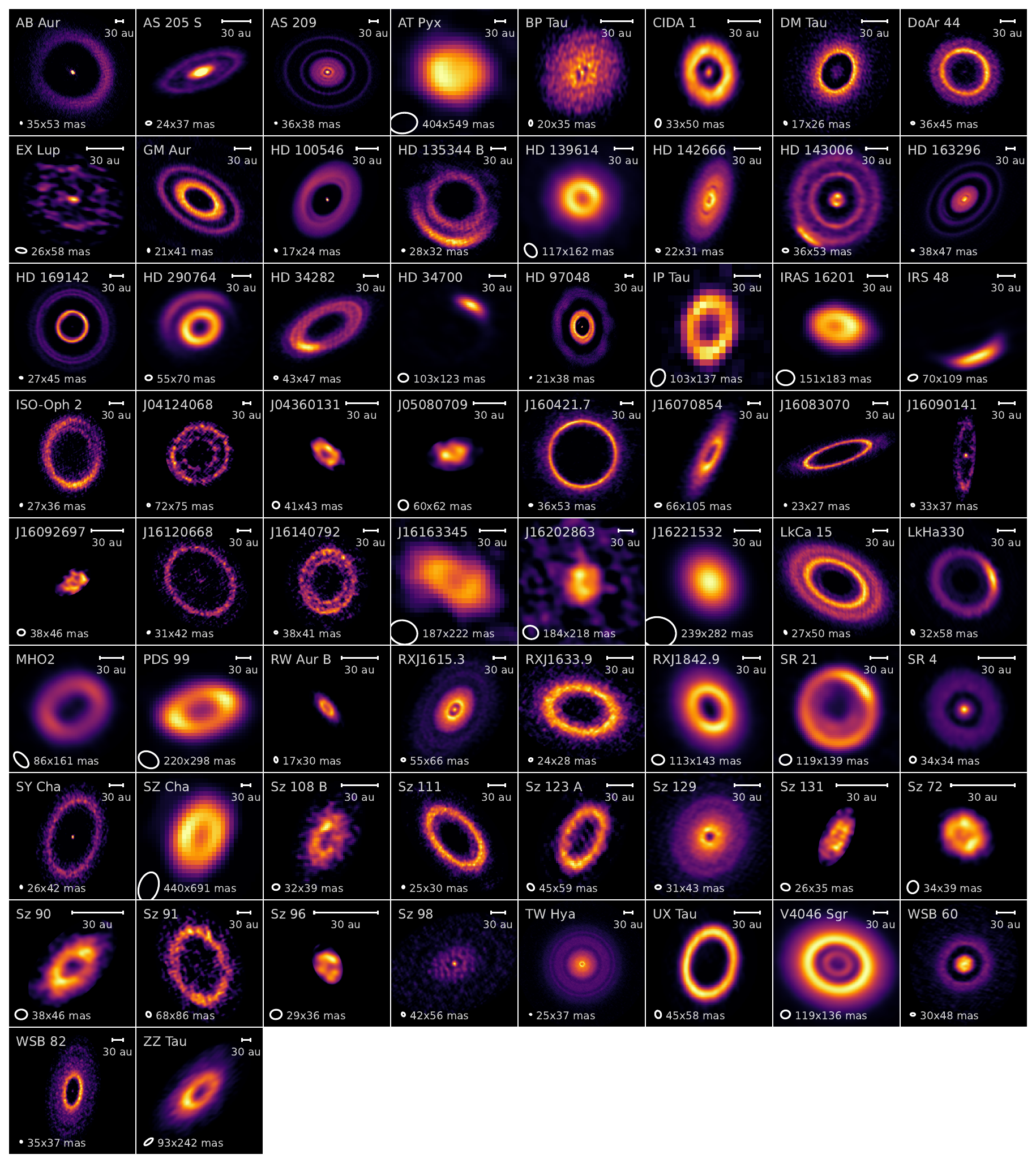}
    \caption{ALMA continuum images of the sample of \textbf{67} disks with inner dust cavities (``transition disks'') for which we do not find a significant proper motion anomaly ($|\Delta \mu|/\sigma_{|\Delta \mu|}<3$) indicative of companions. \textbf{WISPIT 2 is missing from this mosaic because its ALMA image showing an inner dust cavity (priv. comm.) was not publicly available at the time of publication}. The companion separation–mass parameter space in which a companion dominating the astrometric signal can be excluded based on this non-detection is described in Sect. \ref{s_nondetections} (see Figs. \ref{fig_LkCa15} and \ref{fig:all_non_detect}).}
    \label{big_mosaic_no_detections}
\end{figure*}

The references for the ALMA continuum images of Figs. \ref{big_mosaic_detections} and \ref{big_mosaic_no_detections} are: AA Tau (\citealp{2024PASJ...76..437Y}),
AB Aur (\citealp{2024Natur.633...58S}),
AK Sco (2019.1.01210.S, PI: Czekala),
AS 205 S (\citealp{2018ApJ...869L..44K}),
AT Pyx (2021.1.01705.S, PI: Ginski),
BP Tau and Sz 98 (\citealp{2025A&A...694A.147G}), CIDA 1 (\citealp{2021A&A...649A.122P}),
CIDA 9 A (\citealp{2024ApJ...961...28H}),
CQ Tau (\citealp{2021A&A...648A..19W}),
CS Cha (\citealp{2022A&A...664A.151K}),
DM Tau (\citealp{2021ApJ...911....5H}),
AS 209, HD 142666, HD 163296, SR 4, and Sz 129 (\citealp{2018ApJ...869L..42H}),
EX Lup, J16090141-3925119, J16092697-3836269, Sz 131, Sz 72, Sz 90, Sz 96, Sz 100, Sz 108 B, Sz 111, Sz 123 A, and Sz 76 (\citealp{2025A&A...696A.232G}), GG Tau (\citealp{2024A&A...684A.134R}), GM Aur (\citealp{2020ApJ...891...48H}), GW Ori (2021.1.01661.S, PI: Kraus), MHO2 and HD 139614 (2022.1.01302.S, PI: Mulders), HD 142527 (\citealp{2019ASPC..523..637Y}), HD 290764 (\citealp{2023A&A...670A.154W}), HD 34282 and PDS 99 (\citealp{2020ApJ...892..111F}), HD 34700 (2022.1.00760.S, PI: Stadler), HD 100453 (\citealp{2020MNRAS.499.3837G}), HD 100546 (\citealp{2022ApJ...933L...4C}), HD 135344 B (\citealp{2021MNRAS.507.3789C}), HD 143006 (\citealp{2018ApJ...869L..50P}), HD 169142 (\citealp{2019AJ....158...15P}), HP Cha and RY Lup (\citealp{2021MNRAS.502.5779N}), IP Tau (\citealp{2018ApJ...869...17L}), IRAS 04125+2902 (\citealp{2025ApJ...993...90S}), IRAS 04158+2805 (\citealp{2021MNRAS.507.1157R}), IRAS 16072-2057, J16140792-1938292 (2022.1.00646.S, PI: Long), IRAS16201-2410 (\citealp{2025ApJ...981L...4D}), IRS 48 (\citealp{2023ApJ...948L...2Y}), J04124068+2438157 (\citealp{2023ApJ...949...27L}), J04343128+1722201, J04360131+1726120, and J05080709+2427123 (\citealp{2024ApJ...966...59S}), J16120668-3010270 (\citealp{2024ApJ...974..102S}), J16163345-2521505, J16202863-2442087, and J16221532-2511349 (\citealp{2025ApJ...989....9V}), J160421.7-213028 (\citealp{2023A&A...670L...1S}), J16083070-3828268 and RXJ1852.3-3700 (2018.1.00689.S, PI: Muto), J16102955-3922144 (2016.1.00715.S, PI: Facchini), LkCa 15 (\citealp{2022ApJ...937L...1L}), LkHa 330 (\citealp{2022A&A...665A.128P}), MHO6 (\citealp{2021A&A...645A.139K}), MP Mus, RXJ1842.9-3532, and V4046 Sgr (\citealp{2025ApJ...984L...9C}), MWC 758 (\citealp{2018ApJ...860..124D}), PDS 111 (\citealp{2024A&A...688A.149D}), PDS 70 (\citealp{2021ApJ...916L...2B}), RW Aur B (\citealp{2024A&A...692A.155K}), RXJ1615.3-3255 (2016.1.01286.S, Benisty), DoAr 44, ISO-Oph 2, RXJ1633.9-2442, WSB 60, and WSB 82 rho-Oph 38 (\citealp{2021MNRAS.501.2934C}), RY Tau, J16070854-3914075, and HD 97048 (\citealp{2024MNRAS.532.1752R}), SR 21 A (\citealp{2025A&A...693A.286S}), SR 24 S (2022.1.00908.S, PI: Rodriguez Jimenez), SY Cha (\citealp{2023PASJ...75..424O}), SZ Cha (\citealp{2016ApJ...831..125P}), Sz 118 (2022.1.00340.S, PI: Andrews), Sz 84 (\citealp{2021ApJ...908..250H}), Sz 91 (\citealp{2021ApJ...923..128M}), T Cha (\citealp{2018MNRAS.475L..62H}), TW Hya (\citealp{2018ApJ...852..122H}), UX Tau A (2021.1.00994.S, PI: Perez), UZ Tau (2017.1.00388.S, PI: Liu), V892 Tau (\citealp{2021ApJ...915..131L}), ZZ Tau IRS (\citealp{2022ApJ...941...66H}).


\begin{sidewaystable*}
\caption{YSO sources with protoplanetary disks with inner dust cavities (``transition disks'') considered in this work.}
\label{Table_1}
\centering
\setlength{\tabcolsep}{3.3pt}
\renewcommand{\arraystretch}{1.3}
\footnotesize
\begin{tabular}{lccccccccccccc}
\hline\hline
Name & Companion & Companion  & Companion & $T_\mathrm{eff}$ & $L_\star$ & Ref. stellar & $M_\star$ & inc & PA & Cavity & Ref. disk  & $|\Delta \mu|/\sigma_{|\Delta \mu|}$ & RUWE \\
 &  & mass (M$_\mathrm{J}$) & separation (au) & (K) & (L$_\odot$) & param. & (M$_\odot$) & (deg) & (deg) & size (au) &  geom.  &  & \\
\hline\\[-2ex]
AA Tau & Yes & $98^{+200}_{-55}$ & $1.93^{+1.9}_{-1.00}$ & $3762^{+190}_{-190}$ & $0.75^{+0.34}_{-0.34}$ & 1 & $0.47\pm0.23$ & $58.540\pm0.020$ & $93.770\pm0.030$ & 44.0 & a  & 5.11 & 3.33 \\
AB Aur & No &  &  & $9000^{+120}_{-120}$ & $45.7^{+4.6}_{-4.6}$ & 2 & $2.464\pm0.094$ & $23.0\pm1.0$ & $36.0\pm1.0$ & 156.0 & b  & 1.37 & 1.37 \\
AK Sco & Yes & $323^{+970}_{-250}$ & $4.3^{+5.9}_{-3.8}$ & $6000^{+250}_{-250}$ & $5.6^{+1.5}_{-1.5}$ & 2 & $1.69\pm0.45$ & $65.0\pm1.0$ & $53.0\pm5.0$ & 20.0 & this work  & 5.35 & 1.60 \\
AS 205 S & No &  &  & $3970^{+50}_{-30}$ & $0.44^{+0.20}_{-0.20}$ & 3 & $0.69\pm0.12$ & $66.3\pm1.7$ & $109.60\pm0.80$ & 3.4 & c  & 1.00 & 1.75 \\
AS 209 & No &  &  & $4266^{+310}_{-280}$ & $1.42^{+0.83}_{-0.52}$ & 4 & $0.90\pm0.58$ & $34.97\pm0.13$ & $85.76\pm0.16$ & 12.0 & d & 1.02 & 1.48 \\
AT Pyx & No &  &  & $4760^{+160}_{-210}$ & $1.63^{+0.73}_{-0.73}$ & 5 & $1.31\pm0.37$ &  &  & 63.0 &  & 2.12 & 3.53 \\
BP Tau & No &  &  & $3777^{+190}_{-190}$ & $0.98^{+0.10}_{-0.10}$ & 5 & $0.466\pm0.058$ & $38.20\pm0.50$ & $151.1\pm1.0$ & 9.0 & e & 2.23 & 2.20 \\
CIDA 1 & No &  &  & $3125^{+160}_{-160}$ & $0.210^{+0.094}_{-0.094}$ & 1 & $0.171\pm0.085$ & $38.20\pm0.15$ & $11.20\pm0.18$ & 21.0 & f & 2.14 & 1.07 \\

CIDA 9 A & Yes & $203^{+190}_{-100}$ & $0.80^{+0.44}_{-0.26}$ & $3589^{+180}_{-180}$ & $0.049^{+0.022}_{-0.022}$ & 1 & $0.49\pm0.20$ & $46.0\pm1.0$ & $103.0\pm1.0$ & 29.0 & b  & 5.08 & 3.59 \\
CQ Tau & Yes & $309^{+840}_{-190}$ & $0.99^{+4.8}_{-0.31}$ & $6750^{+120}_{-120}$ & $6.61^{+0.66}_{-0.66}$ & 2 & $1.541\pm0.062$ & $35.240\pm0.020$ & $53.870\pm0.020$ & 50.0 & a  & 4.04 & 3.70 \\
CS Cha & Yes & $668^{+670}_{-400}$ & $3.60^{+6.7}_{-0.60}$ & $4900^{+240}_{-240}$ & $2.04^{+0.92}_{-0.92}$ & 1 & $1.48\pm0.40$ & $17.9\pm1.0$ & $82.6\pm1.0$ & 37.0 & g  & 18.48 & 6.41 \\
DM Tau & No &  &  & $3415^{+170}_{-170}$ & $0.24^{+0.11}_{-0.11}$ & 1 & $0.29\pm0.17$ & $35.970\pm0.050$ & $155.600\pm0.080$ & 18.0 & a  & 1.10 & 1.49 \\

DoAr 44 & No &  &  & $4780^{+240}_{-240}$ & $1.91^{+0.19}_{-0.19}$ & 6 & $1.30\pm0.49$ & $20.0\pm1.0$ & $30.0\pm1.0$ & 40.0 & b  & 2.18 & 1.29 \\
EX Lup & No &  &  & $3850^{+180}_{-180}$ & $0.74^{+0.53}_{-0.53}$ & 7 & $0.53\pm0.29$ & $32.40\pm0.90$ & $64.6\pm3.3$ & 28.0 & h  & 0.80 & 1.12 \\

GG Tau & Yes & No solution  & No solution  & $4060^{+200}_{-200}$ & $1.11^{+0.19}_{-0.16}$ & 8 & $0.69\pm0.34$ & $36.0\pm1.0$ & $98.0\pm1.0$ & 224.0 & b  & 12.68 & 6.44 \\

GM Aur & No &  &  & $4202^{+210}_{-210}$ & $0.90^{+0.41}_{-0.41}$ & 1 & $0.84\pm0.41$ & $53.0\pm1.0$ & $56.0\pm1.0$ & 40.0 & b  & 1.71 & 2.53 \\

GW Ori & Yes & $1162^{+1300}_{-710}$ & $2.7^{+6.1}_{-1.4}$ & $5700^{+150}_{-150}$ & $35.5^{+4.2}_{-3.5}$ & 9 & $3.36\pm0.30$ & $40.0\pm1.0$ & $1.0\pm1.0$ & 29.0 & this work  & 8.65 & 4.23 \\

HD 100453 & Yes & $94^{+660}_{-87}$ & $4.4^{+22}_{-4.3}$ & $7250^{+120}_{-120}$ & $6.17^{+0.62}_{-0.62}$ & 2 & $1.61\pm0.11$ & $30.0\pm1.0$ & $149.0\pm1.0$ & 30.0 & b  & 3.76 & 1.08 \\
HD 100546 & No &  &  & $9250^{+120}_{-120}$ & $21.9^{+2.2}_{-2.2}$ & 2 & $2.10\pm0.10$ & $42.0\pm1.0$ & $139.0\pm1.0$ & 25.0 & b & 2.16 & 1.30 \\
HD 135344 B & No &  &  & $6250^{+120}_{-120}$ & $5.13^{+0.51}_{-0.51}$ & 2 & $1.504\pm0.073$ & $20.730\pm0.020$ & $28.920\pm0.090$ & 52.0 & a & 1.68 & 0.91 \\
HD 139614 & No &  &  & $7500^{+120}_{-120}$ & $6.76^{+0.68}_{-0.68}$ & 2 & $1.65\pm0.14$ &  &  & 9.4 &   & 0.83 & 1.05 \\
HD 142527 & Yes & $158^{+1300}_{-110}$ & $5.3^{+8.7}_{-2.7}$ & $6750^{+120}_{-120}$ & $22.4^{+2.2}_{-2.2}$ & 2 & $2.24\pm0.16$ & $27.0\pm1.0$ & $25.0\pm1.0$ & 185.0 & b  & 4.60 & 1.18 \\
HD 142666 & No &  &  & $7250^{+120}_{-120}$ & $13.5^{+1.3}_{-1.3}$ & 2 & $1.87\pm0.18$ & $62.22\pm0.14$ & $162.11\pm0.15$ & 6.0 & d  & 1.12 & 1.04 \\
HD 143006 & No &  &  & $5500^{+120}_{-120}$ & $3.47^{+0.35}_{-0.35}$ & 2 & $1.66\pm0.21$ & $18.690\pm0.090$ & $7.53\pm0.35$ & 41.0 & a & 1.32 & 1.13 \\
HD 163296 & No &  &  & $8750^{+120}_{-120}$ & $15.5^{+1.5}_{-1.7}$ & 2 & $1.91\pm0.11$ & $46.70\pm0.10$ & $133.33\pm0.15$ & 3.0 & d  & 0.33 & 0.92 \\
HD 169142 & No &  &  & $7250^{+120}_{-120}$ & $5.75^{+0.58}_{-0.58}$ & 2 & $1.571\pm0.093$ & $12.0\pm1.0$ & $5.0\pm1.0$ & 24.0 & b  & 0.88 & 0.78 \\
HD 290764 & No &  &  & $7750^{+120}_{-120}$ & $21.9^{+2.2}_{-2.2}$ & 2 & $2.17\pm0.10$ & $30.0\pm1.0$ & $115.4\pm1.0$ & 52.0 & this work  & 2.03 & 1.25 \\
HD 34282 & No &  &  & $9500^{+250}_{-250}$ & $14.5^{+1.4}_{-1.4}$ & 10 & $2.03\pm0.15$ & $59.090\pm0.010$ & $117.150\pm0.010$ & 80.0 & a  & 0.72 & 1.41 \\
HD 34700 & No &  &  & $6000^{+120}_{-120}$ & $22.9^{+2.3}_{-2.3}$ & 2 & $2.70\pm0.40$ & $37.9\pm2.0$ & $87.1\pm1.2$ & 65.0 & i & 1.43 & 1.05 \\
HD 97048 & No &  &  & $11000^{+120}_{-120}$ & $64.6^{+6.5}_{-6.5}$ & 2 & $2.846\pm0.096$ & $41.0\pm1.0$ & $4.0\pm1.0$ & 63.0 & b & 1.32 & 1.43 \\
HP Cha & Yes & $150^{+330}_{-110}$ & $2.6^{+3.3}_{-1.4}$ & $4060^{+200}_{-200}$ & $3.29^{+0.33}_{-0.33}$ & 11 & $0.57\pm0.35$ & $37.0\pm1.0$ & $162.0\pm1.0$ & 50.0 & b & 3.66 & 3.25 \\
IP Tau & No &  &  & $3762^{+190}_{-190}$ & $0.52^{+0.23}_{-0.23}$ & 1 & $0.49\pm0.27$ & $45.0\pm1.0$ & $173.0\pm1.0$ & 25.0 & b & 2.64 & 1.30 \\
\\[-3ex]
\hline
\end{tabular}
\tablefoot{\textbf{98} YSO sources with inner dust cavities (``transition disks'') considered in this work. Stellar masses were derived in this work (Sect. \ref{sec_methodology}). The table is available in electronic form at the CDS (Sect.~\ref{sec_data_availability}). The online version of the table also includes \textit{Gaia} DR3 and DR2 source IDs, \textit{Gaia} DR3 coordinates, $|\Delta \mu|$, $\sigma_{|\Delta \mu|}$, and the remaining \textit{Gaia} DR3 astrometric parameters. References for effective temperatures and luminosities: 1--\citet{2023ASPC..534..539M}, 2--\citealp{2021A&A...650A.182G}, 3--\citet{2005ApJ...623..952E}, 4--\citet{2018ApJ...869L..41A}, 5--\citet{2014ApJ...786...97H}, 6--\citet{2014A&A...568A..18M}, 7--\citet{2017A&A...600A..20A}, 8--\citet{1999ApJ...520..811W}, 9--\citet{2021A&A...652A.133V}, 10--\citet{2020MNRAS.493..234W}, 11--\citet{2017A&A...604A.127M}. References for disk geometries: a--\citet{2025ApJ...984L...9C}, b--\citet{2020ApJ...892..111F}, c--\citet{2018ApJ...869L..44K}, d--\citet{2018ApJ...869L..42H}, e--\citet{2025A&A...694A.147G}, f--\citet{2021A&A...645A.139K}, g--\citet{2022A&A...664A.151K}, h--\citet{2018ApJ...859..111H}, i--\citet{2024A&A...681A..19C}.}
\end{sidewaystable*}

\begin{sidewaystable*}
\centering
\setlength{\tabcolsep}{3.3pt}
\renewcommand{\arraystretch}{1.3}
\footnotesize
\begin{tabular}{lccccccccccccc}
\hline\hline
Name & Companion & Companion  & Companion & $T_\mathrm{eff}$ & $L_\star$ & Ref. stellar & $M_\star$ & inc & PA & Cavity & Ref. disk  & $|\Delta \mu|/\sigma_{|\Delta \mu|}$ & RUWE \\
 &  & mass (M$_\mathrm{J}$) & separation (au) & (K) & (L$_\odot$) & param. & (M$_\odot$) & (deg) & (deg) & size (au) &  geom.  &  & \\
\hline\\[-2ex]
IRAS 04125+2902 & Yes & $150^{+310}_{-120}$ & $2.3^{+5.6}_{-2.0}$ & $3912^{+72}_{-72}$ & $0.466^{+0.047}_{-0.047}$ & 12 & $0.62\pm0.14$ & $31\pm12$ &  & 130.0 & j  & 4.20 & 1.36 \\
IRAS 04158+2805 & Yes & $120^{+77}_{-39}$ & $1.60^{+0.87}_{-0.54}$ & $2880^{+1000}_{-500}$ & $0.050^{+0.050}_{-0.030}$ & 13 & $0.08\pm0.52$ & $62.0\pm3.0$ & $92.0\pm1.0$ & 185.0 & k & 3.99 & 1.70 \\
IRAS 16072-2057 & Yes & $160^{+550}_{-120}$ & $1.5^{+5.6}_{-1.1}$ & $4375^{+180}_{-180}$ & $0.950^{+0.095}_{-0.095}$ & 14 & $1.06\pm0.36$ & $60.0\pm1.0$ & $35.0\pm1.0$ & 54.0 & this work & 3.10 & 2.31 \\
IRAS 16201-2410 & No &  &  & $6000^{+600}_{-600}$ & $0.95^{+0.10}_{-0.10}$ & 15 & $1.14\pm0.20$ & $49.0\pm1.0$ & $82.0\pm1.0$ & 28.0 & l & 2.29 & 1.26 \\
IRS 48 & No &  &  & $9520^{+480}_{-480}$ & $18.4^{+1.8}_{-1.8}$ & 16 & $2.02\pm0.18$ & $50.0\pm1.0$ & $100.0\pm1.0$ & 70.0 & b & 2.60 & 1.04 \\
ISO-Oph 2 & No &  &  & $3850^{+190}_{-190}$ & $0.441^{+0.044}_{-0.044}$ & 17 & $0.57\pm0.31$ & $37.60\pm0.80$ & $0.4\pm1.4$ & 50.0 & m & 0.99 & 1.24 \\
J04124068+2438157 & No &  &  & $3160^{+200}_{-280}$ & $0.126^{+0.013}_{-0.013}$ & 18 & $0.18\pm0.14$ & $15.90\pm0.82$ & $124.1\pm2.8$ & 60.0 & n & 0.20 & 1.22 \\

J04343128+1722201 & Yes & $13^{+100}_{-13}$ & $1.7^{+11}_{-1.5}$ & $3160^{+200}_{-280}$ & $0.115^{+0.011}_{-0.011}$ & 19 & $0.18\pm0.15$ & $68.54\pm0.23$ & $96.67\pm0.24$ & 10.0 & n & 3.07 & 1.46 \\

J04360131+1726120 & No &  &  & $3360^{+130}_{-200}$ & $0.146^{+0.015}_{-0.015}$ & 19 & $0.28\pm0.16$ & $53.5\pm1.5$ & $47.0\pm2.0$ & 12.0 & n & 1.89 & 1.52 \\
J05080709+2427123 & No &  &  & $3160^{+200}_{-280}$ & $0.110^{+0.011}_{-0.011}$ & 20 & $0.18\pm0.15$ & $54.3\pm1.8$ & $98.9\pm2.1$ & 7.0 & n & 2.48 & 1.12 \\

J160421.7-213028 & No &  &  & $4730^{+240}_{-240}$ & $0.84^{+0.38}_{-0.38}$ & 1 & $1.21\pm0.34$ & $8.720\pm0.090$ & $123.24\pm0.15$ & 87.0 & a  & 1.13 & 1.75 \\
J16070854-3914075 & No &  &  & $3125^{+72}_{-72}$ & $0.0078^{+0.0060}_{-0.0048}$ & 7 & $0.127\pm0.063$ & $73.08\pm0.38$ & $153.99\pm0.36$ & 40.0 & o & 1.96 & 1.08 \\
J16083070-3828268 & No &  &  & $4900^{+230}_{-230}$ & $1.88^{+0.85}_{-0.85}$ & 7 & $1.46\pm0.38$ & $72.86\pm0.18$ & $107.68\pm0.18$ & 62.0 & o  & 1.78 & 1.94 \\
J16090141-3925119 & No &  &  & $3270^{+75}_{-75}$ & $0.093^{+0.042}_{-0.042}$ & 7 & $0.241\pm0.084$ & $76.20\pm0.31$ & $176.83\pm0.25$ & 64.0 & o  & 0.16 & 1.40 \\
J16092697-3836269 & No &  &  & $3200^{+74}_{-74}$ & $0.073^{+0.050}_{-0.050}$ & 7 & $0.207\pm0.084$ & $55.2\pm2.5$ & $122.7\pm2.5$ & 11.1 & p & 1.26 & 1.10 \\
J16102955-3922144 & Yes & $27^{+110}_{-20}$ & $1.6^{+3.5}_{-1.3}$ & $3200^{+74}_{-74}$ & $0.106^{+0.048}_{-0.048}$ & 7 & $0.205\pm0.072$ & $57\pm25$ & $106.8\pm9.6$ & 11.0 & o & 3.36 & 1.06 \\
J16120668-3010270 & No &  &  & $3770^{+110}_{-140}$ & $0.250^{+0.050}_{-0.050}$ & 21 & $0.56\pm0.21$ & $37.00\pm0.20$ & $45.10\pm0.90$ & 75.0 & q & 2.14 & 1.31 \\

J16140792-1938292 & No &  &  & $4947^{+110}_{-160}$ & $1.14^{+0.11}_{-0.11}$ & 14 & $1.30\pm0.12$ & $32.0\pm1.0$ & $171.0\pm1.0$ & 72.0 & this work  & 1.95 & 1.65 \\

J16163345-2521505 & No &  &  & $3770^{+110}_{-140}$ & $0.180^{+0.036}_{-0.036}$ & 21 & $0.59\pm0.22$ & $62.9\pm2.7$ & $59.8\pm2.5$ & 40.0 & r  & 1.76 & 1.07 \\
J16202863-2442087 & No &  &  & $3490^{+140}_{-130}$ & $0.230^{+0.046}_{-0.046}$ & 22 & $0.34\pm0.16$ & $32.4\pm9.0$ & $179\pm11$ & 30.0 & r & 2.25 & 1.43 \\
J16221532-2511349 & No &  &  & $3360^{+130}_{-200}$ & $0.140^{+0.028}_{-0.028}$ & 22 & $0.28\pm0.17$ & $56.24\pm0.29$ & $16.27\pm0.33$ & 17.0 & r  & 2.65 & 1.04 \\

LkCa 15 & No &  &  & $4276^{+210}_{-210}$ & $1.10^{+0.49}_{-0.49}$ & 1 & $0.91\pm0.44$ & $50.590\pm0.020$ & $61.570\pm0.010$ & 76.0 & a  & 1.93 & 1.36 \\
LkHa 330 & No &  &  & $6240^{+100}_{-70}$ & $14.4^{+1.4}_{-1.4}$ & 9 & $2.17\pm0.14$ & $27.5\pm1.0$ & $49.2\pm1.0$ & 68.0 & this work  & 0.41 & 1.85 \\
MHO2 & No &  &  & $3470^{+170}_{-170}$ & $0.971^{+0.097}_{-0.097}$ & 23 & $0.46\pm0.23$ & $38.0\pm1.0$ & $120.0\pm1.0$ & 28.0 & b & 2.44 & 1.55 \\
MHO6 & Yes & $26^{+100}_{-17}$ & $1.83^{+3.0}_{-0.97}$ & $3161^{+160}_{-160}$ & $0.081^{+0.036}_{-0.036}$ & 1 & $0.19\pm0.14$ & $64.560\pm0.060$ & $113.550\pm0.050$ & 10.0 & f & 3.01 & 1.22 \\
MP Mus & Yes & $45^{+500}_{-35}$ & $5.1^{+11}_{-3.4}$ & $4920^{+110}_{-160}$ & $1.20^{+0.12}_{-0.12}$ & 24 & $1.33\pm0.11$ & $32.0\pm1.0$ & $10.0\pm1.0$ & 3.0 & s & 4.44 & 0.96 \\

MWC 758 & Yes & $179^{+830}_{-140}$ & $5.8^{+7.0}_{-5.2}$ & $7250^{+120}_{-120}$ & $8.71^{+0.87}_{-0.87}$ & 2 & $1.649\pm0.071$ & $7.27\pm0.23$ & $76.17\pm0.13$ & 62.0 & a & 4.43 & 0.99 \\

PDS 111 & Yes & $265^{+570}_{-180}$ & $3.6^{+3.9}_{-1.3}$ & $5900^{+100}_{-150}$ & $2.52^{+0.75}_{-0.40}$ & 25 & $1.26\pm0.19$ & $58.20\pm0.10$ & $66.20\pm0.10$ & 30.0 & t  & 15.68 & 2.61 \\
PDS 70 & Yes & $34^{+300}_{-34}$ & $4.5^{+20}_{-3.9}$ & $4060^{+200}_{-200}$ & $0.297^{+0.030}_{-0.030}$ & 26 & $0.81\pm0.27$ & $49.5\pm1.0$ & $161.0\pm1.0$ & 74.0 & u  & 3.29 & 1.39 \\
PDS 99 & No &  &  & $4205^{+210}_{-210}$ & $1.13^{+0.11}_{-0.11}$ & 27 & $0.84\pm0.43$ & $55.0\pm1.0$ & $107.0\pm1.0$ & 56.0 & b  & 1.50 & 2.10 \\
RW Aur B & No &  &  & $4020^{+120}_{-50}$ & $0.54^{+0.14}_{-0.11}$ & 5 & $0.71\pm0.17$ & $63.98\pm0.16$ & $39.65\pm0.20$ & 7.0 & v & 0.68 & 16.42 \\
RXJ1615.3-3255 & No &  &  & $3970^{+200}_{-200}$ & $0.907^{+0.091}_{-0.091}$ & 28 & $0.62\pm0.31$ & $47.100\pm0.010$ & $146.140\pm0.020$ & 17.0 & a & 1.77 & 1.55 \\
RXJ1633.9-2442 & No &  &  & $4200^{+210}_{-210}$ & $1.04^{+0.10}_{-0.10}$ & 29 & $0.83\pm0.42$ & $47.9\pm1.1$ & $77.0\pm1.4$ & 36.0 & w & 2.60 & 1.56 \\
\\[-3ex]
\hline
\end{tabular}
\tablefoot{Table A.1 (continued). References for effective temperatures and luminosities: 12--\citet{2024Natur.635..574B}, 13--\citet{2004ApJ...616..998W}, 14--\citet{2025ApJ...978..117C}, 15--\citet{2018ApJ...865..157A}, 16--\citet{2012ApJ...744..116B}, 17--\citet{2006A&A...460..547G}, 18--\citet{2023ApJ...949...27L}, 19--\citet{2024ApJ...966...59S}, 20--\citet{2018AJ....156..271L}, 21--\citet{2023ApJ...945..112F}, 22--\citet{2022AJ....163...24L}, 23--\citet{2010ApJS..186..111L}, 24--\citet{2002AJ....124.1670M}, 25--\citet{2024A&A...688A.149D}, 26--\citet{2016MNRAS.461..794P}, 27--\citet{2006A&A...460..695T}, 28--\citet{1999MNRAS.307..909W}, 29--\citet{2021MNRAS.501.2934C}. References for disk geometries: j--\citet{2024Natur.635..574B}, k--\citet{2020A&A...642A.164V}, l--\citet{2018ApJ...859...32P}, m--\citet{2023MNRAS.522.2611A}, n--\citet{2024ApJ...966...59S}, o--\citet{2022arXiv220408225V}, p--\citet{2025A&A...696A.232G}, q--\citet{2024ApJ...974..102S}, r--\citet{2025ApJ...989....9V}, s--\citet{2025NatAs...9.1176R}, t--\citet{2024A&A...688A.149D}, u--\citet{2021ApJ...916L...2B}, v--\citet{2024A&A...692A.155K}, w--\citet{2021MNRAS.501.2934C}.}

\end{sidewaystable*}

\begin{sidewaystable*}
\centering
\setlength{\tabcolsep}{3.3pt}
\renewcommand{\arraystretch}{1.3}
\footnotesize
\begin{tabular}{lccccccccccccc}
\hline\hline
Name & Companion & Companion  & Companion & $T_\mathrm{eff}$ & $L_\star$ & Ref. stellar & $M_\star$ & inc & PA & Cavity & Ref. disk  & $|\Delta \mu|/\sigma_{|\Delta \mu|}$ & RUWE \\
 &  & mass (M$_\mathrm{J}$) & separation (au) & (K) & (L$_\odot$) & param. & (M$_\odot$) & (deg) & (deg) & size (au) &  geom.  &  & \\
\hline\\[-2ex]
RXJ1842.9-3532 & No &  &  & $4780^{+240}_{-240}$ & $0.769^{+0.077}_{-0.077}$ & 6 & $1.16\pm0.13$ & $39.220\pm0.040$ & $26.350\pm0.060$ & 37.0 & a & 2.40 & 1.10 \\
RXJ1852.3-3700 & Yes & $124^{+570}_{-96}$ & $2.6^{+7.9}_{-2.3}$ & $4780^{+240}_{-240}$ & $0.609^{+0.061}_{-0.061}$ & 6 & $1.07\pm0.13$ & $32.500\pm0.050$ & $117.610\pm0.030$ & 49.0 & a  & 4.06 & 1.58 \\

RY Lup & Yes & $256^{+770}_{-170}$ & $2.5^{+4.9}_{-1.8}$ & $4900^{+230}_{-230}$ & $1.84^{+0.83}_{-0.83}$ & 7 & $1.43\pm0.37$ & $67.31\pm0.36$ & $108.91\pm0.30$ & 67.0 & o  & 6.82 & 3.12 \\

RY Tau & Yes & $798^{+940}_{-410}$ & $2.62^{+1.6}_{-0.65}$ & $5945^{+140}_{-140}$ & $12.0^{+7.0}_{-4.2}$ & 9 & $2.21\pm0.61$ & $65.0\pm1.0$ & $23.0\pm1.0$ & 27.0 & b & 6.66 & 13.28 \\
SR 21 A & No &  &  & $5950^{+300}_{-300}$ & $7.0^{+2.0}_{-1.5}$ & 9 & $1.88\pm0.52$ & $16.0\pm1.0$ & $14.0\pm1.0$ & 56.0 & b & 1.64 & 1.14 \\
SR 24 S & Yes & $327^{+190}_{-63}$ & $1.74^{+1.8}_{-0.29}$ & $4060^{+200}_{-200}$ & $1.94^{+0.19}_{-0.19}$ & 30 & $0.85\pm0.15$ & $46.0\pm1.0$ & $23.0\pm1.0$ & 35.0 & b  & 8.79 & 8.24 \\

SR 4 & No &  &  & $4074^{+190}_{-180}$ & $1.19^{+0.69}_{-0.44}$ & 4 & $0.70\pm0.31$ & $22.0\pm2.0$ & $18.0\pm5.0$ & 13.0 & d  & 0.90 & 1.46 \\

SY Cha & No &  &  & $4060^{+200}_{-200}$ & $0.55^{+0.25}_{-0.25}$ & 1 & $0.75\pm0.35$ & $51.650\pm0.030$ & $165.770\pm0.040$ & 35.0 & a  & 0.77 & 1.37 \\
SZ Cha & No &  &  & $4900^{+240}_{-240}$ & $1.65^{+0.74}_{-0.74}$ & 1 & $1.37\pm0.35$ & $51.0\pm1.0$ & $160.0\pm1.0$ & 72.0 & this work  & 2.38 & 2.94 \\

Sz 100 & Yes & $61^{+44}_{-30}$ & $1.10^{+0.81}_{-0.26}$ & $3057^{+70}_{-70}$ & $0.106^{+0.048}_{-0.048}$ & 7 & $0.144\pm0.049$ & $42.9\pm1.7$ & $66.6\pm5.2$ & 26.0 & o & 8.65 & 3.88 \\
Sz 108 B & No &  &  & $3125^{+72}_{-72}$ & $0.098^{+0.081}_{-0.081}$ & 7 & $0.174\pm0.071$ & $54.87\pm0.60$ & $158.96\pm0.66$ & 13.0 & p & 1.51 & 1.21 \\

Sz 111 & No &  &  & $3705^{+170}_{-170}$ & $0.210^{+0.094}_{-0.094}$ & 7 & $0.51\pm0.28$ & $54.38\pm0.46$ & $43.6\pm2.5$ & 55.0 & o & 2.44 & 1.30 \\

Sz 118 & Yes & $176^{+470}_{-150}$ & $2.5^{+6.9}_{-2.2}$ & $4350^{+200}_{-200}$ & $0.70^{+0.31}_{-0.31}$ & 7 & $1.03\pm0.29$ & $65.95\pm0.62$ & $172.85\pm0.41$ & 64.0 & o  & 4.10 & 1.58 \\
Sz 123 A & No &  &  & $3705^{+170}_{-170}$ & $0.136^{+0.061}_{-0.061}$ & 7 & $0.55\pm0.25$ & $51.8\pm1.8$ & $154.4\pm5.2$ & 39.0 & o & 0.96 & 1.34 \\
Sz 129 & No &  &  & $4060^{+190}_{-190}$ & $0.42^{+0.19}_{-0.19}$ & 7 & $0.79\pm0.29$ & $30.80\pm0.42$ & $149.11\pm0.78$ & 10.0 & o  & 2.20 & 1.14 \\
Sz 131 & No &  &  & $3415^{+79}_{-79}$ & $0.151^{+0.058}_{-0.058}$ & 7 & $0.306\pm0.099$ & $62.85\pm0.84$ & $155.6\pm3.0$ & 6.4 & p & 1.92 & 1.10 \\

Sz 72 & No &  &  & $3560^{+160}_{-160}$ & $0.28^{+0.12}_{-0.12}$ & 7 & $0.38\pm0.21$ & $31.5\pm1.7$ & $47.9\pm3.5$ & 7.8 & p  & 1.81 & 1.29 \\
Sz 76 & Yes & $36^{+100}_{-28}$ & $1.6^{+3.6}_{-1.3}$ & $3270^{+75}_{-75}$ & $0.172^{+0.077}_{-0.077}$ & 7 & $0.225\pm0.074$ & $66.3\pm1.5$ & $85.6\pm1.5$ & 4.0 & o  & 4.16 & 1.25 \\
Sz 84 & Yes & $40^{+78}_{-19}$ & $1.25^{+1.0}_{-0.47}$ & $3125^{+160}_{-160}$ & $0.139^{+0.063}_{-0.063}$ & 1 & $0.17\pm0.11$ & $75.1\pm1.7$ & $165.7\pm3.2$ & 12.0 & x  & 5.25 & 1.90 \\

Sz 90 & No &  &  & $4060^{+190}_{-190}$ & $0.42^{+0.28}_{-0.28}$ & 7 & $0.79\pm0.26$ & $56.62\pm0.55$ & $135.87\pm0.75$ & 9.6 & p & 2.07 & 1.36 \\
Sz 91 & No &  &  & $3705^{+170}_{-170}$ & $0.200^{+0.090}_{-0.090}$ & 7 & $0.51\pm0.28$ & $47.0\pm1.0$ & $21.0\pm1.0$ & 67.0 & this work  & 1.44 & 1.14 \\
Sz 96 & No &  &  & $3705^{+170}_{-170}$ & $0.42^{+0.32}_{-0.32}$ & 7 & $0.45\pm0.28$ & $48.9\pm3.4$ & $23.3\pm3.9$ & 6.2 & p & 1.68 & 1.05 \\
Sz 98 & No &  &  & $4060^{+190}_{-190}$ & $1.53^{+0.15}_{-0.15}$ & 7 & $0.671\pm0.077$ & $47.1\pm1.0$ & $111.6\pm1.0$ & 16.4 & e & 0.40 & 1.62 \\
T Cha & Yes & $136^{+620}_{-96}$ & $5.3^{+5.5}_{-2.6}$ & $5570^{+280}_{-280}$ & $1.20^{+0.12}_{-0.12}$ & 31 & $1.08\pm0.16$ & $73.0\pm1.0$ & $113.0\pm1.0$ & 34.0 & b & 7.26 & 1.38 \\
TW Hya & No &  &  & $4205^{+210}_{-210}$ & $0.301^{+0.030}_{-0.030}$ & 6 & $0.89\pm0.13$ & $7.0\pm1.0$ & $155.0\pm1.0$ & 3.0 & b  & 2.38 & 1.18 \\
UX Tau A & No &  &  & $5490^{+130}_{-210}$ & $8.9^{+3.1}_{-2.9}$ & 9 & $2.29\pm0.52$ & $40.0\pm1.0$ & $167.0\pm1.0$ & 31.0 & b & 2.84 & 1.62 \\

UZ Tau E & Yes & $120^{+140}_{-71}$ & $1.75^{+1.5}_{-0.63}$ & $3574^{+180}_{-180}$ & $0.347^{+0.035}_{-0.035}$ & 5 & $0.38\pm0.20$ & $56.10\pm0.40$ & $90.40\pm0.40$ & 10.0 & y & 10.72 & 5.32 \\

V4046 Sgr & No &  &  & $4060^{+200}_{-200}$ & $0.493^{+0.049}_{-0.049}$ & 32 & $0.77\pm0.37$ & $33.360\pm0.010$ & $76.020\pm0.020$ & 31.0 & a & 0.40 & 0.86 \\
V892 Tau & Yes & $606^{+1400}_{-350}$ & $2.64^{+4.8}_{-0.80}$ & $11500^{+1500}_{-800}$ & $1.73^{+2.0}_{-0.82}$ & 33, 34 & $3.0\pm1.0$ & $55.5\pm1.0$ & $52.7\pm1.0$ & 27.0 & z & 13.10 & 3.55 \\

WISPIT 2 & No &  &  & $4400^{+50}_{-50}$ & $0.699^{+0.021}_{-0.021}$ & 35 & $1.087\pm0.028$ &  &  & 68.0* &   & 2.49 & 1.17 \\

WSB 60 ISO-Oph 196 & No &  &  & $3197^{+160}_{-160}$ & $0.33^{+0.15}_{-0.15}$ & 1 & $0.215\pm0.085$ & $28.0\pm1.0$ & $172.0\pm1.0$ & 32.0 & b & 1.10 & 0.97 \\
WSB 82 rho-Oph 38 & No &  &  & $5030^{+250}_{-250}$ & $4.51^{+0.45}_{-0.45}$ & 29 & $1.70\pm0.50$ & $61.20\pm0.50$ & $173.0\pm1.0$ & 50.0 & w  & 0.93 & 2.07 \\
ZZ Tau IRS & No &  &  & $3230^{+160}_{-160}$ & $0.52^{+0.24}_{-0.24}$ & 1 & $0.249\pm0.064$ & $60.160\pm0.070$ & $134.730\pm0.090$ & 50.0 & aa & 0.43 & 6.64 \\
\\[-3ex]
\hline
\end{tabular}
\tablefoot{Table A.1 (continued). References for effective temperatures and luminosities: 30--\citet{2006A&A...452..245N}, 31--\citet{2009A&A...501.1013S}, 32--\citet{2004A&A...421.1159S}, 33--\citet{2025A&A...703A.210A}, 34--\citet{2018A&A...620A.128V}, 35--\citet{2025ApJ...990L...8V}. References for disk geometries: x--\citet{2021ApJ...908..250H}, y--\citet{2019ApJ...882...49L}, z--\citet{2024A&A...687A.311A}, aa--\citet{2021AJ....161..264H}. Asterisk (*) indicates near-IR cavity.}
\end{sidewaystable*}

\begin{figure*}
\begin{subfigure}{1\textwidth}

  \includegraphics[width=0.25\textwidth]{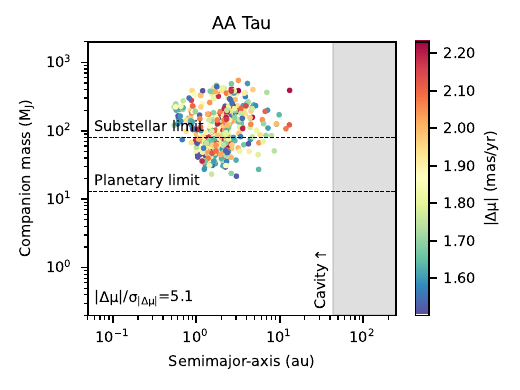}
  \hspace*{-0.30cm}\includegraphics[width=0.25\textwidth]{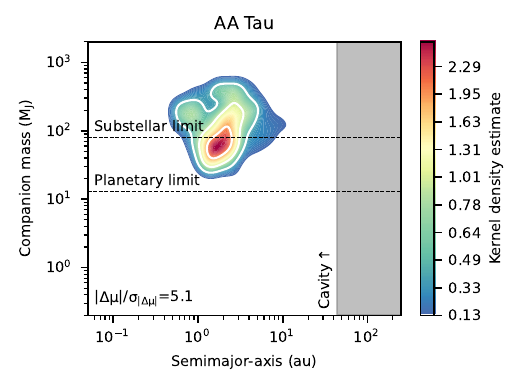}
  \hspace*{0.30cm}\includegraphics[width=0.25\textwidth]{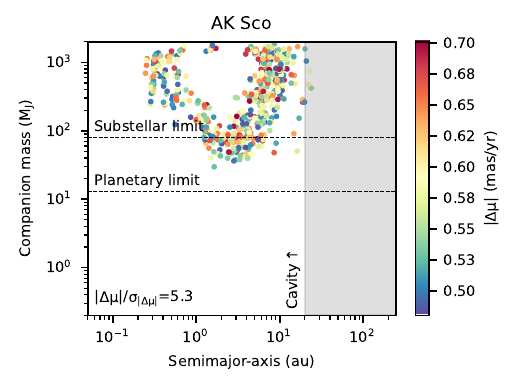}
  \hspace*{-0.30cm}\includegraphics[width=0.25\textwidth]{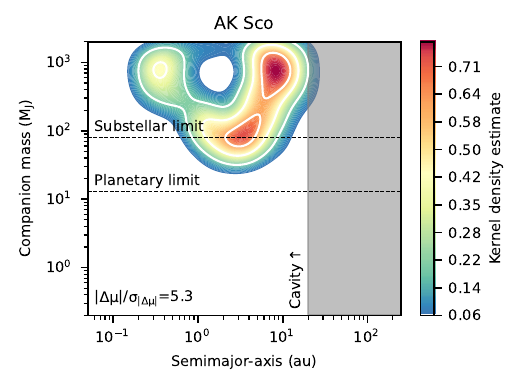}

  \vspace{-5pt}\includegraphics[width=0.25\textwidth]{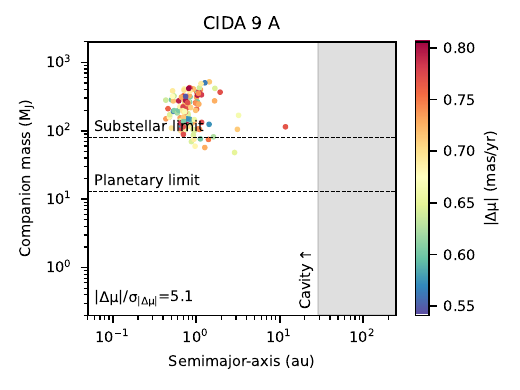}
  \hspace*{-0.30cm}\includegraphics[width=0.25\textwidth]{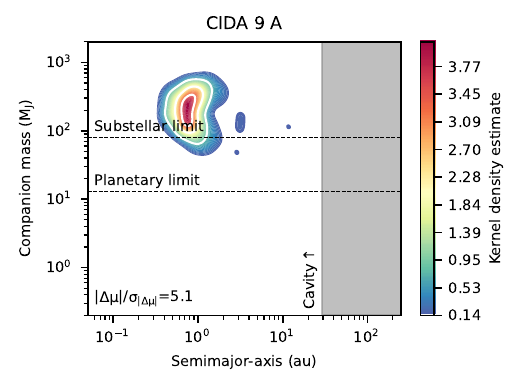}
  \hspace*{0.30cm}\includegraphics[width=0.25\textwidth]{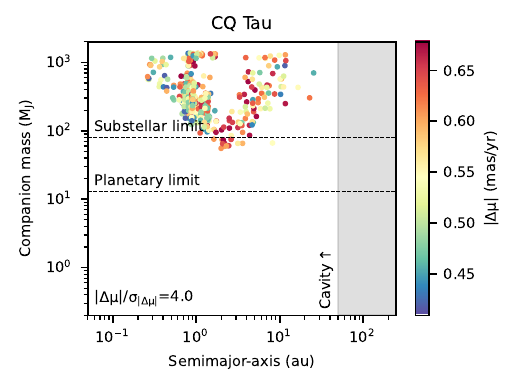}
  \hspace*{-0.30cm}\includegraphics[width=0.25\textwidth]{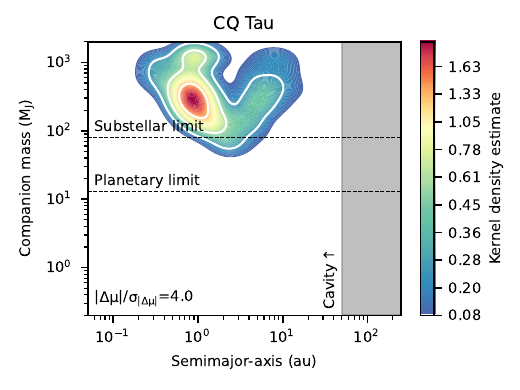}

  \vspace{-5pt}\includegraphics[width=0.25\textwidth]{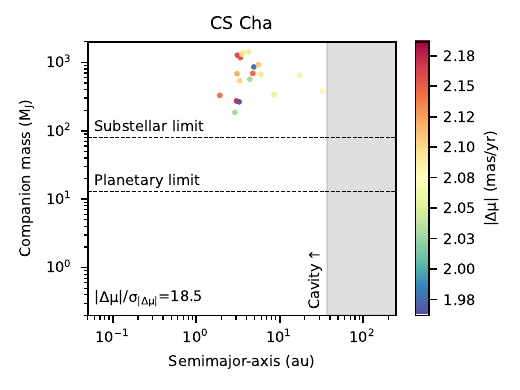}
  \hspace*{-0.30cm}\includegraphics[width=0.25\textwidth]{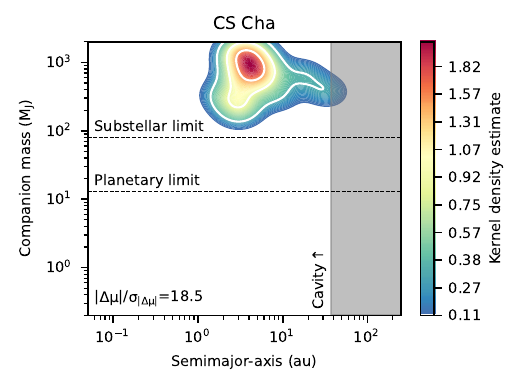}
  \hspace*{0.30cm}\includegraphics[width=0.25\textwidth]{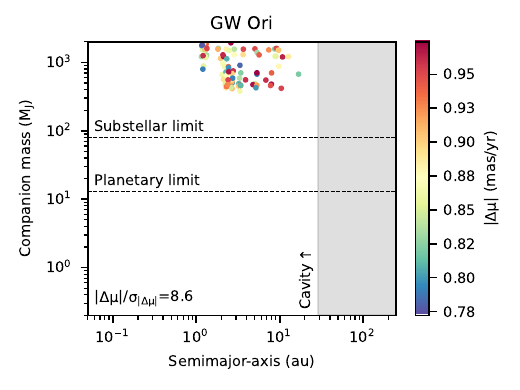}
  \hspace*{-0.30cm}\includegraphics[width=0.25\textwidth]{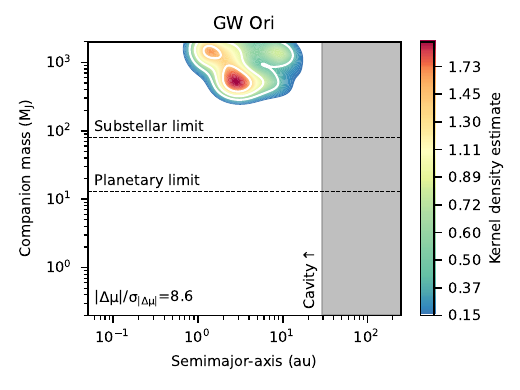}

  \vspace{-5pt}\includegraphics[width=0.25\textwidth]{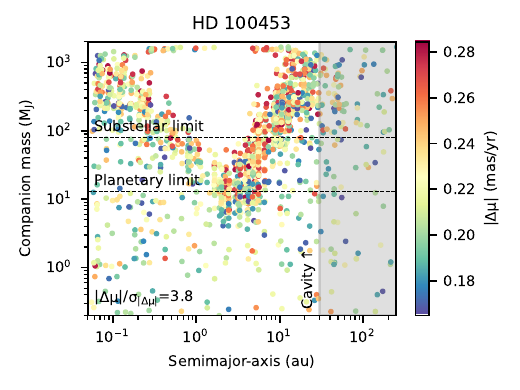}
  \hspace*{-0.30cm}\includegraphics[width=0.25\textwidth]{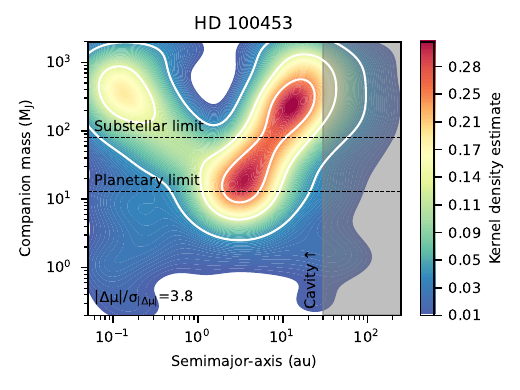}
  \hspace*{0.30cm}\includegraphics[width=0.25\textwidth]{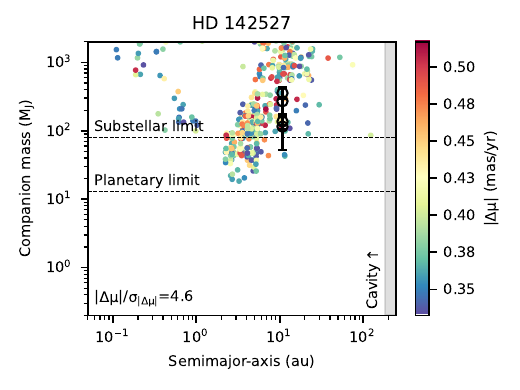}
  \hspace*{-0.30cm}\includegraphics[width=0.25\textwidth]{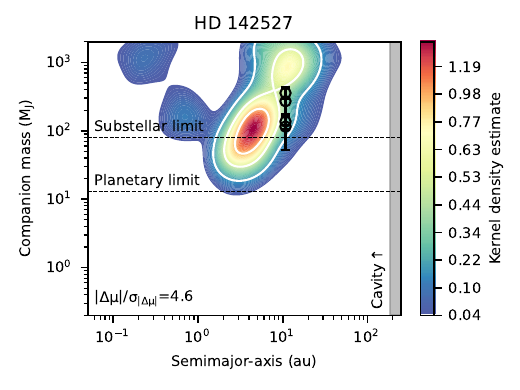}

  \vspace{-5pt}\includegraphics[width=0.25\textwidth]{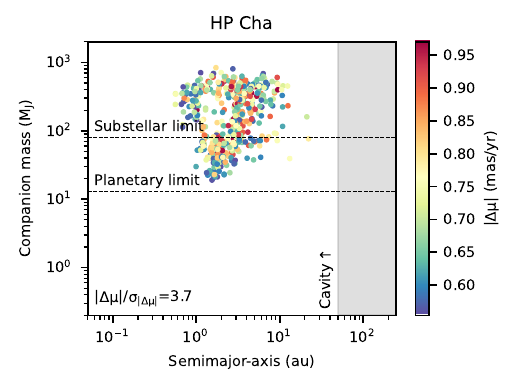}
  \hspace*{-0.30cm}\includegraphics[width=0.25\textwidth]{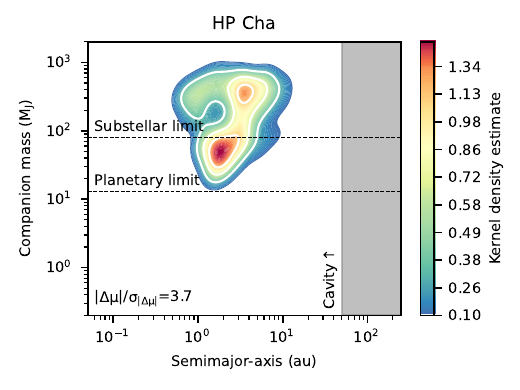}
  \hspace*{0.30cm}\includegraphics[width=0.25\textwidth]{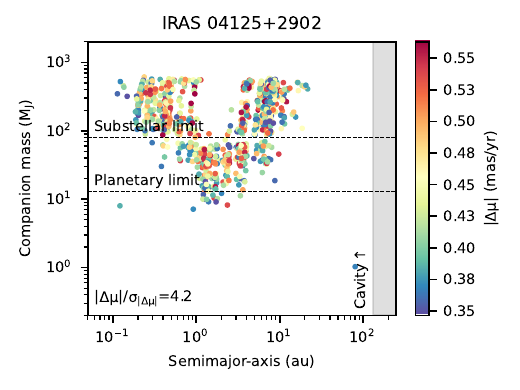}
  \hspace*{-0.30cm}\includegraphics[width=0.25\textwidth]{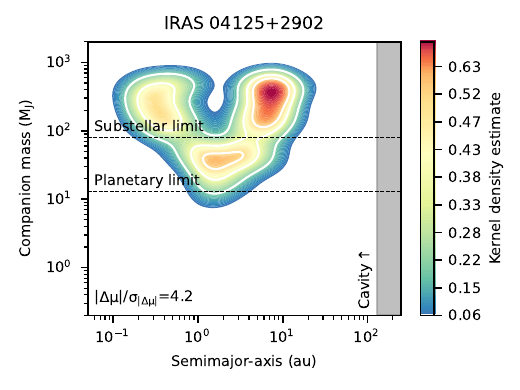}

  \vspace{-5pt}\includegraphics[width=0.25\textwidth]{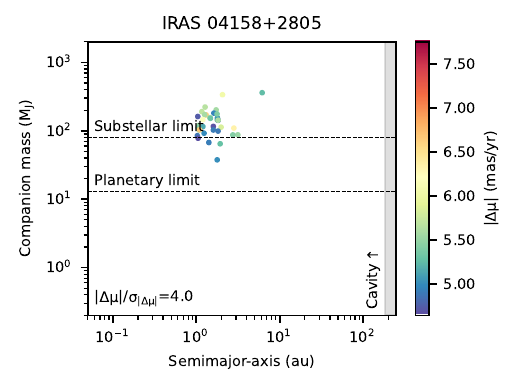}
  \hspace*{-0.30cm}\includegraphics[width=0.25\textwidth]{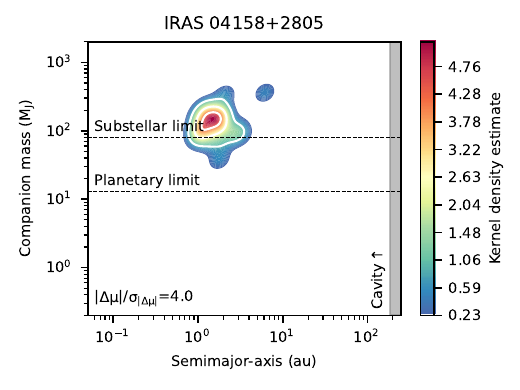}
  \hspace*{0.30cm}\includegraphics[width=0.25\textwidth]{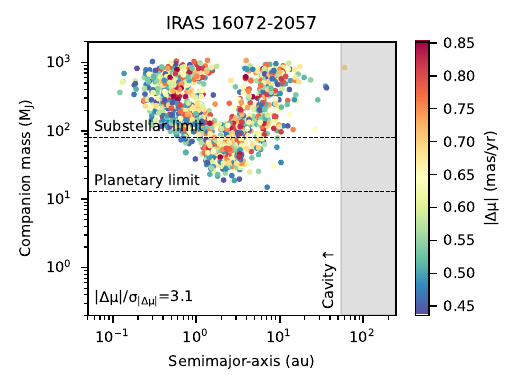}
  \hspace*{-0.30cm}\includegraphics[width=0.25\textwidth]{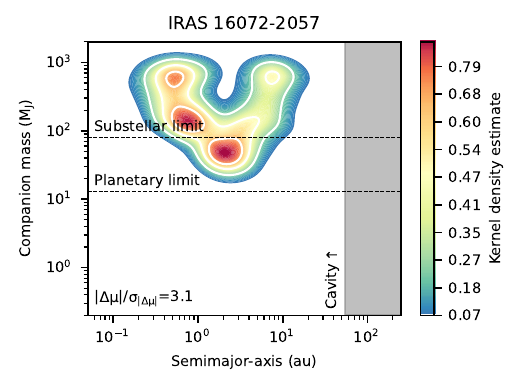}

  \vspace{-5pt}\includegraphics[width=0.25\textwidth]{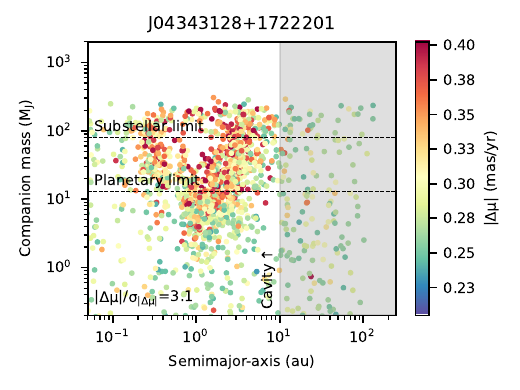}
  \hspace*{-0.30cm}\includegraphics[width=0.25\textwidth]{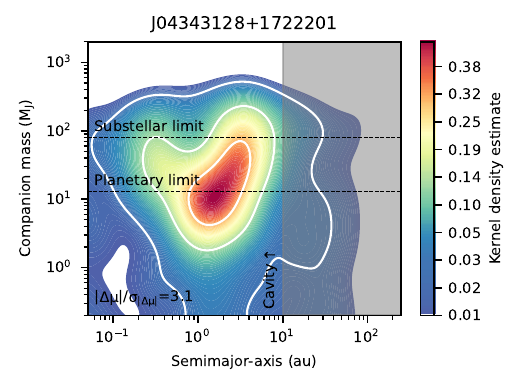}
  \hspace*{0.30cm}\includegraphics[width=0.25\textwidth]{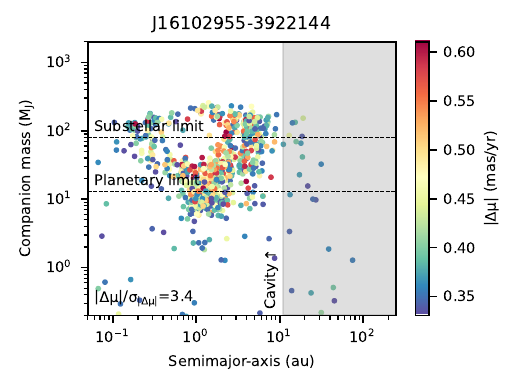}
  \hspace*{-0.30cm}\includegraphics[width=0.25\textwidth]{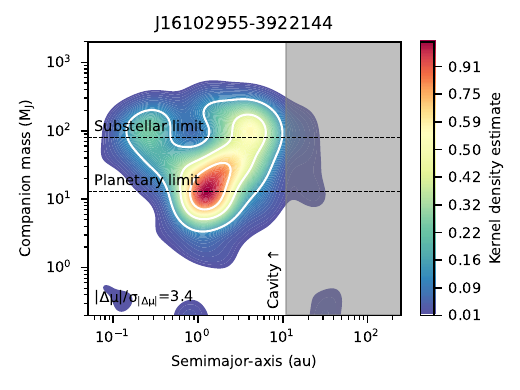}

\end{subfigure}
\caption{Orbital separation and companion mass that would produce the observed \textit{Gaia} astrometry for the 31 sources with a significant proper motion anomaly ($|\Delta \mu|/\sigma_{|\Delta \mu|} \geq 3$). GG Tau is not shown here because its companion could not be modelled.}\label{fig:all_detect}
\end{figure*}

\begin{figure*}
\ContinuedFloat
\begin{subfigure}{1\textwidth}

  \vspace{-5pt}\includegraphics[width=0.25\textwidth]{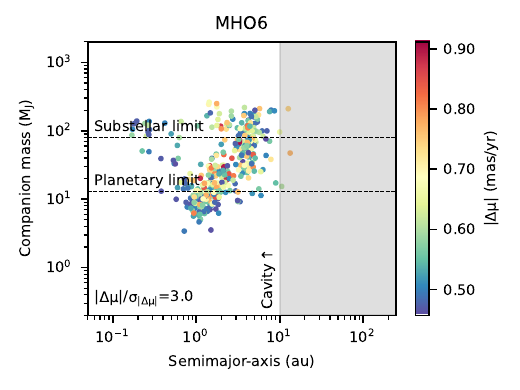}
  \hspace*{-0.30cm}\includegraphics[width=0.25\textwidth]{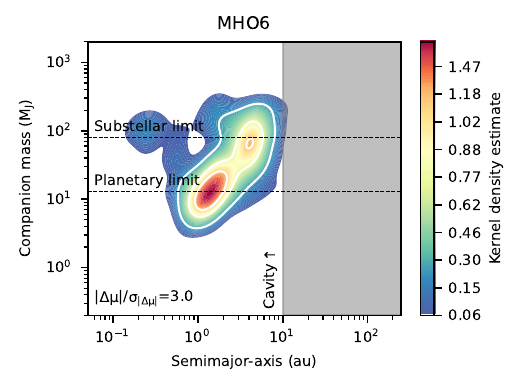}
  \hspace*{0.30cm}\includegraphics[width=0.25\textwidth]{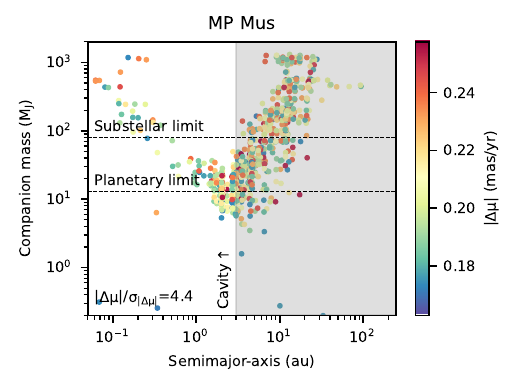}
  \hspace*{-0.30cm}\includegraphics[width=0.25\textwidth]{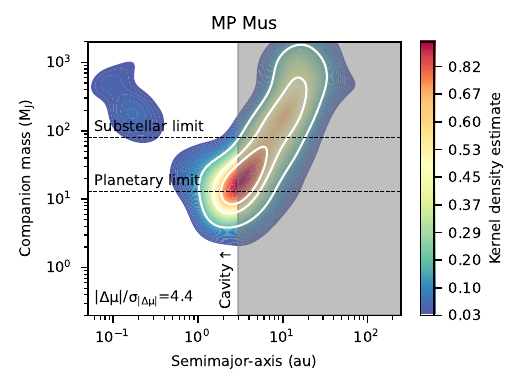}

  \vspace{-5pt}\includegraphics[width=0.25\textwidth]{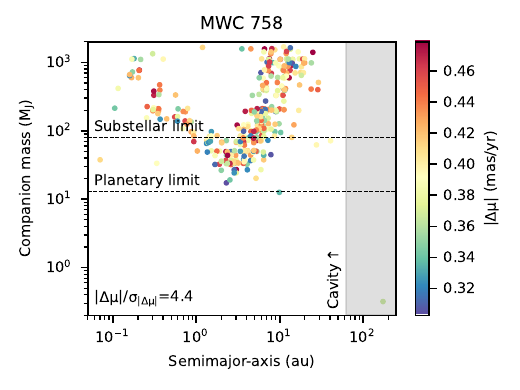}
  \hspace*{-0.30cm}\includegraphics[width=0.25\textwidth]{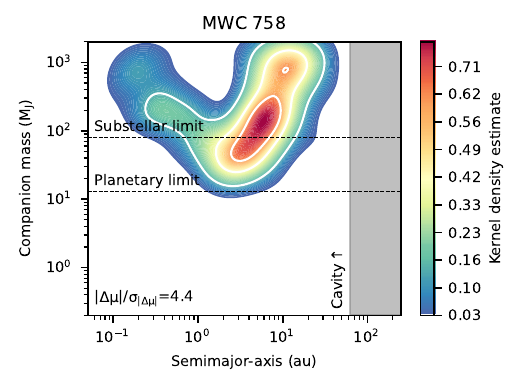}
  \hspace*{0.30cm}\includegraphics[width=0.25\textwidth]{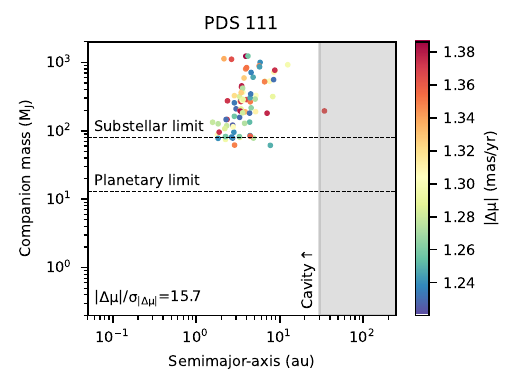}
  \hspace*{-0.30cm}\includegraphics[width=0.25\textwidth]{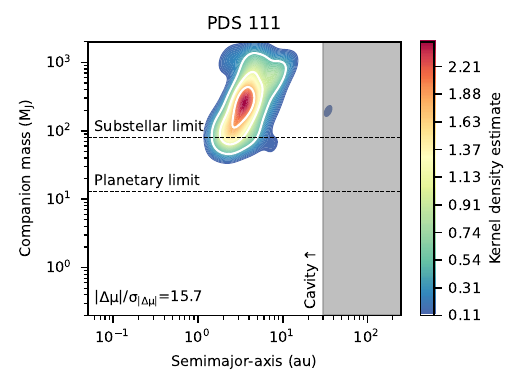}

  \vspace{-5pt}\includegraphics[width=0.25\textwidth]{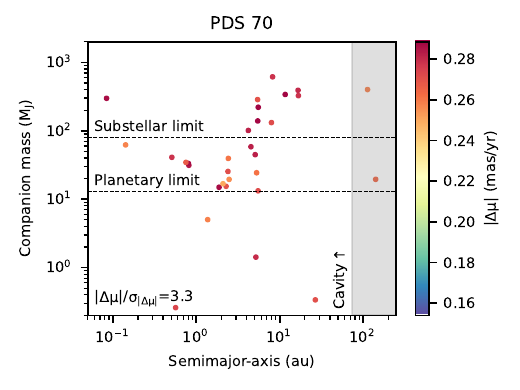}
  \hspace*{-0.30cm}\includegraphics[width=0.25\textwidth]{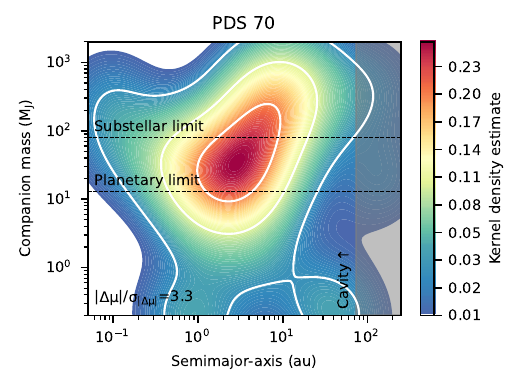}
  \hspace*{0.30cm}\includegraphics[width=0.25\textwidth]{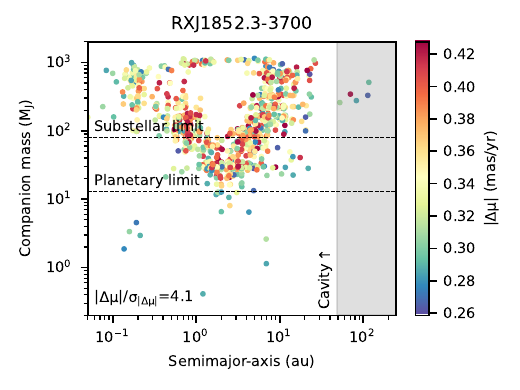}
  \hspace*{-0.30cm}\includegraphics[width=0.25\textwidth]{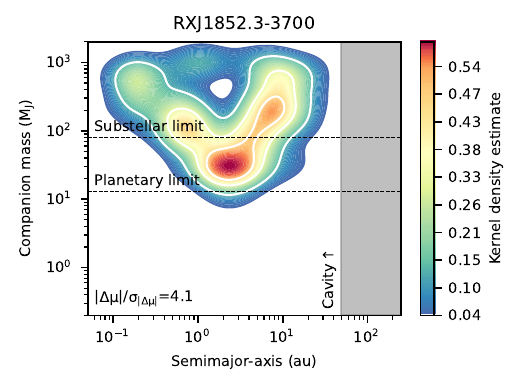}

  \vspace{-5pt}\includegraphics[width=0.25\textwidth]{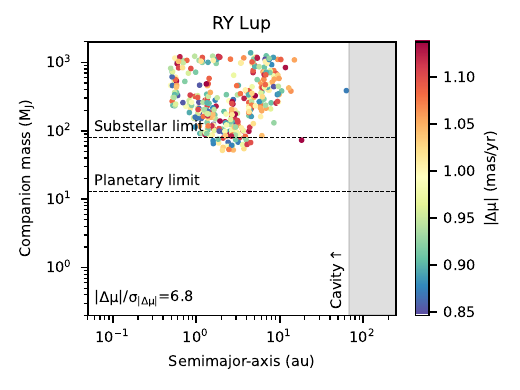}
  \hspace*{-0.30cm}\includegraphics[width=0.25\textwidth]{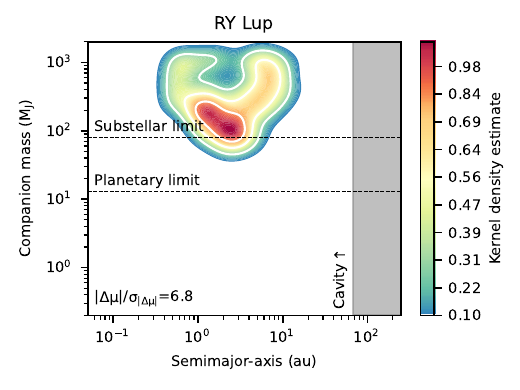}
  \hspace*{0.30cm}\includegraphics[width=0.25\textwidth]{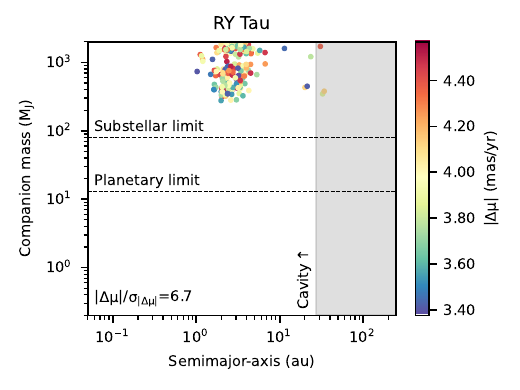}
  \hspace*{-0.30cm}\includegraphics[width=0.25\textwidth]{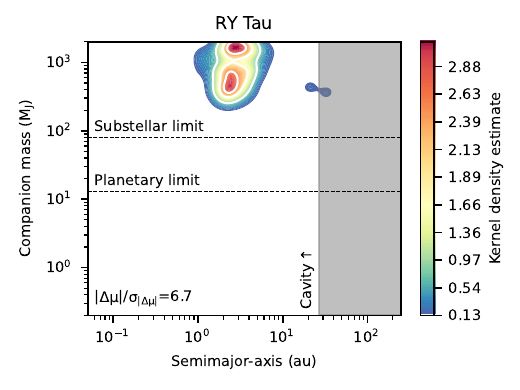}

  \vspace{-5pt}\includegraphics[width=0.25\textwidth]{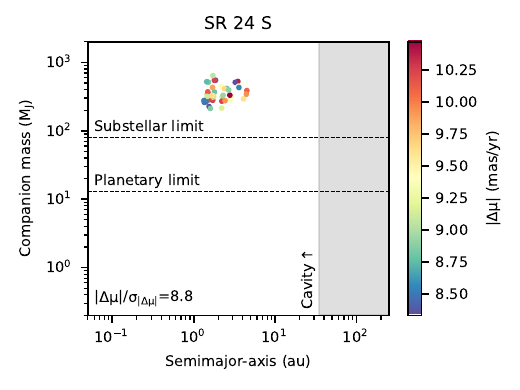}
  \hspace*{-0.30cm}\includegraphics[width=0.25\textwidth]{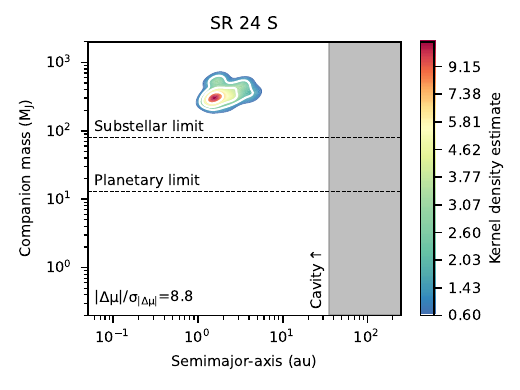}
  \hspace*{0.30cm}\includegraphics[width=0.25\textwidth]{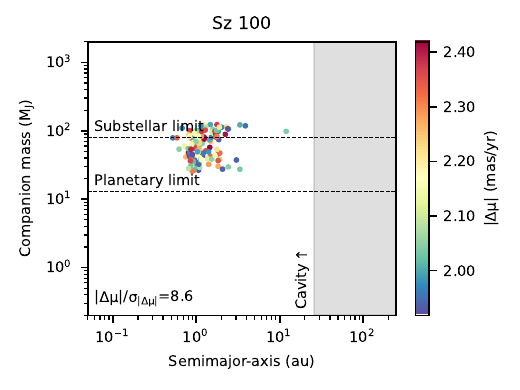}
  \hspace*{-0.30cm}\includegraphics[width=0.25\textwidth]{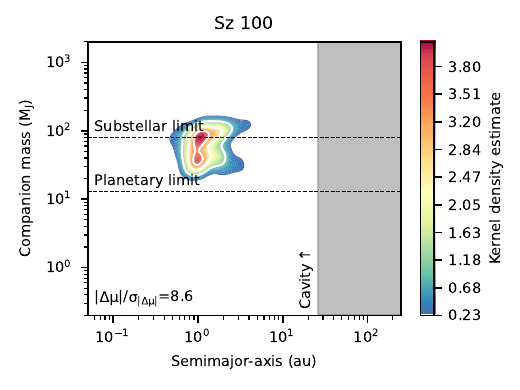}

  \vspace{-5pt}\includegraphics[width=0.25\textwidth]{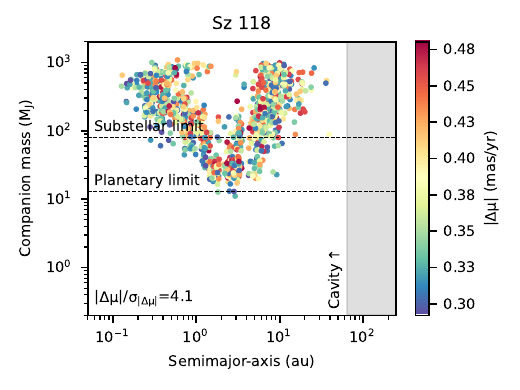}
  \hspace*{-0.30cm}\includegraphics[width=0.25\textwidth]{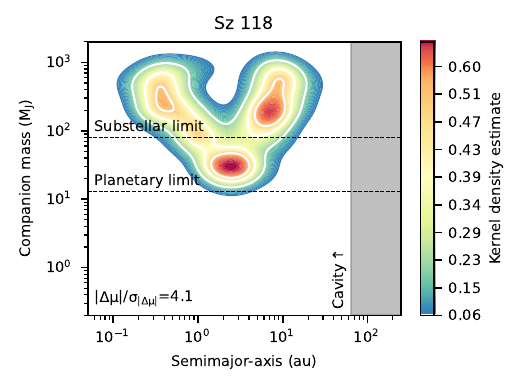}
  \hspace*{0.30cm}\includegraphics[width=0.25\textwidth]{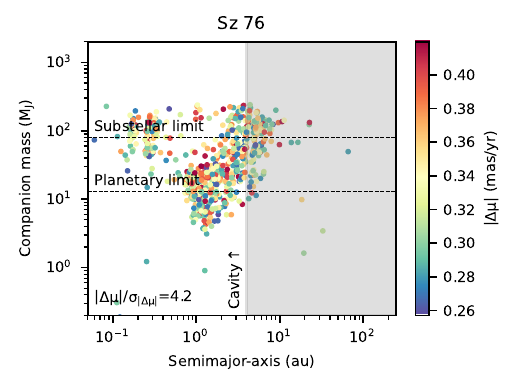}
  \hspace*{-0.30cm}\includegraphics[width=0.25\textwidth]{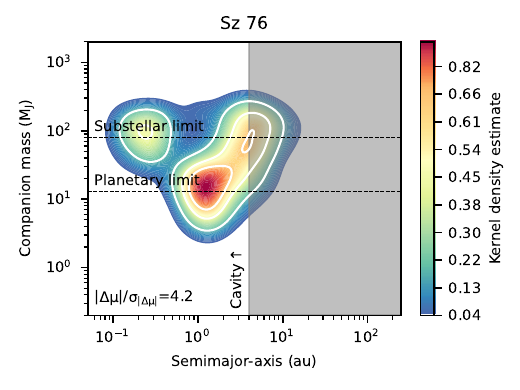}

  \vspace{-5pt}\includegraphics[width=0.25\textwidth]{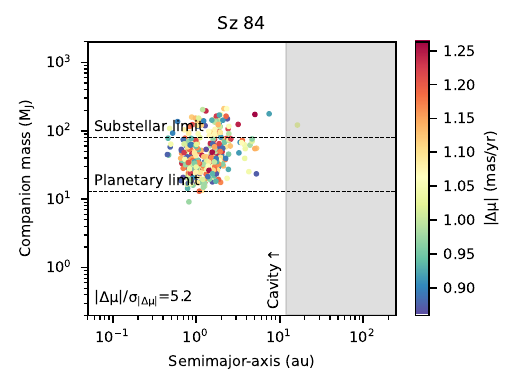}
  \hspace*{-0.30cm}\includegraphics[width=0.25\textwidth]{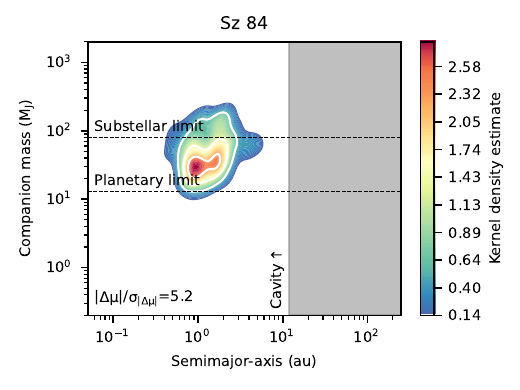}
  \hspace*{0.30cm}\includegraphics[width=0.25\textwidth]{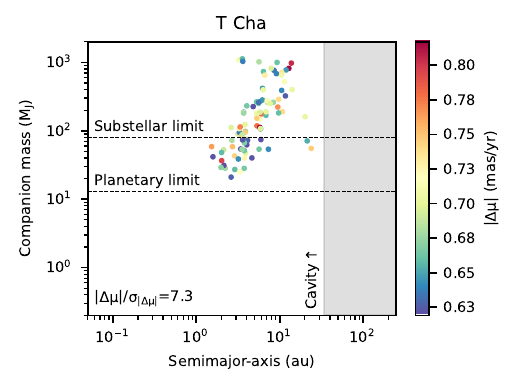}
  \hspace*{-0.30cm}\includegraphics[width=0.25\textwidth]{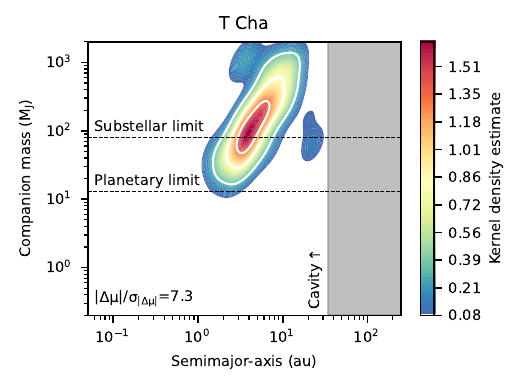}

\end{subfigure}
\captionsetup{justification=centering}
\caption{continued.}
\end{figure*}

\begin{figure*}
\ContinuedFloat
\begin{subfigure}{1\textwidth}  

  \vspace{-5pt}\includegraphics[width=0.25\textwidth]{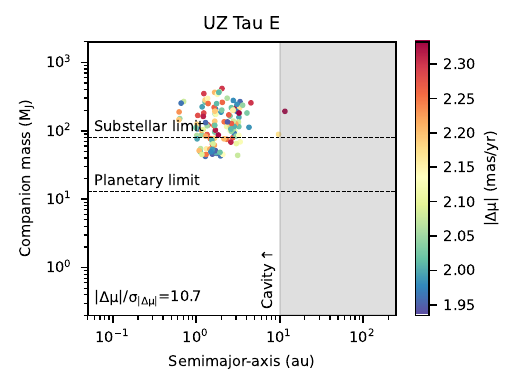}
  \hspace*{-0.30cm}\includegraphics[width=0.25\textwidth]{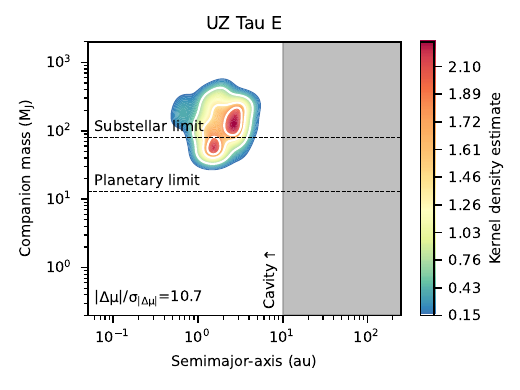}
  \hspace*{0.30cm}\includegraphics[width=0.25\textwidth]{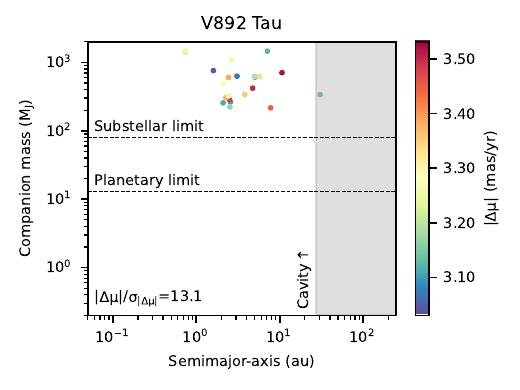}
  \hspace*{-0.30cm}\includegraphics[width=0.25\textwidth]{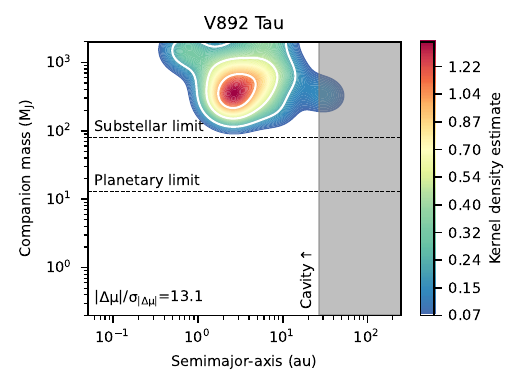}

\end{subfigure}
\caption{continued.}
\end{figure*}

\begin{figure*}
\begin{subfigure}{1\textwidth}

  \includegraphics[width=0.25\textwidth]{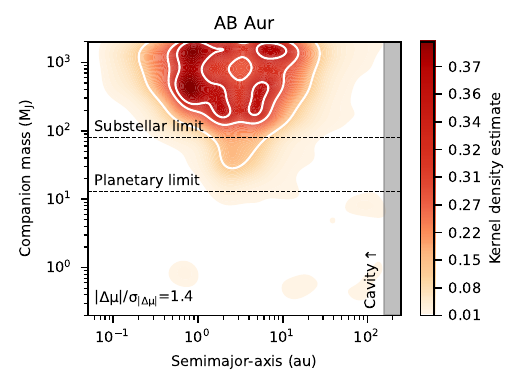}
  \hspace*{-0.15cm}\includegraphics[width=0.25\textwidth]{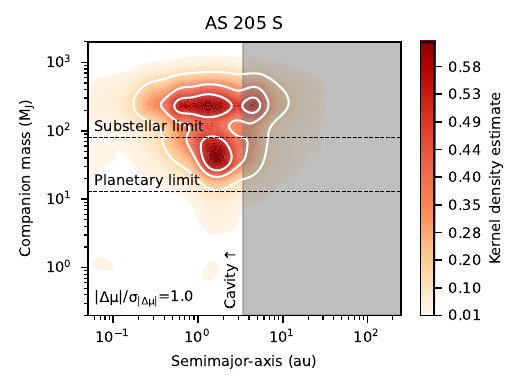}
  \hspace*{-0.15cm}\includegraphics[width=0.25\textwidth]{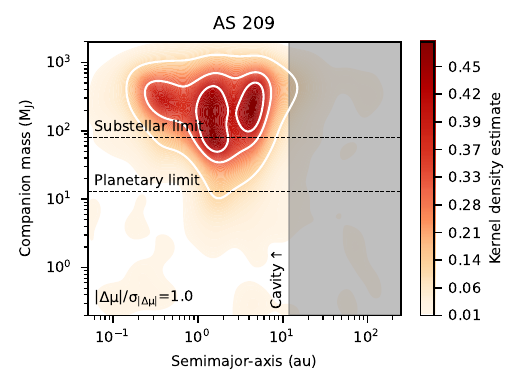}
  \hspace*{-0.15cm}\includegraphics[width=0.25\textwidth]{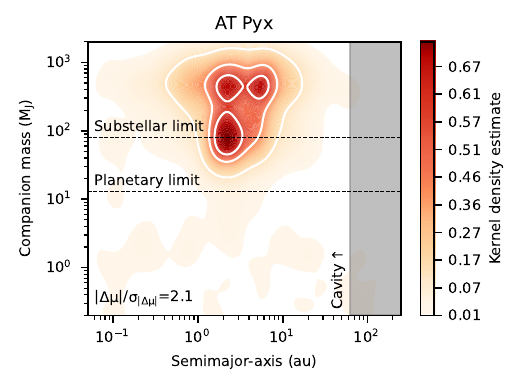}

  \vspace{-5pt} 
  \includegraphics[width=0.25\textwidth]{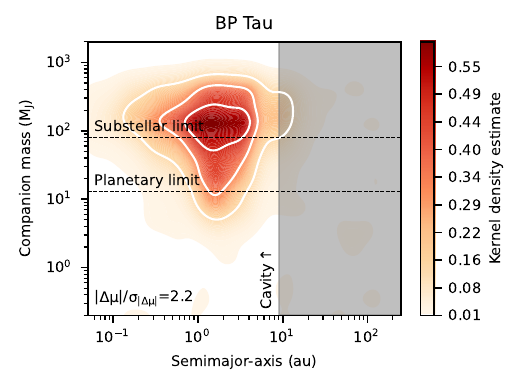}
  \hspace*{-0.15cm}\includegraphics[width=0.25\textwidth]{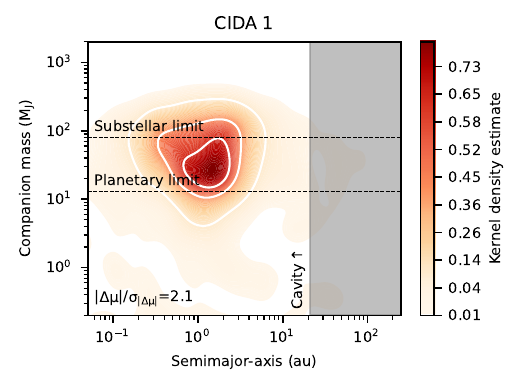}
  \hspace*{-0.15cm}\includegraphics[width=0.25\textwidth]{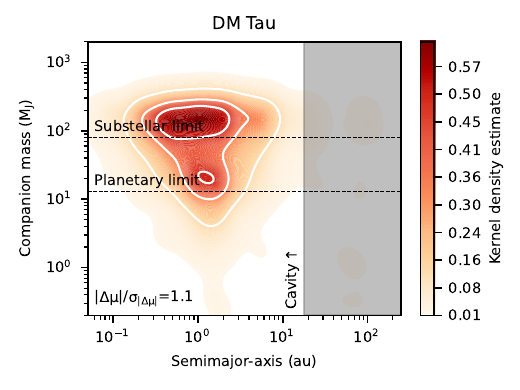}
  \hspace*{-0.15cm}\includegraphics[width=0.25\textwidth]{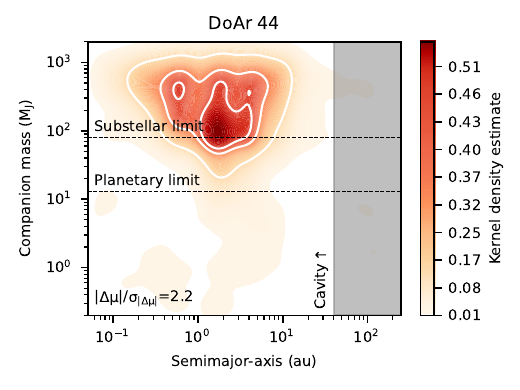}

  \vspace{-5pt}\includegraphics[width=0.25\textwidth]{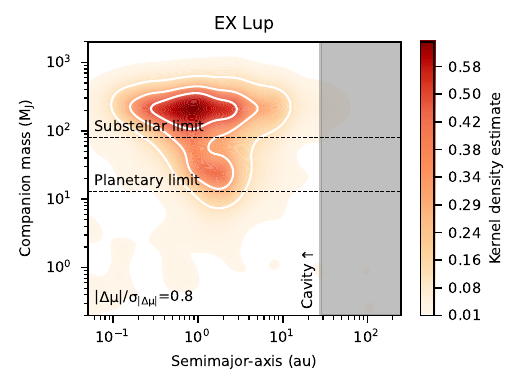}
  \hspace*{-0.15cm}\includegraphics[width=0.25\textwidth]{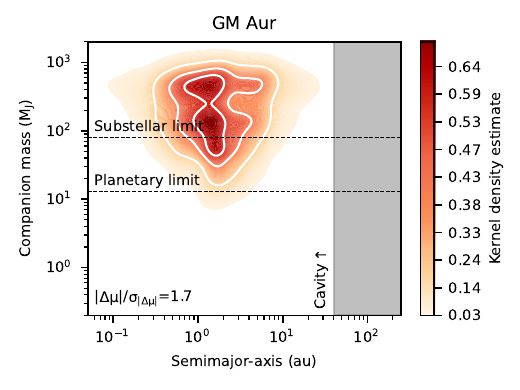}
  \hspace*{-0.15cm}\includegraphics[width=0.25\textwidth]{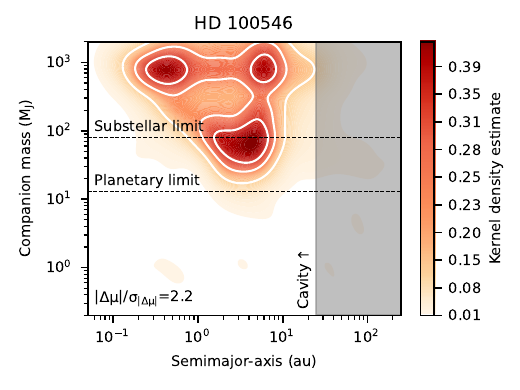}
  \hspace*{-0.15cm}\includegraphics[width=0.25\textwidth]{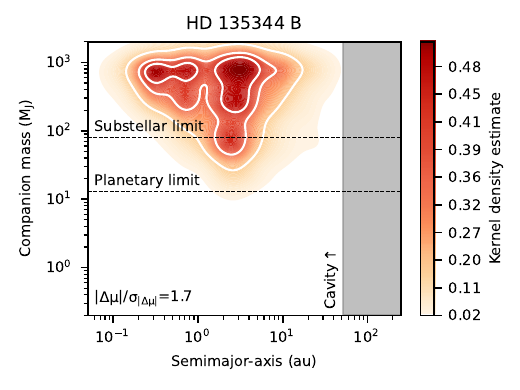}

  \vspace{-5pt}\includegraphics[width=0.25\textwidth]{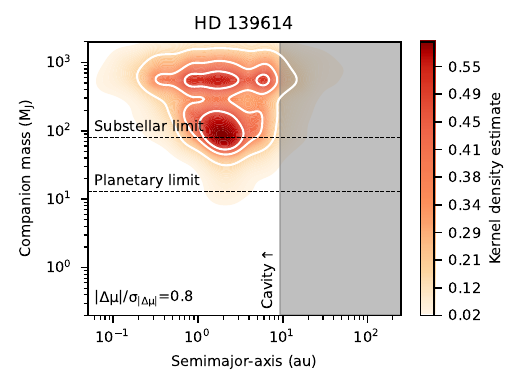}
  \hspace*{-0.15cm}\includegraphics[width=0.25\textwidth]{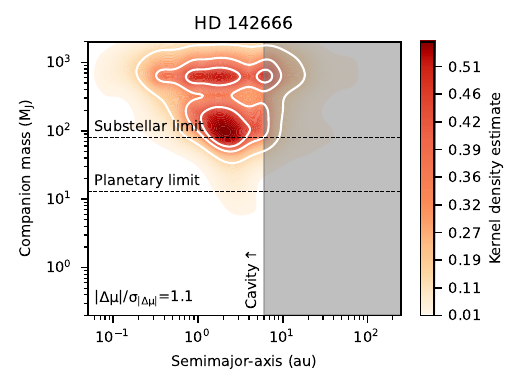}
  \hspace*{-0.15cm}\includegraphics[width=0.25\textwidth]{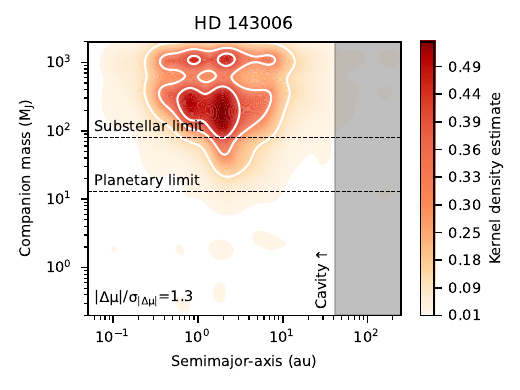}
  \hspace*{-0.15cm}\includegraphics[width=0.25\textwidth]{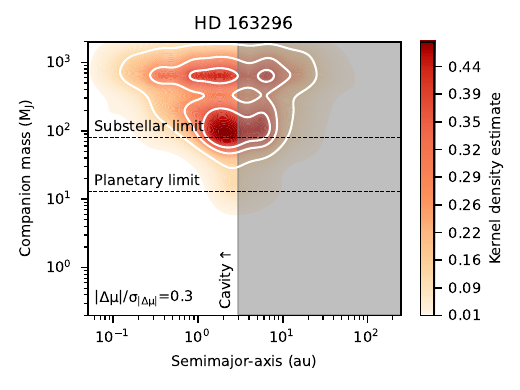}

  \vspace{-5pt} \includegraphics[width=0.25\textwidth]{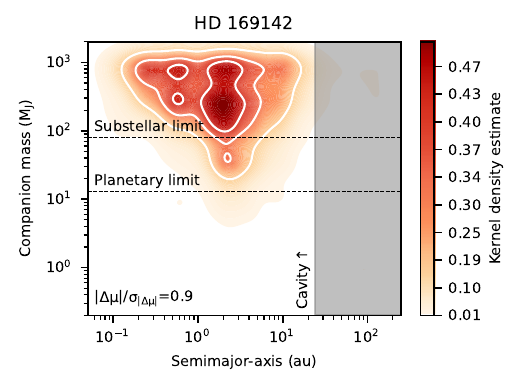}
  \hspace*{-0.15cm}\includegraphics[width=0.25\textwidth]{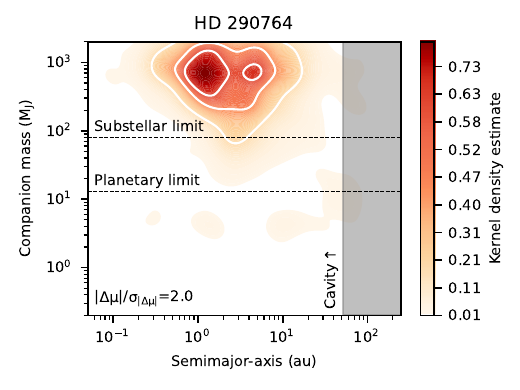}
  \hspace*{-0.15cm}\includegraphics[width=0.25\textwidth]{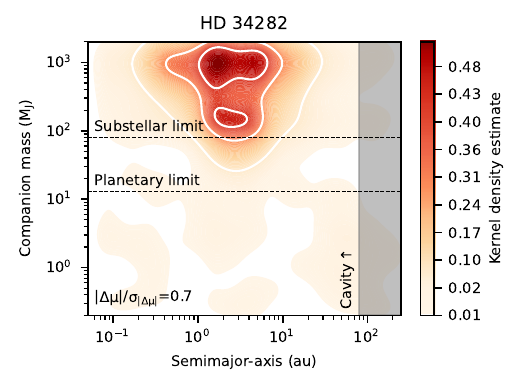}
  \hspace*{-0.15cm}\includegraphics[width=0.25\textwidth]{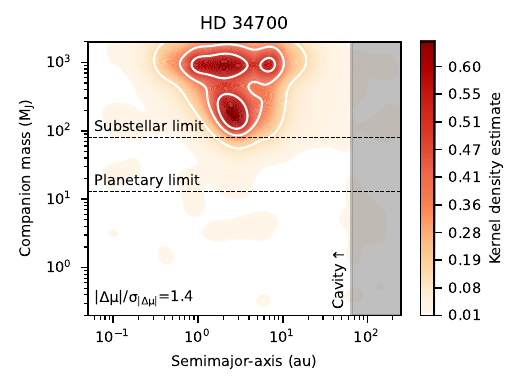}

  \hspace*{-0.15cm}\vspace{-5pt} \includegraphics[width=0.25\textwidth]{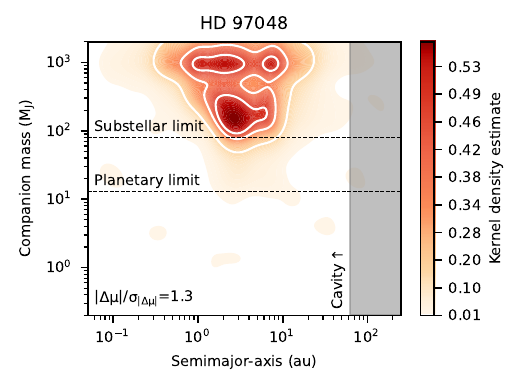}
  \hspace*{-0.15cm}\includegraphics[width=0.25\textwidth]{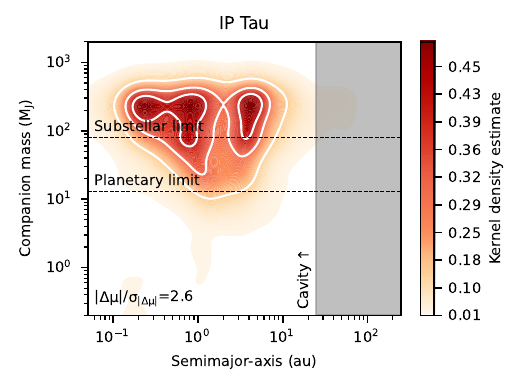}
  \hspace*{-0.15cm}\includegraphics[width=0.25\textwidth]{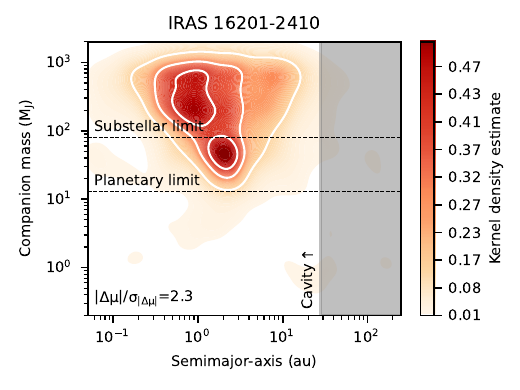}
  \hspace*{-0.15cm}\includegraphics[width=0.25\textwidth]{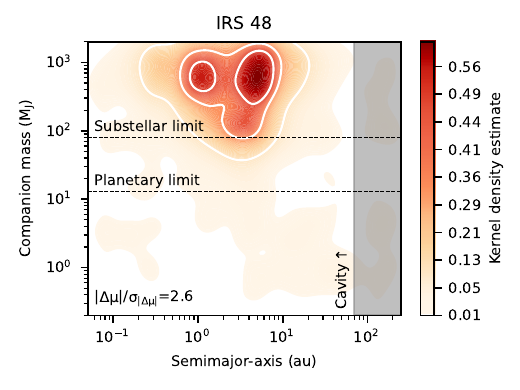}
  
  \vspace{-5pt} \includegraphics[width=0.25\textwidth]{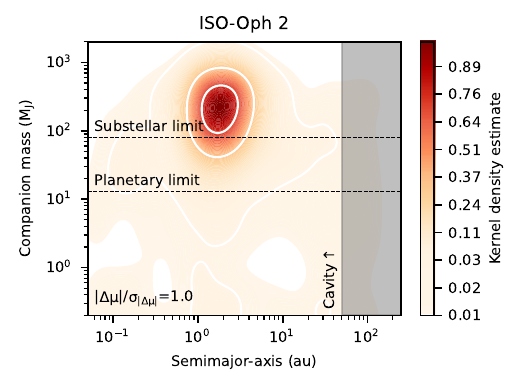}
  \hspace*{-0.15cm}\includegraphics[width=0.25\textwidth]{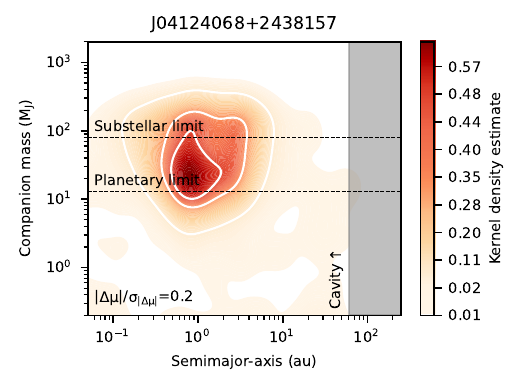}
  \hspace*{-0.15cm}\includegraphics[width=0.25\textwidth]{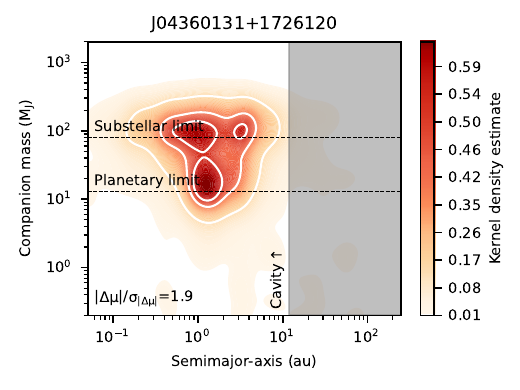}
  \hspace*{-0.15cm}\includegraphics[width=0.25\textwidth]{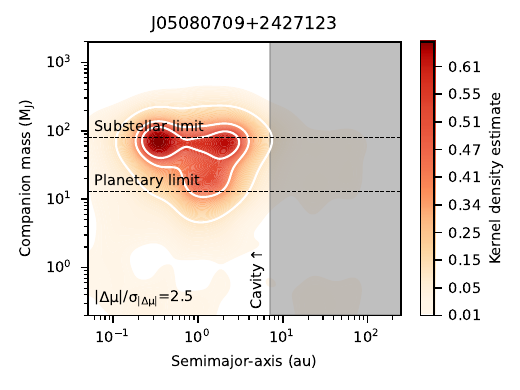}

\end{subfigure}
\caption{Parameter space of orbital separations and companion masses that would produce either a significant \textit{Gaia} proper motion anomaly or a high RUWE value, for sources without a significant proper motion anomaly ($|\Delta \mu|/\sigma_{|\Delta \mu|} < 3$). These figures indicate the companion separation–mass parameter space in which a companion dominating the astrometric signal can be excluded.}\label{fig:all_non_detect}
\end{figure*}

\begin{figure*}
\ContinuedFloat
\begin{subfigure}{1\textwidth}

  \includegraphics[width=0.25\textwidth]{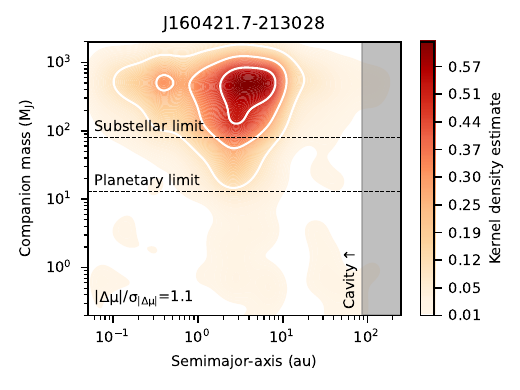}
  \hspace*{-0.15cm}\includegraphics[width=0.25\textwidth]{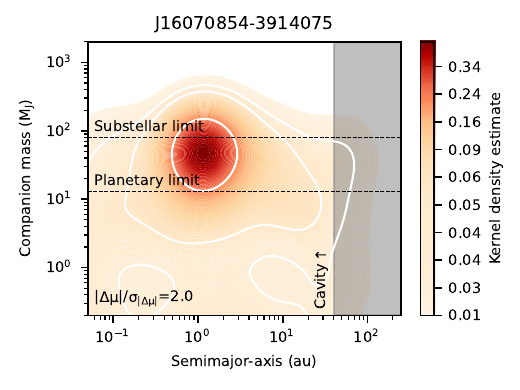}
  \hspace*{-0.15cm}\includegraphics[width=0.25\textwidth]{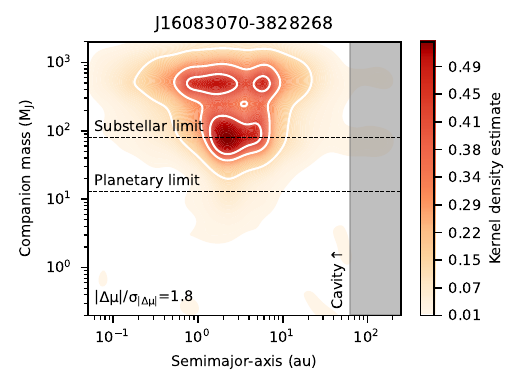}
  \hspace*{-0.15cm}\includegraphics[width=0.25\textwidth]{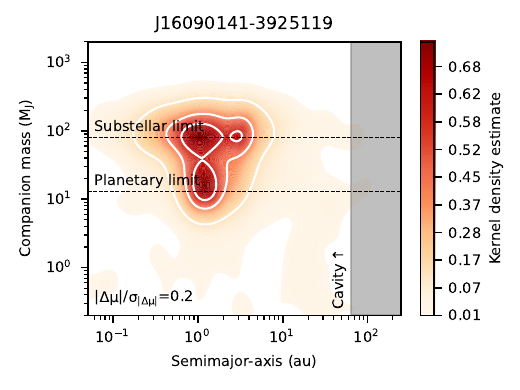}
  
  \vspace{-5pt} \includegraphics[width=0.25\textwidth]{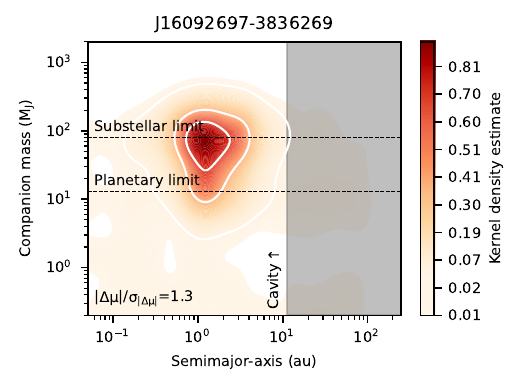}
  \hspace*{-0.15cm} \includegraphics[width=0.25\textwidth]{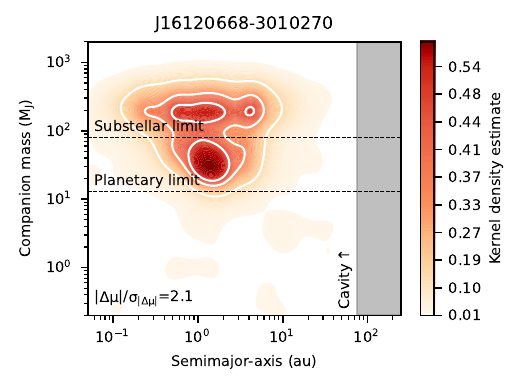}
  \hspace*{-0.15cm}\includegraphics[width=0.25\textwidth]{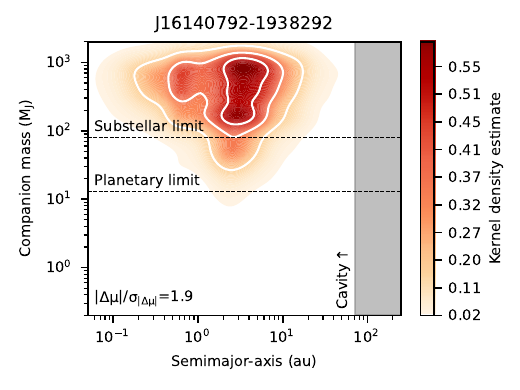}
  \hspace*{-0.15cm}\includegraphics[width=0.25\textwidth]{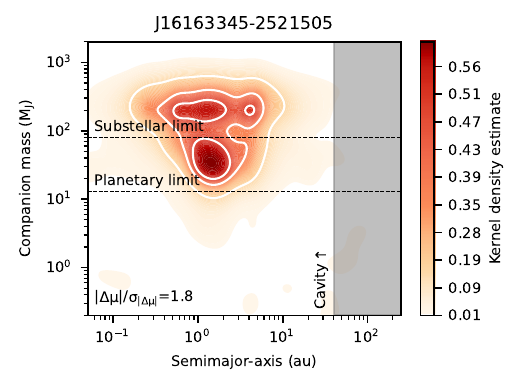}

  \vspace{-5pt} \includegraphics[width=0.25\textwidth]{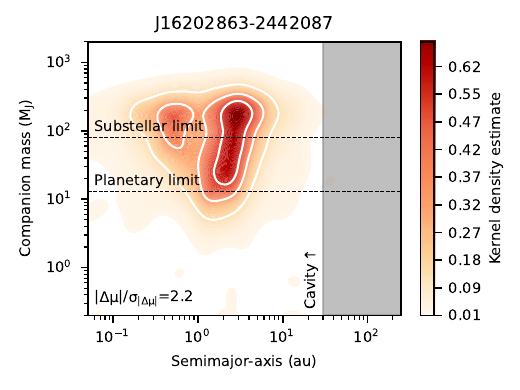}
  \hspace*{-0.15cm}\includegraphics[width=0.25\textwidth]{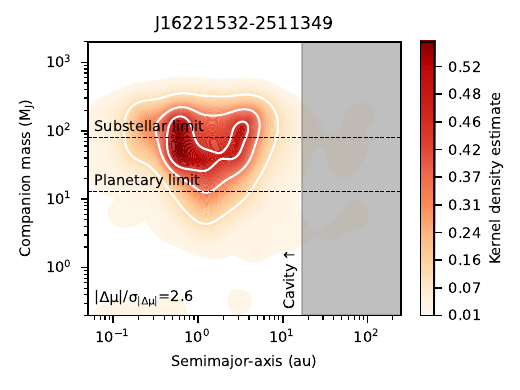}
  \hspace*{-0.15cm}\includegraphics[width=0.25\textwidth]{Plots/TDs_after_referee/LkCa_15_density.pdf}
  \hspace*{-0.15cm}\includegraphics[width=0.25\textwidth]{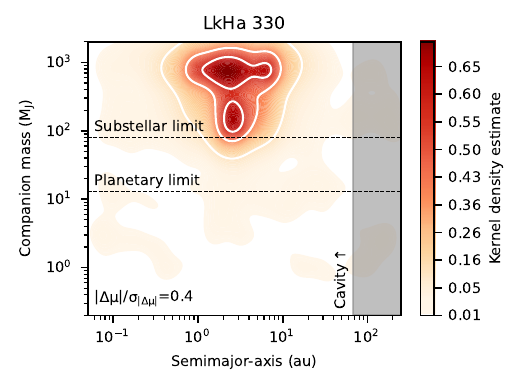}

  \vspace{-5pt} \includegraphics[width=0.25\textwidth]{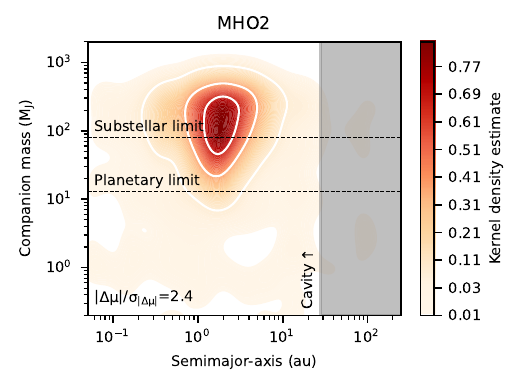}
  \hspace*{-0.15cm}\includegraphics[width=0.25\textwidth]{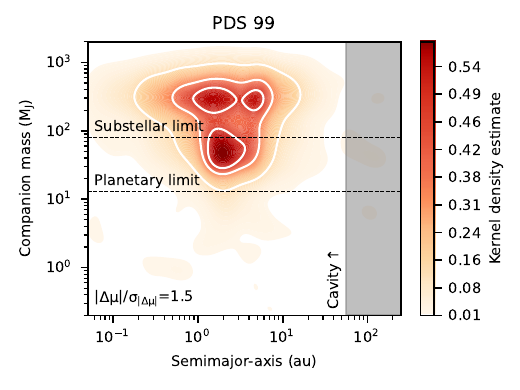}
  \hspace*{-0.15cm}\includegraphics[width=0.25\textwidth]{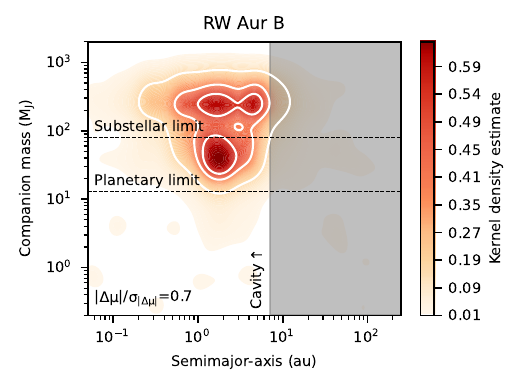}
  \hspace*{-0.15cm}\includegraphics[width=0.25\textwidth]{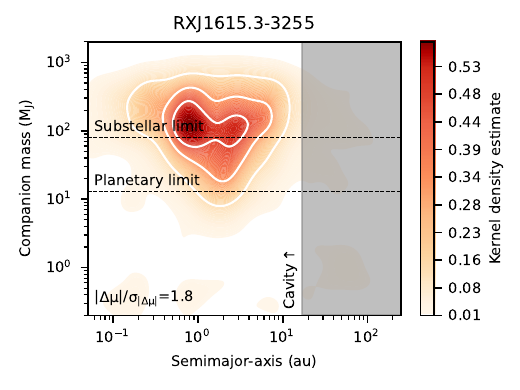}

  \vspace{-5pt} \includegraphics[width=0.25\textwidth]{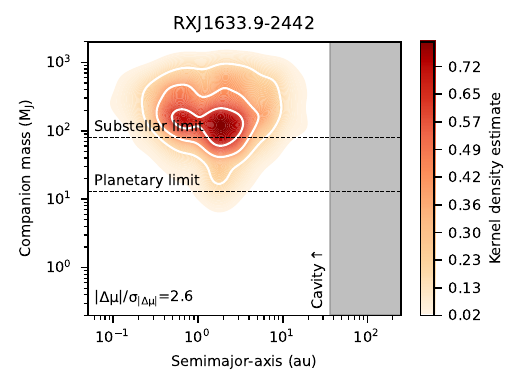}
  \hspace*{-0.15cm}\includegraphics[width=0.25\textwidth]{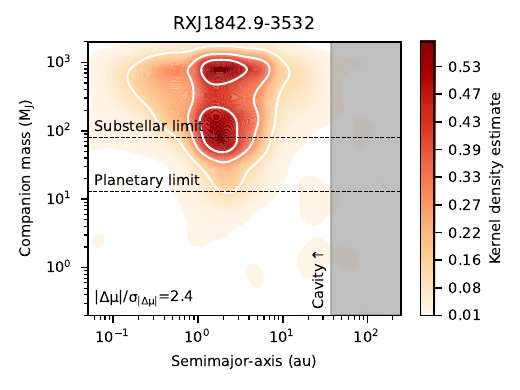}
  \hspace*{-0.15cm}\includegraphics[width=0.25\textwidth]{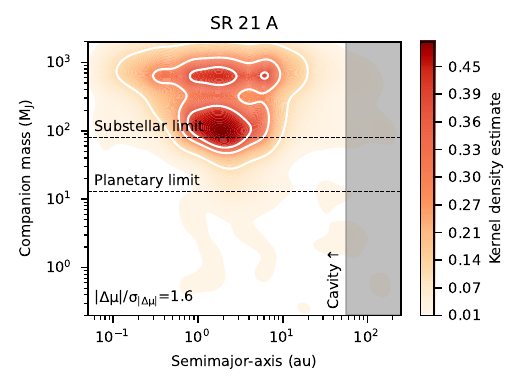}
  \hspace*{-0.15cm}\includegraphics[width=0.25\textwidth]{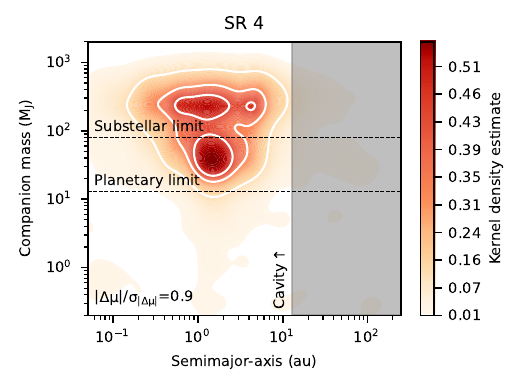}

  \vspace{-5pt} \includegraphics[width=0.25\textwidth]{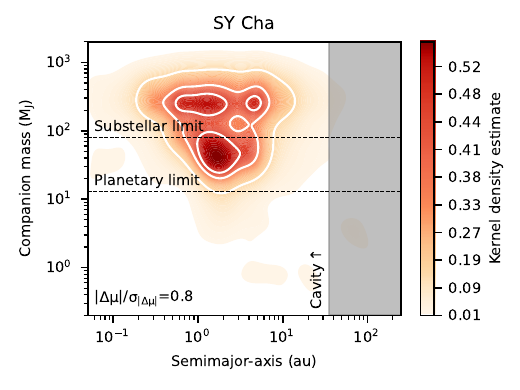}
  \hspace*{-0.15cm}\includegraphics[width=0.25\textwidth]{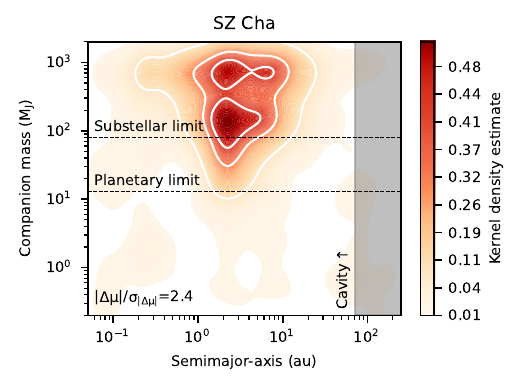}
  \hspace*{-0.15cm}\includegraphics[width=0.25\textwidth]{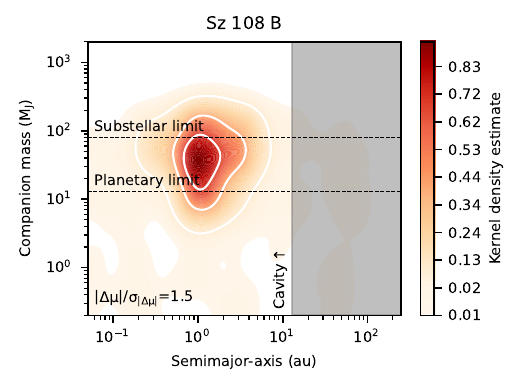}
  \hspace*{-0.15cm}\includegraphics[width=0.25\textwidth]{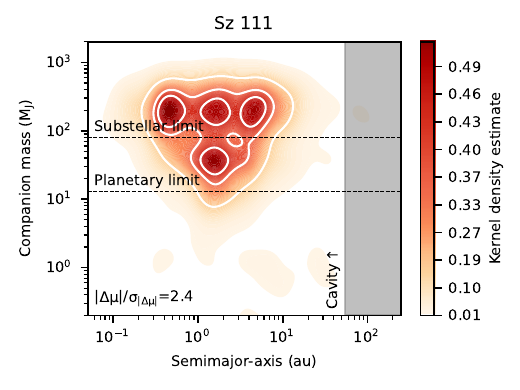}

  \vspace{-5pt}\includegraphics[width=0.25\textwidth]{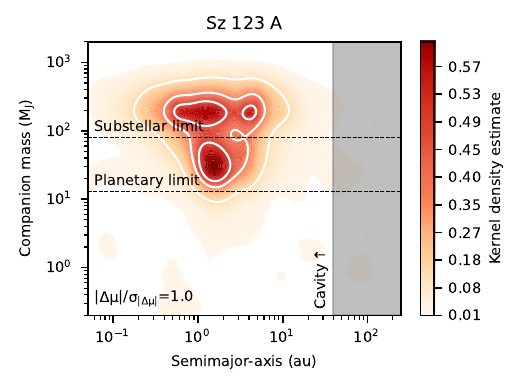}
  \hspace*{-0.15cm}\includegraphics[width=0.25\textwidth]{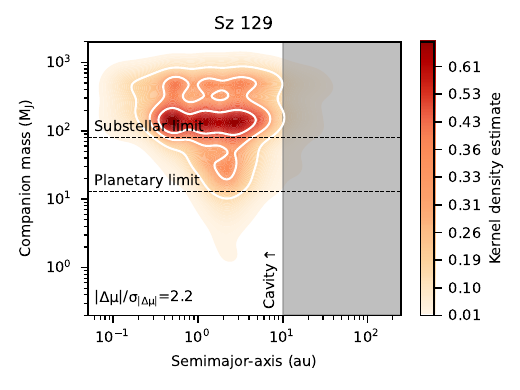}
  \hspace*{-0.15cm}\includegraphics[width=0.25\textwidth]{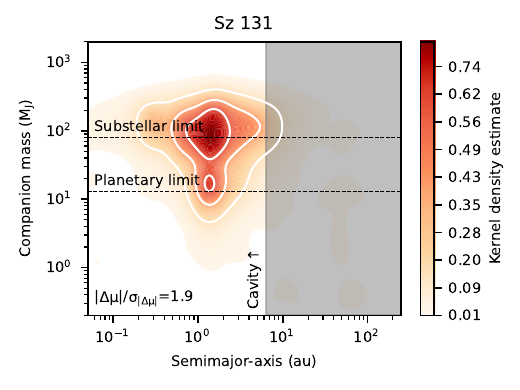}
  \hspace*{-0.15cm}\includegraphics[width=0.25\textwidth]{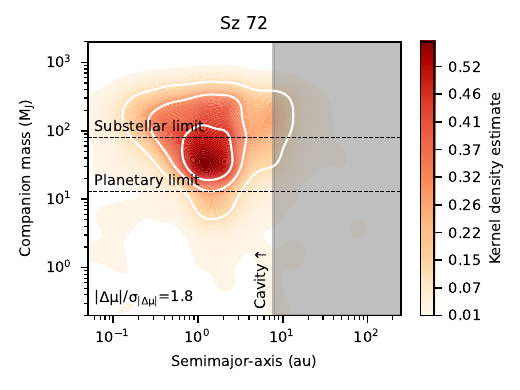}

\end{subfigure}
\caption{continued.}
\end{figure*}

\begin{figure*}
\ContinuedFloat
\begin{subfigure}{1\textwidth}

  \includegraphics[width=0.25\textwidth]{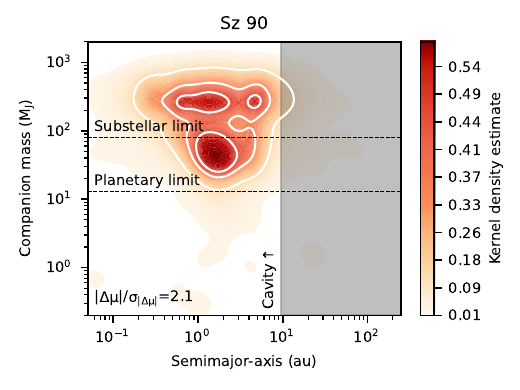}
  \hspace*{-0.15cm}\includegraphics[width=0.25\textwidth]{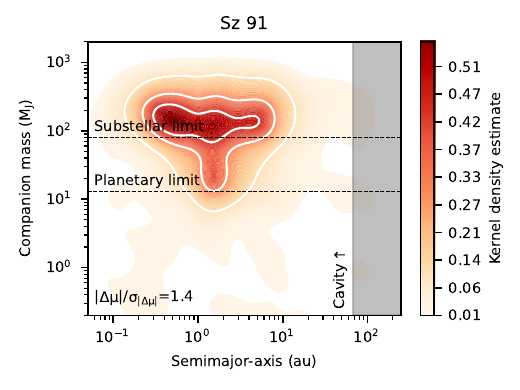}
  \hspace*{-0.15cm}\includegraphics[width=0.25\textwidth]{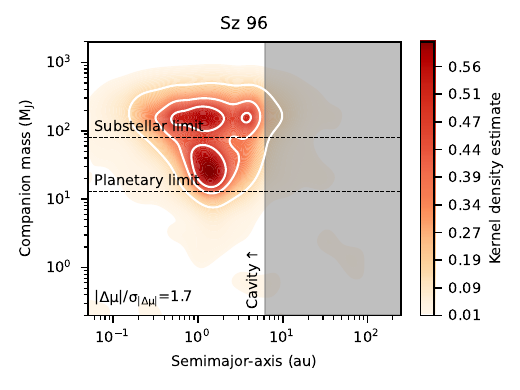}
  \hspace*{-0.15cm}\includegraphics[width=0.25\textwidth]{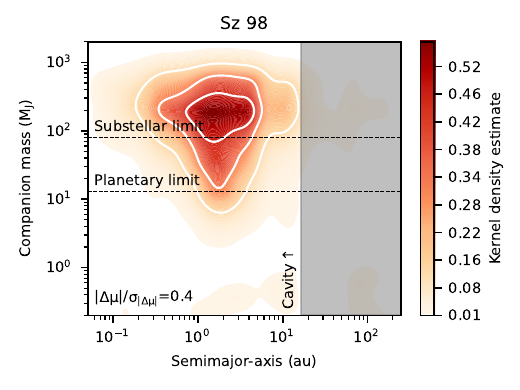}

  \vspace{-5pt} \includegraphics[width=0.25\textwidth]{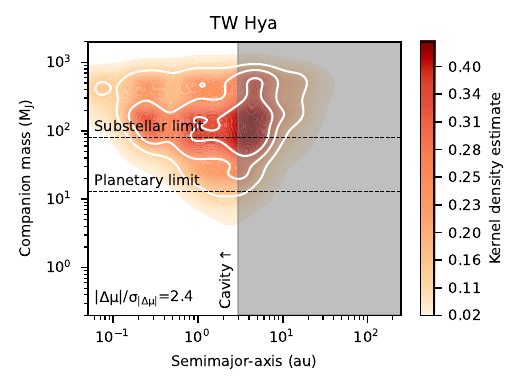}
  \hspace*{-0.15cm}\includegraphics[width=0.25\textwidth]{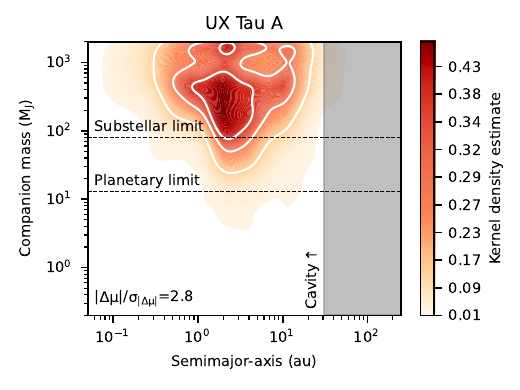}
  \hspace*{-0.15cm}\includegraphics[width=0.25\textwidth]{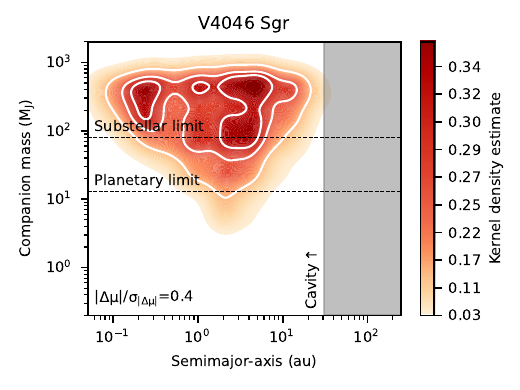}
  \hspace*{-0.15cm}\includegraphics[width=0.25\textwidth]{Plots/TDs_after_referee/WISPIT_2_density.pdf}

  \vspace{-5pt} \includegraphics[width=0.25\textwidth]{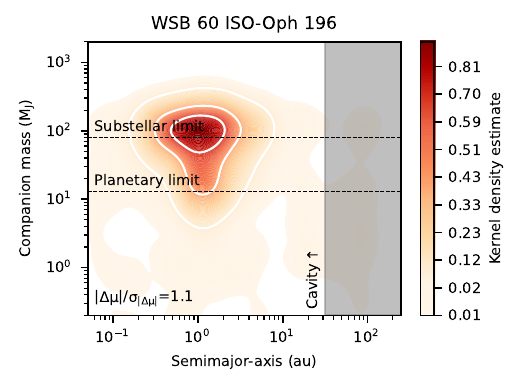}
  \hspace*{-0.15cm}\includegraphics[width=0.25\textwidth]{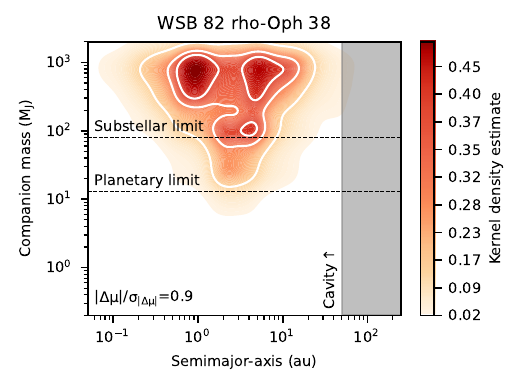}
  \hspace*{-0.15cm}\includegraphics[width=0.25\textwidth]{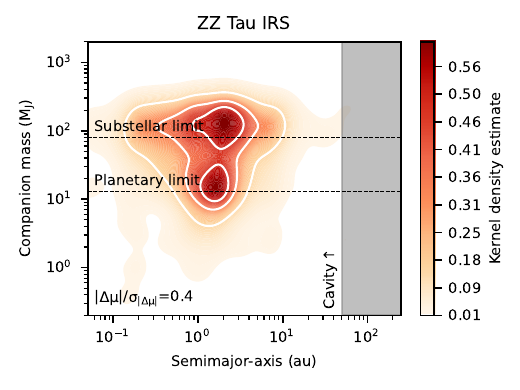}

\end{subfigure}
\caption{continued.}
\end{figure*}

\section{Details of the hydrodynamic simulations}\label{AppendixC}

This Appendix details the hydrodynamical simulations presented in Sect. \ref{S_other_sources} and Fig.~\ref{Richard_plot}.

We investigate the impact of a massive, eccentric disk produced by a companion by conducting 2D locally-isothermal hydrodynamic simulations with FARGO3D \citep{Benitez-Llambay2016}. The disk aspect ratio is set to $h = 0.05 (R/R_{\rm p,0})^{0.25}$, where $R_{\rm p, 0}$ is the planet's initial location and $R$ is the distance from the star. The surface density is set to $\Sigma(R) = \Sigma_0 (R/R_{\rm p, 0})^{-1} \exp(-R/R_{\rm c})$, where $R_{\rm c} = 30\,R_{\rm p,0}$ is used by default. The normalisation, $\Sigma_0$, is set by specifying the Toomre $Q$ parameter at $R_{\rm p, 0}$. For a fiducial value of $Q = 20$, our choices correspond to a disk-to-star mass ratio within the simulation domain of 0.09. We have explored $Q$ in the range $[10, 500]$, $R_{\rm c}$ in the range $[10, 30]\,R_{\rm p,0}$  and companion-to star masses in the range $[0.005, 0.2]$, which corresponds to approximately 5 to 200~$M_{\rm J}$ for a 1~$M_\odot$ star.

The simulations are computed on a grid of size $N_R \times N_\phi = 1500 \times 2048$. The radial cells are logarithmically-spaced from 0.2 $R_{\rm p,0}$ to 30 $R_{\rm p, 0}$. We use an $\alpha$ viscosity with $\alpha=10^{-4}$. We apply damping boundary conditions with a damping zone 1.15 times larger/smaller than the inner/outer radius. The disk self-gravity is included using the scheme of \citet{Baruteau2008}. To ensure that Newton's 3rd law is properly accounted for, we include the indirect term \citep[e.g.][]{Crida2025}. The potentials due to the planet and disk are both softened over a length scale $0.04\,R$, as the self-gravity module requires a softening law $\propto R$. The simulations are run for 1000 orbits at $R_{\rm p,0}$. 

Since the simulations are conducted in a frame centred on the star, the motion of the star is tracked by computing the centre of mass of the system in that frame. The proper motion anomaly is then estimated by sampling the centre of mass at 50 uniformly chosen points over the observational \textbf{time} baseline and computing the average motion during that baseline. To convert the signal to physical units, $R_{\rm p,0}$ is scaled to the appropriate distance and the star is assumed to be at a distance of 150~pc.

\end{appendix}

\end{document}